\documentclass[pteplogo]{ptephy_v1}
\usepackage{afterpage}
\usepackage{lineno}
\usepackage{alphalph}

\preprintnumber{XXXX-XXXX} 

\begin{document}

\title{ Measurement of the $\nu_{\mu}$ charged-current cross sections on water, hydrocarbon, iron, and their ratios with the T2K on-axis detectors  }

	\author[51]{ {K.} {Abe}}
\affil{University Autonoma Madrid, Department of Theoretical Physics, 28049 Madrid, Spain}
	\author[52]{ {R.} {Akutsu}}
\affil{University of Bern, Albert Einstein Center for Fundamental Physics, Laboratory for High Energy Physics (LHEP), Bern, Switzerland}
	\author[22]{ {A.} {Ali}}
\affil{Boston University, Department of Physics, Boston, Massachusetts, U.S.A.}
	\author[49,30]{ {C.} {Andreopoulos}}
\affil{University of British Columbia, Department of Physics and Astronomy, Vancouver, British Columbia, Canada}
	\author[30]{ {L.} {Anthony}}
\affil{University of California, Irvine, Department of Physics and Astronomy, Irvine, California, U.S.A.}
	\author[17]{ {M.} {Antonova}}
\affil{IRFU, CEA Saclay, Gif-sur-Yvette, France}
	\author[27]{ {S.} {Aoki}}
\affil{University of Colorado at Boulder, Department of Physics, Boulder, Colorado, U.S.A.}
	\author[2]{ {A.} {Ariga}}
\affil{Colorado State University, Department of Physics, Fort Collins, Colorado, U.S.A.}
	\author[28]{ {Y.} {Ashida}}
\affil{Duke University, Department of Physics, Durham, North Carolina, U.S.A.}
	\author[54]{ {Y.} {Awataguchi}}
\affil{Ecole Polytechnique, IN2P3-CNRS, Laboratoire Leprince-Ringuet, Palaiseau, France }
% ...



%
%
%


\begin{abstract}%
	We report a measurement of the flux-integrated $\nu_{\mu}$ charged-current cross sections on water, hydrocarbon, and iron in the T2K on-axis neutrino beam with a mean neutrino energy of 1.5~GeV. The measured cross sections
	on water, hydrocarbon, and iron are $\sigma^{\rm{H_{2}O}}_{\rm{CC}}$ = (0.840$\pm 0.010$(stat.)$^{+0.10}_{-0.08}$(syst.))$\times$10$^{-38}$~cm$^2$/nucleon, $\sigma^{\rm{CH}}_{\rm{CC}}$ = (0.817$\pm 0.007$(stat.)$^{+0.11}_{-0.08}$(syst.))$\times$10$^{-38}$~cm$^2$/nucleon, and $\sigma^{\rm{Fe}}_{\rm{CC}}$ = (0.859$\pm 0.003$(stat.) $^{+0.12}_{-0.10}$(syst.))$\times$10$^{-38}$~cm$^2$/nucleon respectively, for a restricted phase space of induced muons: $\theta_{\mu}<45^{\circ}$ and $p_{\mu}>$0.4~GeV/$c$ in the laboratory frame. The measured cross section ratios are ${\sigma^{\rm{H_{2}O}}_{\rm{CC}}}/{\sigma^{\rm{CH}}_{\rm{CC}}}$ = 1.028$\pm 0.016${(stat.)}$\pm 0.053$(syst.), ${\sigma^{\rm{Fe}}_{\rm{CC}}}/{\sigma^{\rm{H_{2}O}}_{\rm{CC}}}$ = 1.023$\pm 0.012$(stat.)$\pm 0.058$(syst.), and ${\sigma^{\rm{Fe}}_{\rm{CC}}}/{\sigma^{\rm{CH}}_{\rm{CC}}}$ = 1.049$\pm 0.010$(stat.)$\pm 0.043$(syst.).
These results, with an unprecedented precision for the measurements of neutrino cross sections on water in the studied energy region, show good agreement with the current neutrino interaction models used in the T2K oscillation analyses.

\end{abstract}

\subjectindex{C04, C32}

\maketitle

\clearpage
\section{Introduction}

The Tokai-to-Kamioka (T2K) experiment~\cite{Abe:2011ks} is a long-baseline neutrino oscillation experiment that started taking physics data in 2010. The T2K experiment studies properties of neutrino oscillations via disappearance of muon (anti-)neutrinos and appearance of electron (anti-)neutrinos from a nearly pure muon (anti-)neutrino beam, which is produced by the J-PARC accelerator complex. The neutrino beam characteristics and neutrino-nucleus interactions are measured with a suite of near detectors, which are situated 280~m from the production target, consisting of the so-called INGRID~\cite{Abe:2011xv} and ND280~\cite{pod,fgd,Abgrall:2010hi,Allan:2013ofa,Aoki:2012mf}. The INGRID is placed at the center of the neutrino beam (on-axis), while the ND280 is at an off-axis of 2.5$^{\circ}$. The neutrino oscillation patterns are observed with the 2.5$^{\circ}$ off-axis far detector, Super-Kamiokande~\cite{Fukuda:2002uc}, which is located 295~km away from the production target.
In order to precisely measure neutrino oscillations, understanding of the neutrino interactions with nuclei is essential.
In the current T2K oscillation analysis~\cite{Abe:2018wpn}, data samples of charged-current candidates in which the interaction vertex is found in one of two Fine-Grained Detectors, FGD1 or FGD2~\cite{fgd}, are used to constrain the neutrino flux prediction and cross section models.
The former detector consists of 100\% plastic scintillators (hydrocarbon) and the latter consists of a mixture of plastic scintillators and water, while the far detector consists of 100\% water. 
The neutrino interaction model is used to extrapolate the near detector spectra to the (oscillated) far detector spectra in a few significant ways. First, the T2K off-axis near detector angular acceptance is more limited than the far detector. Second, the near detector event rate also includes significant interactions on materials other than the far detector (water) target. Finally, the interaction model is tuned at the near detector to predict the far detector energy spectra and this parameterization can be incomplete. Therefore testing the interaction model with different target materials and at various range of the neutrino energies are valuable to the T2K oscillation analysis.
However, there are only a few publications of the neutrino cross sections on water so far~\cite{cc1pifgd,cc0pipod,scifi}. Two exclusive channels of charged current interactions are measured by the ND280~\cite{cc1pifgd,cc0pipod} with approximately 15\% uncertainties with a mean neutrino energy of 0.6~GeV. There is only one measurement of axial vector mass~\cite{scifi} with 10\% uncertainty with a mean neutrino energy above 1~GeV.

A new water-target neutrino detector, named the Water Module~\cite{Koga:2015iqa}, has been constructed for the precise measurements of neutrino interactions on water with a mean neutrino energy of 1.5~GeV.
In this article, by using the Water Module and the other T2K detectors including the Proton Module~\cite{ccincpm} and INGRID~\cite{Abe:2011xv}, we measure the $\nu_{\mu}$ charged-current (CC) cross sections on water, hydrocarbon, iron, and their ratios. 
Dominant errors of the absolute cross section measurements come from the uncertainty of the T2K neutrino beam prediction, which largely cancels out when performing measurements on their cross section ratios. 
This method was established in the previous measurement of a cross section ratio between hydrocarbon and iron by using the Proton Module and INGRID~\cite{ccincpm}. In this article, measurements of neutrino interaction on water with the Water Module are conducted for the first time.
In addition, in order to reduce the dependence on the Monte Carlo implemented model of neutrino-nucleus interactions in extracting the cross section values, a method for unfolding the total cross section as a function of muon scattering angles is implemented. Hereafter, we will describe the detector configuration, the Monte Carlo simulation, the used data sample, the event selection, the method to extract the cross sections, systematic uncertainties, and the results.

\section{Detector configuration}

\begin{figure}[htb]
   \begin{minipage}{0.50\hsize}
           \centering
           \includegraphics[width=5.0cm, bb=100 50 580 509]{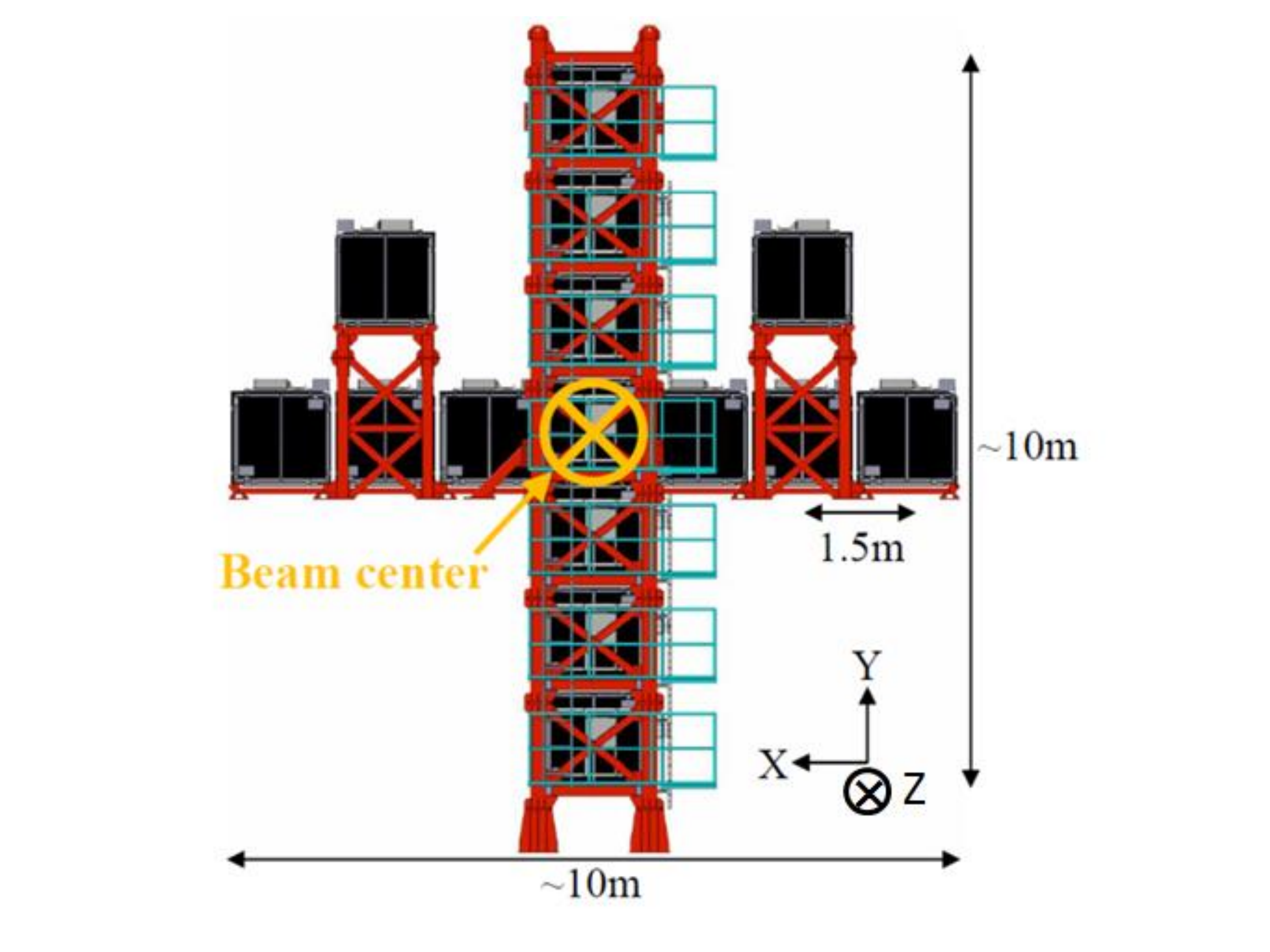}
   \end{minipage}
   \begin{minipage}{0.50\hsize}
    \begin{center}
           \includegraphics[width=5.0cm, bb=100 50 580 509]{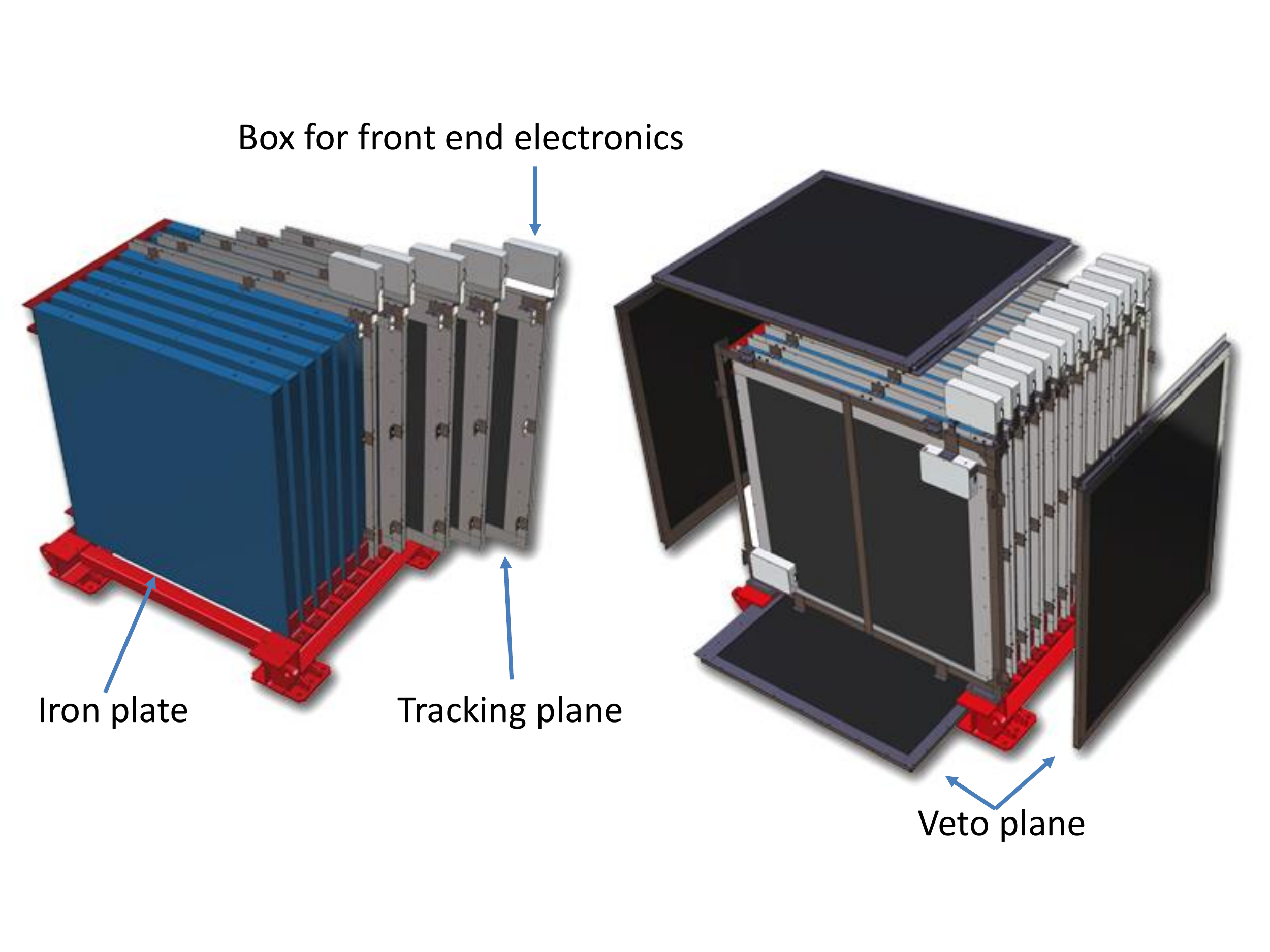}
    \end{center}
   \end{minipage}
   \caption{ Schematic view of the INGRID detector (left), and one of the modules (right). The coordinate system used in this article is shown in the left figure. }
  \label{ fig:INGRID }
\end{figure}

We use the three detectors, INGRID, Proton Module, and Water Module as iron (Fe), hydrocarbon (CH), and water ($\rm H_{2}O$) interaction targets, respectively. Table~\ref{ tab:wmperformance } shows specifications of the three detectors.
INGRID consists of fourteen identical modules arranged in a cross shape; each module has a sandwich structure comprised of nine iron planes and eleven tracking planes as shown in Fig.~\ref{ fig:INGRID }. INGRID has been operating since 2009 to monitor the neutrino beam rate, its direction, and stability in real-time.
The tracking planes are formed from two layers of scintillator, each of which is composed of twenty-four bars that are oriented either horizontally or vertically. The thickness of the iron planes is 6.5~cm and the thickness of the scintillator is 1.0~cm. The iron planes, which play a role as the neutrino interaction target in this analysis, make up 96\% of the total fiducial mass of the module. There are veto planes surrounding the module designed for tracking the charged particles entering into the detector. More detailed information about the INGRID can be found in \cite{Abe:2011xv}. In this analysis, the central horizontal INGRID module is used as the iron target. The three horizontal INGRID modules surrounding the beam center are used for muon identification for the Proton Module and Water Module.

\begin{figure}[htb]
    \begin{tabular}{c}
     \begin{minipage}{0.50\hsize}
     	  \hspace*{-10mm}
             \includegraphics[width=8.0cm, bb=0 0 692 569]{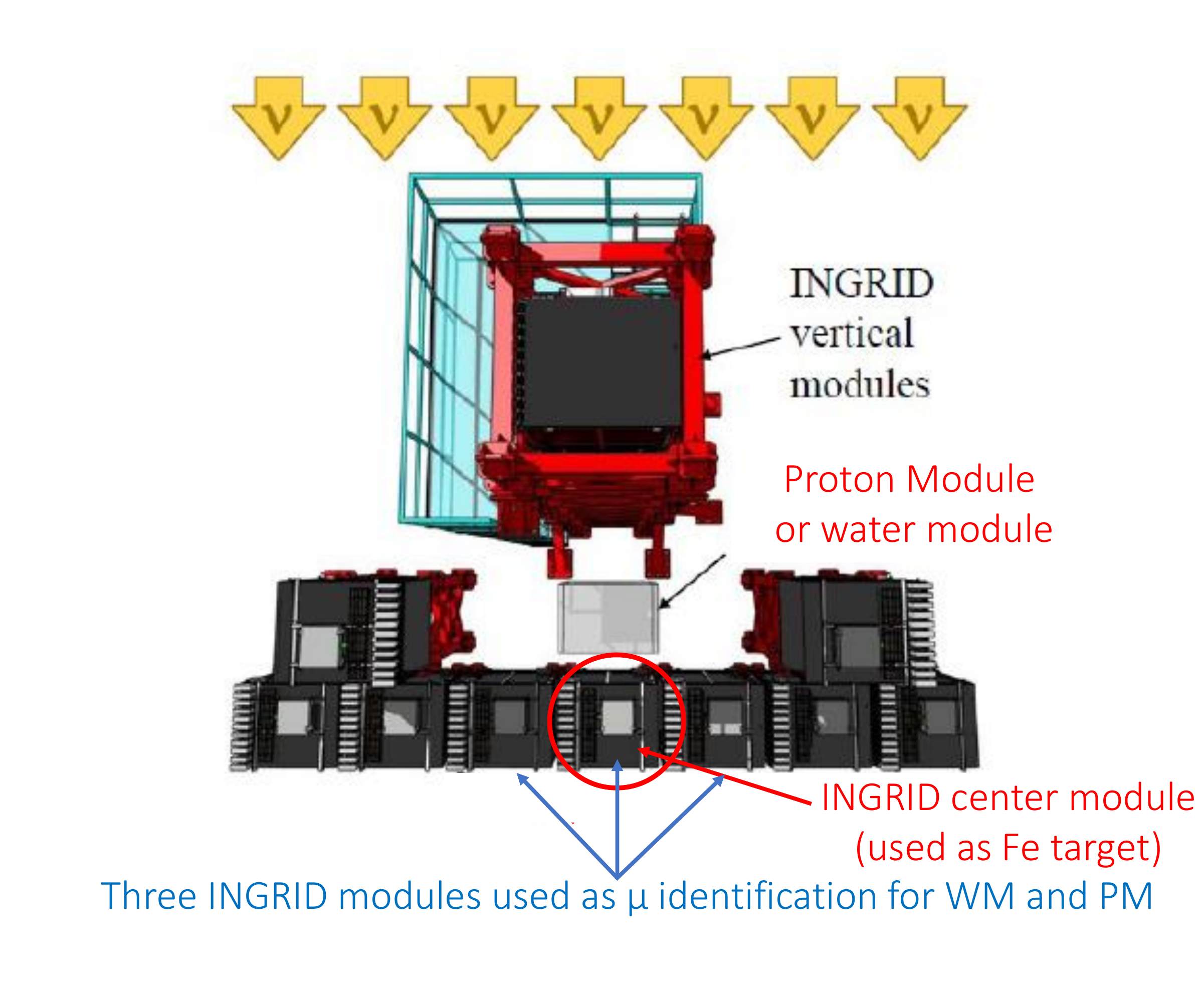}
     \end{minipage}
     \begin{minipage}{0.50\hsize}
       \begin{center}
             \includegraphics[width=8.0cm, bb=0 0 692 569]{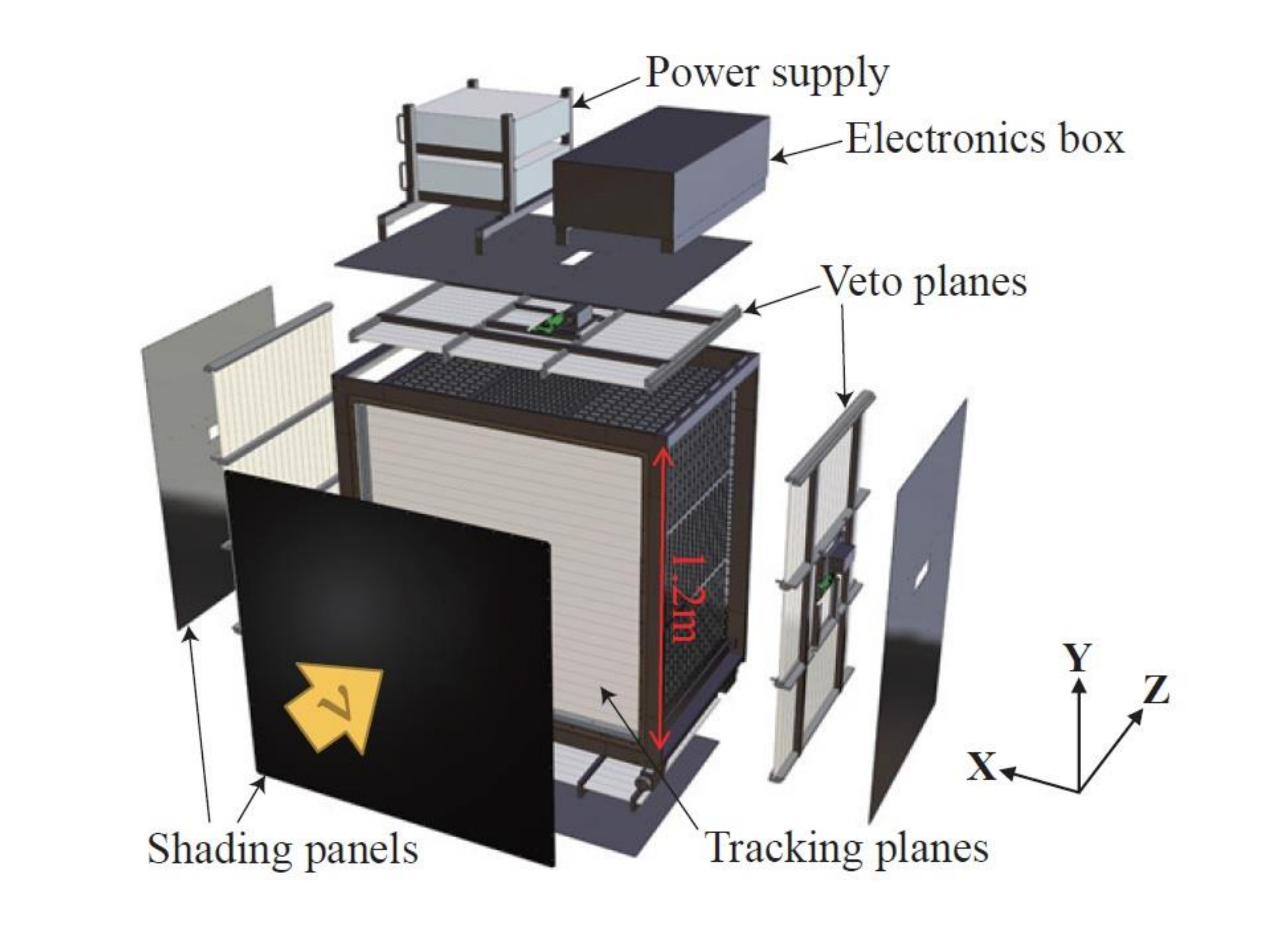}
       \end{center}
     \end{minipage}
    \end{tabular}
	\caption{ Top view of the Water Module, Proton Module, and INGRID (left) and schematic view of the Proton Module (right)~\cite{ccincpm}. }
	\label{fig:INGRID_view}
\end{figure}

The Proton Module is a plastic scintillator target detector located between the horizontal and vertical INGRID modules, as shown in Fig.~\ref{fig:INGRID_view}. It was built for the measurement of the neutrino cross section on hydrocarbon and it had been located at the on-axis position from November 2010 to May 2016.
It consists of thirty-four tracking planes with each plane being an array of thirty-two scintillator bars that are oriented either horizontally or vertically. Two types (SciBar-type and INGRID-type) of scintillator bars, which have different sizes, are used in the inner and outer sections of each tracking plane. Hydrocarbon in the scintillators of the tracking planes serves as the neutrino interaction target and composes 98\% of the total fiducial mass of the Proton Module. Similar to the INGRID modules, the Proton Module is composed of veto planes surrounding the tracking planes of the detector.
More detailed information about the Proton Module can be found in \cite{ccincpm}.

\begin{figure}[!htb]
\begin{minipage}{0.50\hsize}
        \centering
	\hspace*{-25mm}
        \includegraphics[width=9.0cm, bb=0 0 692 569]{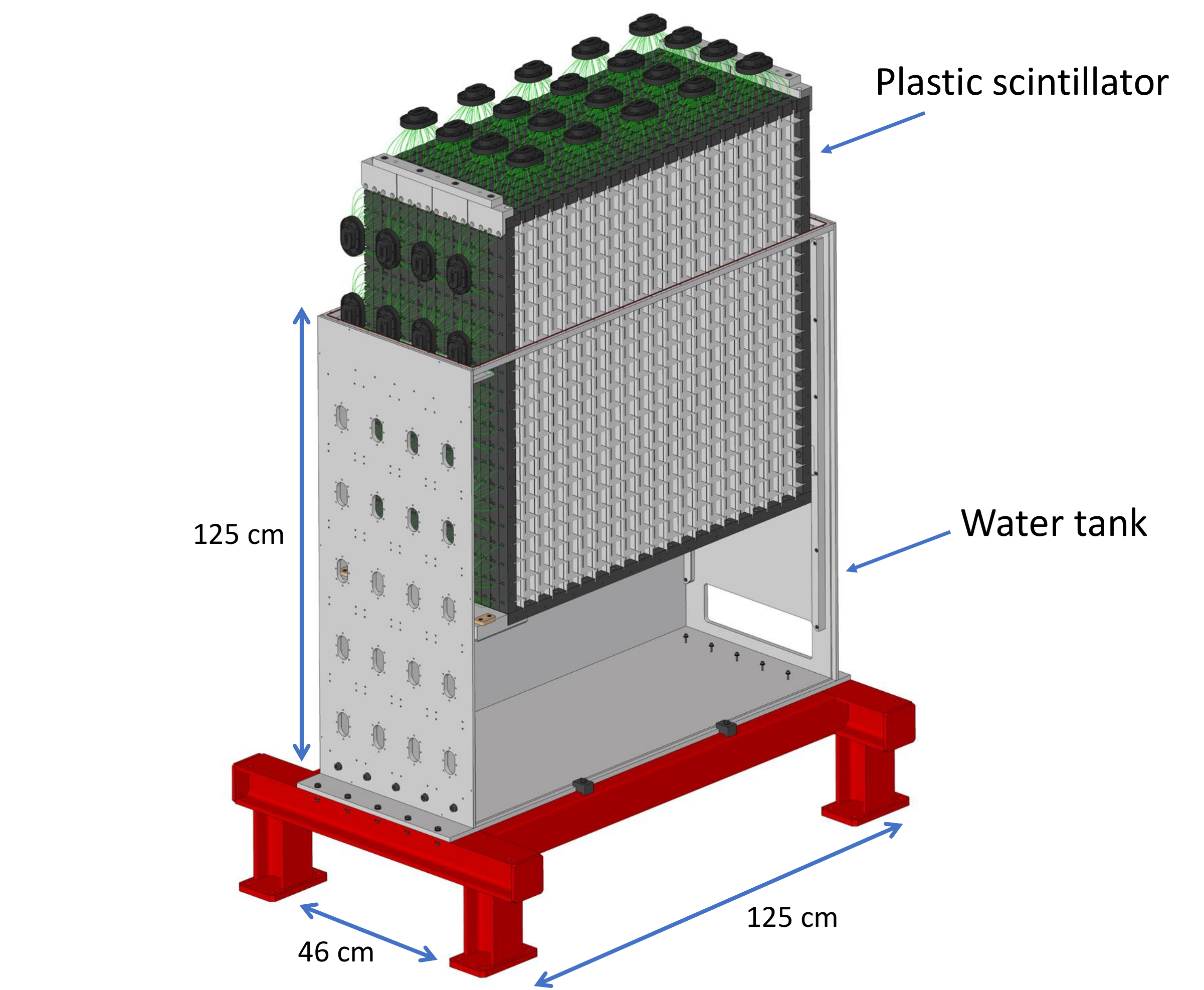}
\end{minipage}
\begin{minipage}{0.50\hsize}
        \centering
	\hspace*{-10mm}
        \includegraphics[width=11.0cm, bb=0 0 667 314]{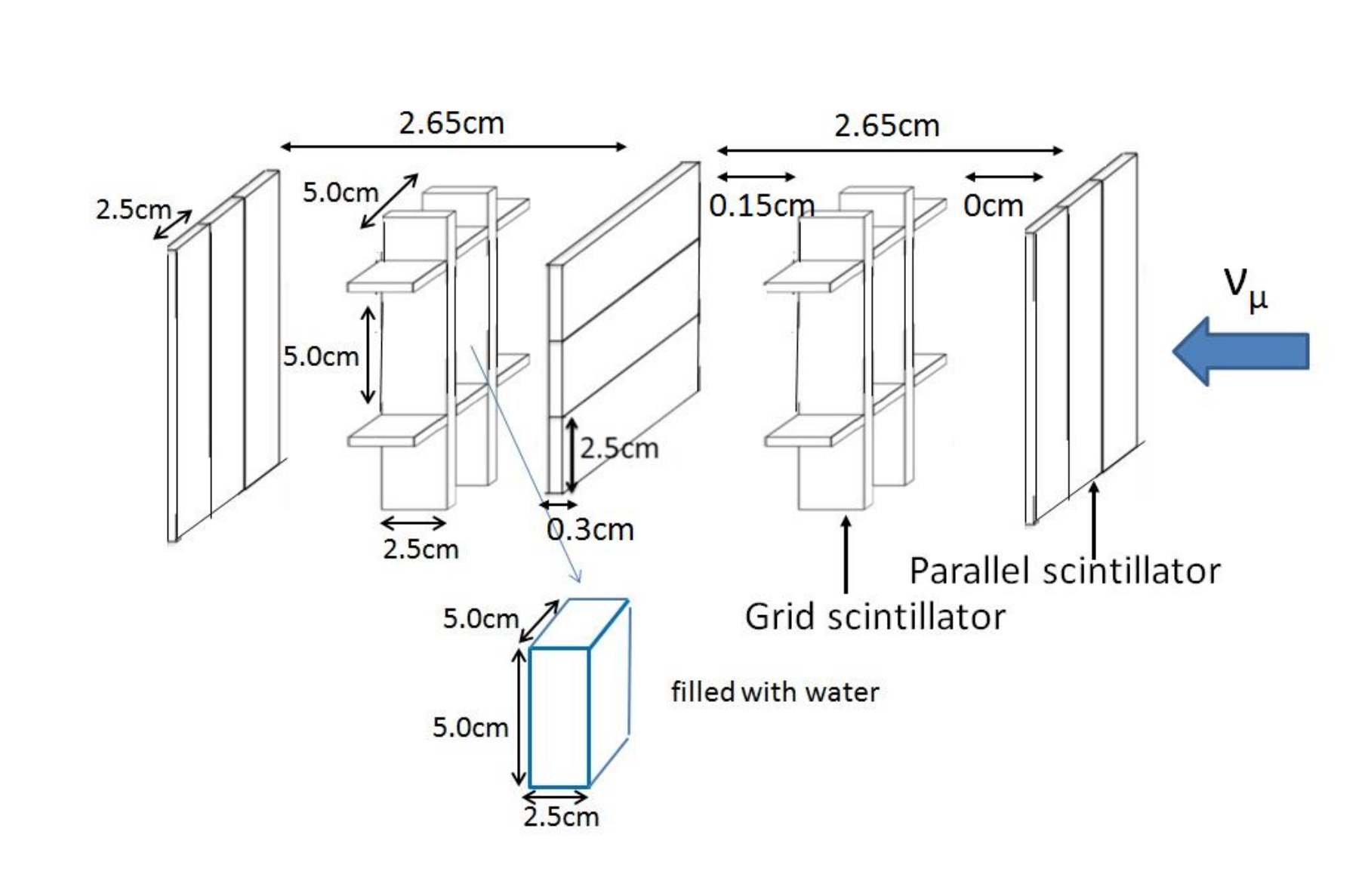}
\end{minipage}
   \caption{ Schematic view of the Water Module (left) and the layout of scintillators (right). }
   \label{ fig:watermodule }
\end{figure}

The Water Module is a neutrino detector with the interaction target region composed of 80\% water and 20\% plastic scintillators. The high fraction of water in the detector, in fact higher than the previous water-target neutrino detectors~\cite{pod,fgd}, is essential to reduce the backgrounds induced by the neutrino interactions on non-water materials.
The Water Module has been located at the on-axis position between the INGRID horizontal modules and vertical modules since June 2016, replacing the Proton Module.
The Water Module consists of a stainless steel tank filled with water and sixteen scintillator tracking planes immersed in the water, as shown in Fig.~\ref{ fig:watermodule }. The eight tracking planes are placed alternately in the {\it x}-direction and {\it y}-direction along the {\it z}-direction so that three-dimensional tracks may be reconstructed. 
Each tracking plane is an array of eighty scintillator bars.   
The forty bars, called parallel scintillators, are placed along the {\it xy}-direction. The other forty bars, called grid scintillators, are placed along the {\it z}-direction with a grid-like structure in order to achieve a large angular acceptance. 
The plastic scintillators of dimension 100~cm (length) $\times$ 2.5~cm (width) $\times$ 0.3~cm (thickness) were produced in the Fermi National Accelerator Laboratory~\cite{Beznosko:2005ba}. The scintillators are made of polystyrene, infused with PPO (1\%) and POPOP (0.03\%). The manufactured scintillator, co-extruded with a white reflective coating of $\rm{TiO_{2}}$ infused in polystyrene, has a rectangular cross section with a groove to house a wavelength shifting (WLS) fiber (Kuraray Y-11~\cite{fiber}).
The WLS fiber is glued onto the scintillator with optical cement (ELJEN TECHNOLOGY EJ-500~\cite{cement}). The surface of the scintillator is painted by a black cement of acrylic silicon to prevent optical crosstalk between the scintillators. 
Each layer of scintillator bars is affixed to a mechanical frame which sits inside a water tank.
Spaces between scintillators are filled with water.
 Scintillation light from the scintillator is collected by the WLS fiber and detected by a multi-pixel photon counter (MPPC)~\cite{mppc}, similar to that for the INGRID and Proton Module.
 While the Hamamatsu S10362-13-050C MPPC was used in the INGRID and Proton Module, a newer type of MPPC, S13660 with higher gain, lower noise rate, crosstalk rate, and after-pulse rate, is used in the Water Module. The same Trip-t electronics~\cite{Tript2007} are used for all the three detectors. 
 To record data from the neutrino beam, delivered typically in eight bunches with a cycle of 581~ns for each 2.48~sec, a trigger from the J-PARC accelerator is provided to each detector. The integrated charge and hit timing of all channels are digitized and recorded with 2.5~photoelectrons (p.e.) threshold for each beam bunch. 


\begin{table}[!htb]
	\begin{center}
		\caption{ Summary of detector specifications of the Water Module, Proton Module, and one of the INGRID modules. The target masses are calculated inside the fiducial volumes, which correspond to the effective target masses and are specially tuned for this analysis, as described in Sec.~\ref{ sec:eventselec }.  }
		\label{ tab:wmperformance }
     	  \hspace*{-5mm}
		\scalebox{0.80}{
		\begin{tabular}{llll}
			\hline
			\hline
		        Parameter                                    &   Water Module                &  Proton Module & INGRID module             \\
			\hline
			Target mass in fiducial volume (ton)         &   0.10                        &  0.16            & 2.1               \\
			Main target materials and fraction           &   $\rm H_{2}O$ (80\%), $\rm CH$ (19\%)  &  CH (98\%)      &  Fe (96\%)        \\
			Dimension of a scintillator ($\rm cm^{3}$)   &   100$\times$2.5$\times$0.3   & 120$\times$2.5$\times$1.3     (SciBar-type), & 120$\times$5$\times$1   \\
			                                             &                               & 120$\times$5$\times$1 (INGRID-type)  &                         \\
		        Dimension of an iron plane ($\rm cm^{3}$)    &   -                           & -                                  & 124$\times$124$\times$6.5   \\
			The number of readout channels               &   1280                        & 1204           & 616       \\
			MPPC serial number            &   S13660     & S10362-13-050C & S10362-13-050C      \\
			MPPC gain stability                          &   10\%                        & 10\%           & 10\%      \\
			MPPC dark noise rate (hits/module/bunch)     &   0.2                         & 12             &  6        \\
			Mean scintillator light yield for MIP        &   16                          & 56 (SciBar-type),& 23   \\
			(p.e. per scintillator thickness)            &                               & 23 (INGRID-type) &      \\
		        Angular acceptance respect to beam axis      &   0$^{\circ}$ to 90$^{\circ}$ & 0$^{\circ}$ to 75$^{\circ}$           & 0$^{\circ}$ to 60$^{\circ}$          \\
			Period located at on-axis position            &   July 2016-     & November 2010-May 2016 & 2009-      \\
			\hline
			\hline
		\end{tabular}
		}
	\end{center}
\end{table}

\section{Monte Carlo simulation} \label{ sec:mc }

\begin{figure}[!htb]
\begin{minipage}{0.50\hsize}
        \centering
	\hspace*{-15mm}
        \includegraphics[width=6.0cm,  bb=20 0 405 350]{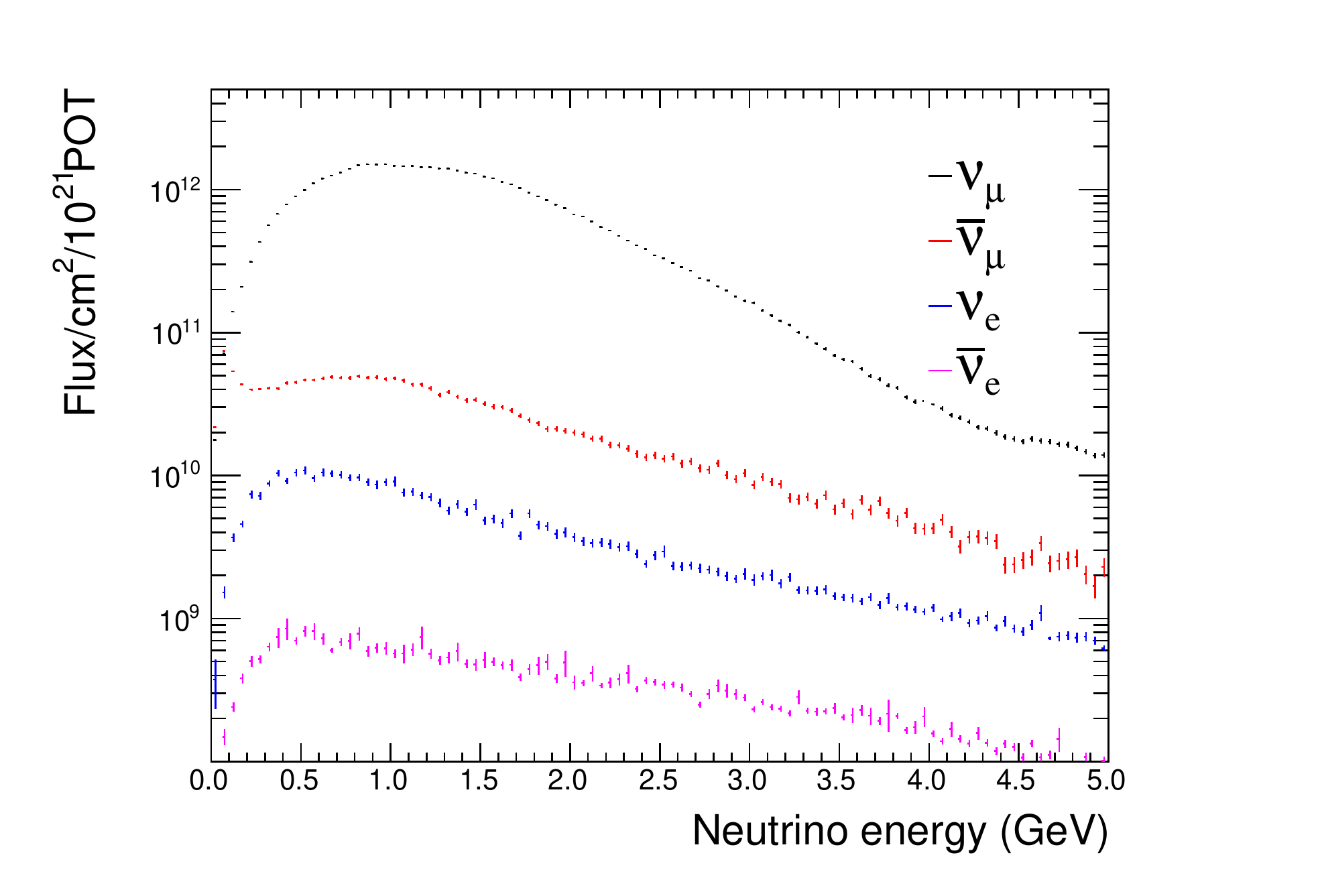}
\end{minipage}
\begin{minipage}{0.50\hsize}
        \centering
	\hspace*{-10mm}
        \includegraphics[width=6.0cm,  bb=20 0 405 350]{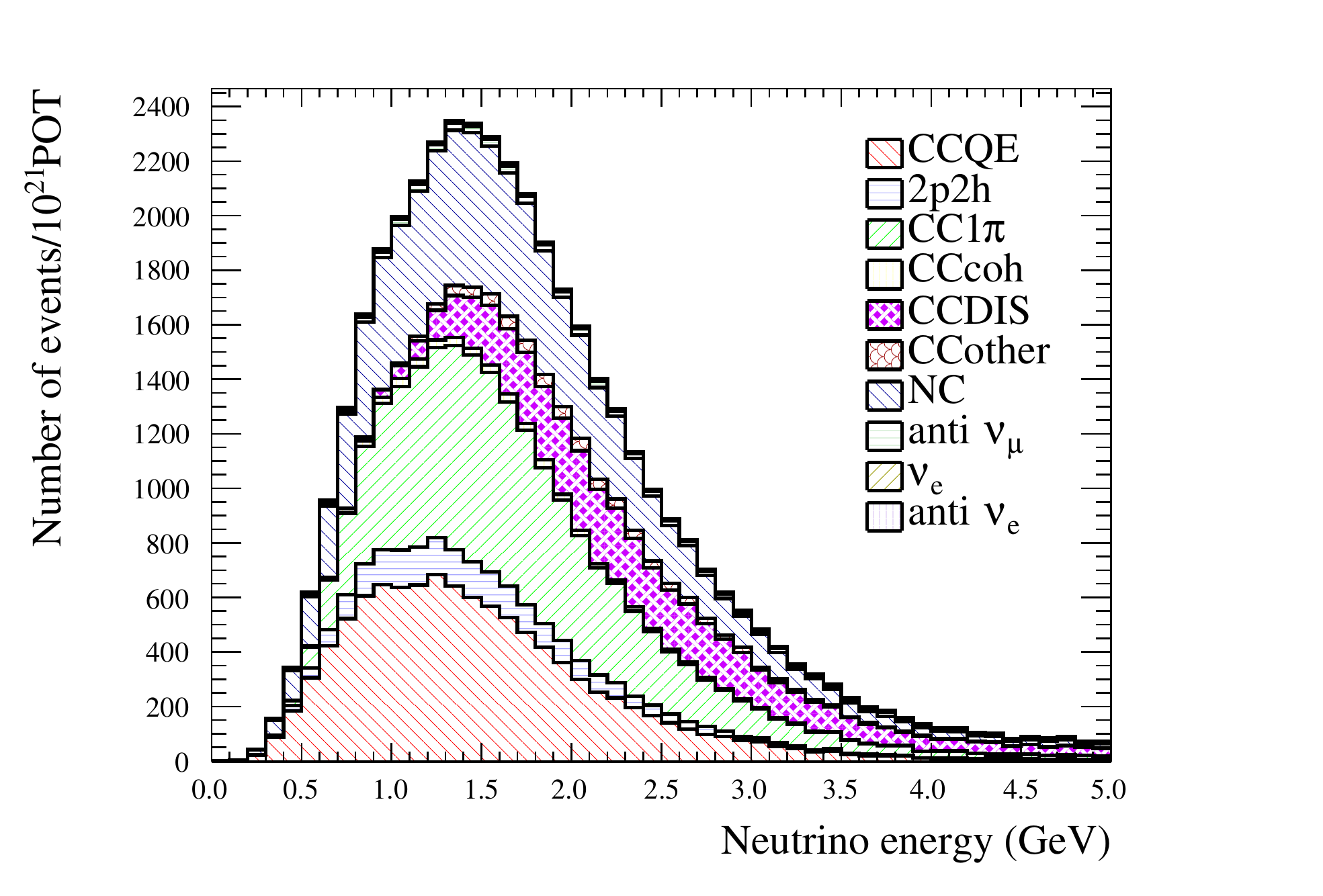}

\end{minipage}
\caption{ Neutrino flux per $10^{21}$ Protons on Target (POT) predicted by JNUBEAM in the muon neutrino beam mode at the position of the simulated Water Module (left) and the energy of neutrinos that interact with the $\rm H_{2}O$ target inside the fiducial volume of the Water Module predicted by NEUT version 5.3.3 (right). In the right figure, the category of CCDIS includes both CC multi-pion and DIS production. }
  \label{ fig:flux }
\end{figure}

A Monte Carlo (MC) simulation is used for the estimation of background contamination and signal detection efficiency. Three pieces of software are used for the chain of simulation: JNUBEAM~\cite{jnubeam} for the neutrino flux prediction, NEUT~\cite{NEUT} for the neutrino interactions with nuclei, and a GEANT4~\cite{geant4}-based detector simulation. JNUBEAM simulates the interaction of 30~GeV primary protons on a graphite target, the propagation of the secondary and tertiary produced mesons in the magnetic fields induced by the magnetic horns and their decays in the decay volume.
The simulation uses the proton beam profiles measured by the J-PARC neutrino beam line and is tuned with external hadron production measurements, mainly from the NA61/SHINE experiment~\cite{na61det,na612009}.
We can select either a muon neutrino beam or a muon anti-neutrino beam by changing the current polarity of the focusing magnetic horns. In this analysis, data collected in the former beam configuration is used.
The simulated on-axis neutrino beam has a mean energy of 1.5~GeV and a 1$\sigma$ standard deviation between $-$0.75~GeV and +0.85~GeV, as shown in Fig.~\ref{ fig:flux }. 

\begin{figure}[!htb]
   \vspace*{-50mm}
   \hspace*{+30mm}
   \centering
   \includegraphics[width=15.0cm, bb=0 0 691 434]{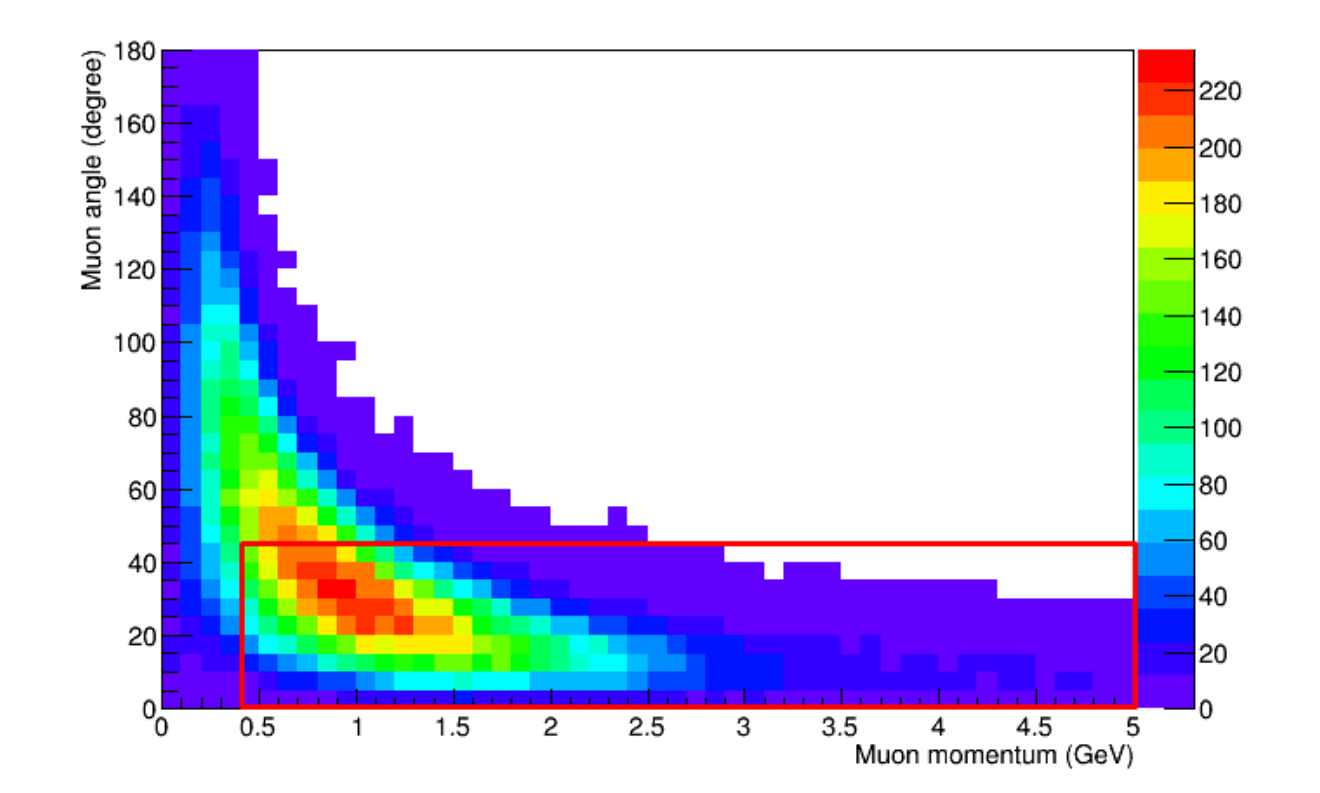}
   \caption{ Scattering angle and momentum of muons produced by CC interactions on the $\rm H_{2}O$ target predicted by NEUT. The red rectangle highlighted includes the signal region where $p_{\mu}<45^{\circ}$ and $p_{\mu}>$0.4~GeV/$c$.} 
   \label{ fig:muon_ptheta }
\end{figure}

For a given flux of incoming neutrinos, NEUT simulates the neutrino interactions with nuclei, including initial and final state interactions inside the nuclei, in order to provide the four-momenta of all induced particles. In this analysis, the version 5.3.3 of NEUT is used. CC quasi elastic (CCQE)-like, neutral-current (NC) elastic, CC and NC single pion production (1$\pi$), deep inelastic scattering (DIS), multi-pion production, and coherent interactions are simulated. The CCQE-like interactions, characterized by the inclusion of a single charged lepton and no mesons in the final state, are simulated with a relativistic Fermi Gas model (RFG)~\cite{cite:rfg}, random phase approximation (RPA)~\cite{cite:rpa}, and multi-nucleon (2p2h) interactions~\cite{cite:mecnieves}. 
In addition to the nominal NEUT model, we test the sensitivity of the analysis to determining alternate, available models~\cite{cite:sf}.
Table~\ref{ tab:neut_model } shows the nominal settings for each of the interaction models and tunable parameters in NEUT in this analysis. 
More details about the underlying neutrino interaction models implemented in NEUT that are used in the analysis can be found in \cite{Abe:2017vif}.
Figure~\ref{ fig:flux } shows the energy of neutrinos that interacted with the target nuclei of the Water Module simulated by NEUT. The main modes of the CC interactions are CCQE\footnote{Here, 2p2h interactions are not included.}, CC1$\pi$, CC multi-pion and DIS production. The fraction of NC interaction is 30\% of all interactions. 
Figure~\ref{ fig:muon_ptheta } shows the momentum and scattering angle distributions in the laboratory frame for muons produced by $\nu_{\mu}$ CC interactions. 
In this analysis, due to the limited acceptance of the horizontal INGRID modules to be used for muon identification for the Water Module and Proton Module as described in Sec.~\ref{ sec:eff }, we define the signal with a restricted phase space of muon kinematics, particularly CC interactions with $\theta_{\mu}<$45$^{\circ}$ and $p_{\mu}>$0.4~GeV/$c$ in the laboratory frame. 
The cross section of the signal per nucleon is predicted by NEUT to be slightly different amongst $\rm H_{2}O$, CH, and Fe, as shown in Table~\ref{ tab:xsecexp }. This is due to the target dependence of the total cross section of the CC coherent interaction, which is proportional to the square of atomic number, and the difference in the fraction of neutrons and protons per nucleus for the targets considered.

GEANT4 simulates the behavior of the secondary particles induced by the neutrino-nucleus interactions in the detector. The version v9r2p01n00 of GEANT4 and the physics list of QGSP BERT are used for the simulation. The geometry of the three detectors and the walls of the detector hall are modeled in GEANT4 based on the measurements performed during the detector construction. Responses of scintillator, MPPC, and electronics are modeled based on the measurements, as shown in Table~\ref{ tab:wmperformance }. The energy deposited in the scintillators estimated by GEANT4 is converted to the observed number of p.e.\ by multiplying it by a constant determined from measurements with minimum ionization particles (MIP), performed during the detector operation. 
The following effects are taken into account: the quenching effect of the scintillator; position-dependent light collection efficiency of WLS fibers; attenuation and propagation time of the light in the WLS fiber; crosstalk between grid scintillators; MPPC noise; MPPC crosstalk and after-pulses; MPPC saturation; noise from electronics; gate width of the electronics; and statistical fluctuation of photon counting. For the physics analysis, the neutrino flux and interactions on detector targets, plastic scintillators, and main mechanical structures of the detector and the walls of the detector hall are simulated for the three detectors. Backgrounds from cosmic rays are negligible, as described in Sec.~\ref{ sec:sys_det }, and are not simulated for the physics analysis.

\begin{table}[!htb]
	\centering
	\caption{Nominal models of the neutrino-nucleus interactions implemented in NEUT used in this analysis.}
		\label{ tab:neut_model }
	\hspace*{-10mm}
	\scalebox{0.90}{	\begin{tabular}{lll}
			\hline
			\hline
			Mode     &  Nominal model                                                         & Parameter                                            \\
			\hline
			\hline
			CCQE-like            &  Dipole type axial form factor                                             &  $M_{A}^{QE}=1.15 $~GeV/$c^{2}$.                               \\ \cline{2-3}
			                     &  RFG model by Smith-Moniz~\cite{cite:rfg}                 &  $E_{b}$= 25, 27, 33~MeV and                         \\
			                     &  with binding energy ($E_{b}$) and Fermi surface momentum ($p_{F}$)        &  $p_{F}$= 217, 225, 250~MeV/c for                    \\
			                     &                                                                            &  $^{12}$C , $^{16}$O, and $^{56}$Fe, respectively.   \\ \cline{2-3}
					     &  RPA model by Nieves {\it et al.}~\cite{cite:rpa}        &  RPA is applied for $^{12}$O and $^{16}$C.           \\
			                     &                                                                                             &  RPA is not applied for $^{56}$Fe.                   \\ \cline{2-3}
					     &  2p2h model by Nieves {\it et al.}~\cite{cite:mecnieves}  &  Normalization                                       \\
			\hline
			1$\pi$               &  Model by Rein-Sehgal~\cite{cite:reinsehgal}                                  & $C^{A}_{5}(0)=1.01$,                         \\
			                     &                                                                            & $M_{A}^{RES}=0.95$~GeV/$c^{2}$,   \\
			                     &                                                                            & $\rm Isospin \frac{1}{2} bg= 1.30$.  \\
			\hline
			DIS                  &  PYTHIA~\cite{cite:pytia},  Parton distribution function by                & Energy dependent normalization                  \\
		 	                     &  GRV98 with Bodek and Yang correction~\cite{Gluck:1998xa}~\cite{Bodek:2003wc}~\cite{Bodek:2004pc}                   &                                                      \\
			\hline
			Coherent             &  Model by Berger-Sehgal~\cite{cite:cccohbarbarseghal}                       & Normalization                                                      \\
			\hline
			\hline
		\end{tabular} }
\end{table}

\begin{table}[!htb]
	\centering
	\caption{ Flux-integrated CC cross sections per nucleon for $\nu_{\mu}$ on Fe, CH, and $\rm H_{2}O$ simulated by NEUT. Neutrino interaction parameters used for the simulation are listed in Table~\ref{ tab:neut_model }. Because RPA for Fe is not implemented in NEUT at present, the expectation of $\sigma_{Fe}$ with RPA is not listed. }
		\label{ tab:xsecexp }
		\begin{tabular}{cccc}
			\hline
			\hline
			  Cross section & NEUT expectation with RPA & NEUT expectation without RPA                       \\
			\hline
			
			 $\sigma_{\rm H_{2}O}$                       &  $  0.819 \times 10^{-38} ~\rm{ cm^{2}}$  &  $  0.860 \times 10^{-38} ~\rm{ cm^{2}}$        \\
			 $\sigma_{\rm CH}$                           &  $  0.832 \times 10^{-38} ~\rm{ cm^{2}}$  &  $  0.875 \times 10^{-38} ~\rm{ cm^{2}}$        \\
			 $\sigma_{\rm Fe}$                           &     not available                         &  $  0.904 \times 10^{-38} ~\rm{ cm^{2}}$       \\
			 $\sigma_{\rm H_{2}O}/\sigma_{\rm CH}$       &  0.984                                    &  0.983                                         \\
			 $\sigma_{\rm Fe}/\sigma_{\rm H_{2}O}$       &     not available                         &  1.051                                         \\
			 $\sigma_{\rm Fe}/\sigma_{\rm CH}$           &     not available                         &  1.033                                         \\
			
			\hline                                                                                                         
			\hline                                                                                                         
		\end{tabular}
\end{table}


\section{Data samples}

In this article, the data samples recorded by both the INGRID and Proton Module were taken from November 2010 to May 2013. The total number of protons on target (POT) is $5.89\times10^{20}$ with the neutrino-mode beam.
In July 2016, after the Water Module construction and its commissioning were completed, the Water Module replaced the Proton Module for physics data taking. A total POT of 7.25$\times10^{20}$ POT were collected with the neutrino-mode beam by the Water Module and INGRID during a period between October 2016 and April 2017. 

\section{Event selections} \label{ sec:eventselec }
 
\begin{figure}[!htb]
   \hspace*{-20mm}
        \centering
        \includegraphics[width=9cm, bb=0 0 578 502]{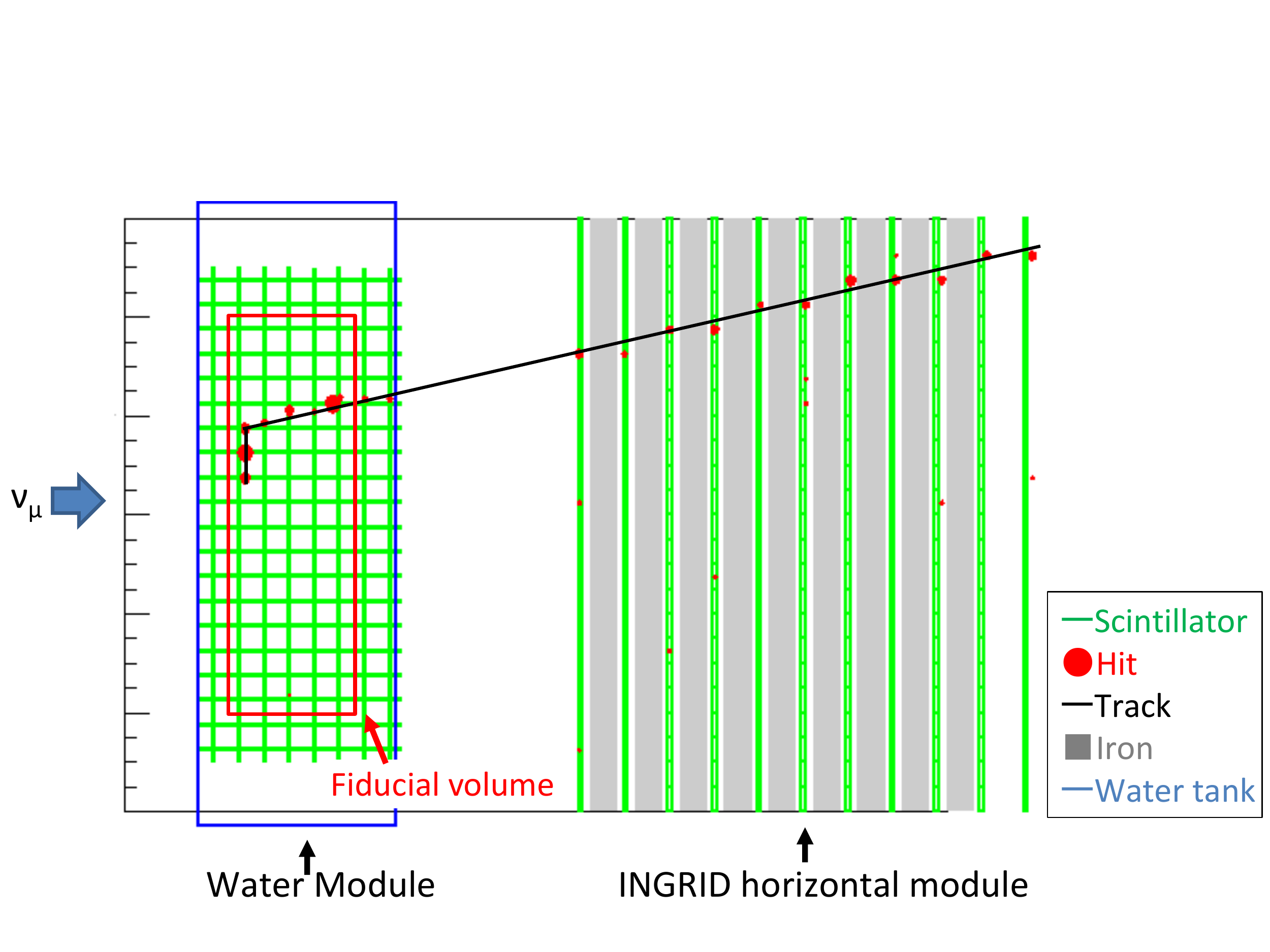}
\caption{ An event display of a typical signal event passing the event selection criteria for the Water Module. }
  \label{ fig:eventdisp }
\end{figure}

In this analysis, we define the signal with a restricted phase space of muon kinematics, particularly CC interactions with $\theta_{\mu}<$45$^{\circ}$ and $p_{\mu}>$0.4~GeV/$c$.
The main signature of the CC interactions is the presence of a muon-like track produced inside the detector.
Neutrino interactions originating from outside the detectors, CC interactions with non-target materials inside the detectors (mainly scintillators for the studied case with the Water Module), $\overline{\nu}_\mu, \nu_{e}, \overline{\nu}_{e}$ CC interactions, and NC interactions are the main sources of backgrounds in this analysis.
The backgrounds from the NC interactions do not produce muons. In order to identify the muons originating from the Water Module and Proton Module, events on the Water Module or Proton Module are required to have a track which penetrates at least two iron planes in one of the three horizontal INGRID modules near the beam center. This method for muon identification limits the phase space of the induced muon, because we reject the CC interactions with low momentum muons, which do not penetrate the iron planes, and high angle muons, which do not enter the three INGRID modules.
The event selections applied to the three detectors are similar to that from a previous analysis~\cite{ccincpm}, achieving a similar selection performance for the cross section measurements in the three targets.
Figure~\ref{ fig:eventdisp } shows an event display of a typical signal event passing the event selection criteria for the Water Module.

\subsection{Event selections for the Water Module}

\subsubsection{Time clustering}
Scintillator channels having charges larger than 2.5~p.e.\ are defined as a $``$hit". Hits are clustered with the following criteria: if there are more than three hits within 100~nsec in the Water Module, all the hits within 50~nsec from the average time are grouped into a single cluster.

\subsubsection{Two-dimensional track reconstruction}
The two-dimensional tracks in the {\it x}-{\it z} and {\it y}-{\it z} views are reconstructed independently by using a cellular automaton algorithm~\cite{scibar} to cluster the hits. More details about the algorithm can be found in the reference. The hits in the neighbor scintillator planes are defined as a $``$cell". Based on $\chi^{2}$ values given by the linear fitting of the relevant hits, it is judged if the pair of two cells having a common hit are merged into a new cell. This is repeated for all cells until no new cell is found and the long cells which have more than three hits are defined as tracks.

\subsubsection{Two dimensional track matching with the horizontal INGRID modules}
When two-dimensional tracks are reconstructed in the same beam bunch for both the Water Module and the three horizontal INGRID modules near the beam center, an attempt is made to match one to the other. The tracks are matched if they meet the following requirements:
\begin{itemize}
  \item The upstream edge of the reconstructed track in the three INGRID modules is in the most upstream two layers of the INGRID modules.
  \item The difference between the reconstructed angle of the three INGRID modules and Water Module tracks with respect to the {\it z}-axis must be less than 35$^{\circ}$.
  \item At the halfway point between the three INGRID modules and Water Module, the distance between the three INGRID modules and Water Module track is less than 150~mm.
\end{itemize}

\subsubsection{Three-dimensional track matching}
Three-dimensional tracks are formed amongst the pairs of two-dimensional INGRID matched tracks in the {\it x}-{\it z} plane and in the {\it y}-{\it z} plane imposing that the difference between the two measurements of the {\it z} coordinates of the most upstream hits to be less than or equal to one plane of the parallel scintillators. If there are multiple candidates, we select a pair with the smallest difference of the most upstream hit point {\it z}. If there are still multiple candidates after the selection, we select a pair with the smallest difference of the most downstream hit point {\it z}.  

Only events which have at least one INGRID-matched track are used for the analysis. 
Because the horizontal INGRID modules are located downstream of the Water Module, the angular acceptance is limited. In addition, the momentum acceptance is limited because the track is required to penetrate at least two iron planes of the INGRID modules for the matching.

\subsubsection{Vertexing}
After the three-dimensional track reconstruction, the most upstream {\it z} coordinate of each INGRID matched three-dimensional track is identified as a reconstructed vertex. If a pair of INGRID-matched three-dimensional tracks meet the following conditions they are identified as tracks coming from a common vertex:
\begin{itemize}
	\item The difference between the most upstream {\it z} coordinate of the two tracks in the {\it x}-{\it z} view, added to the same difference in the {\it y}-{\it z} view, has to be less than three planes of the parallel scintillators.
	\item The distance between the upstream {\it z} coordinate  of the two tracks in the {\it x}-{\it y} plane is less than 150~mm.
\end{itemize}
These cuts are applied to every vertex since each one is expected to correspond to a single neutrino interaction. The vertex position is re-defined as that of the longest INGRID-matched track amongst those that belong to the common vertex. The longest INGRID-matched track is defined as a muon-like track.

\subsubsection{Beam timing cut}
\begin{figure}[!htb]
   \hspace*{-5mm}
   \centering
   \includegraphics[width=10cm, bb=0 0 350 250]{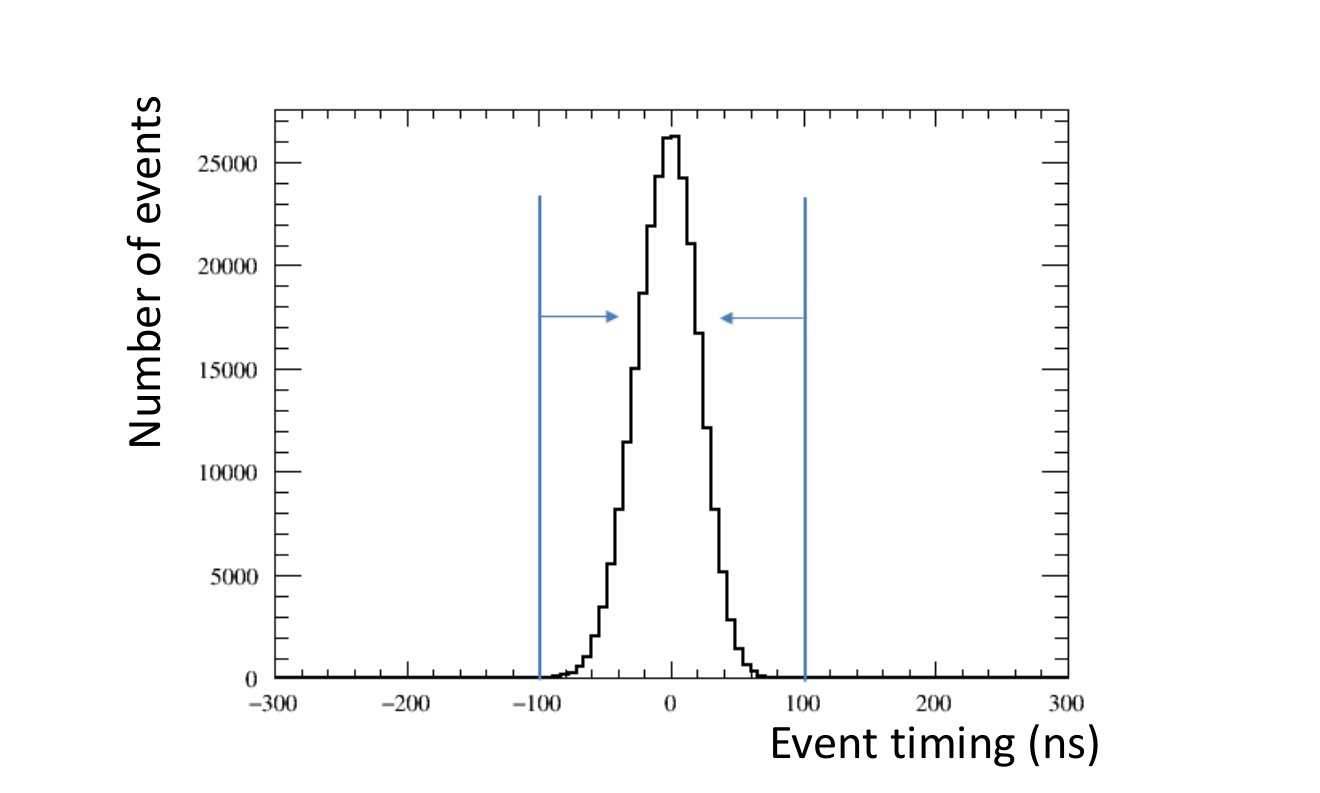}
   \caption{ Timing difference between the selected events and the expected beam bunch time, after the vertexing cut. }
   \label{ fig:timing }
\end{figure}

To reduce non-beam backgrounds, such as cosmic rays, only events within 100~nsec of the expected beam bunch timing are selected, as shown in Fig.~\ref{ fig:timing }. The individual event timing is defined as the time recorded by the MPPC channel with the largest light yield.

\subsubsection{Upstream veto cut and fiducial cut}
Two cuts are applied based on the position of the vertex to reduce beam-induced backgrounds from neutrino interactions outside the Water Module, mainly from the walls of the detector hall and the INGRID vertical modules. If the upstream point of a track is in the first or second plane of the parallel scintillators, then that event is rejected. 
The fiducial volume is defined as the central part of the Water module with dimensions of 70~cm (in {\it x}-coordinate) $\times$70~cm (in {\it y}-coordinate) $\times$ 21~cm (in {\it z}-coordinate).

\begin{figure}[!htb]
\vspace*{-20mm}
\hspace*{-20mm}
\begin{minipage}{7.5cm}
        \centering
        \includegraphics[width=12cm, bb=0 0 503 338]{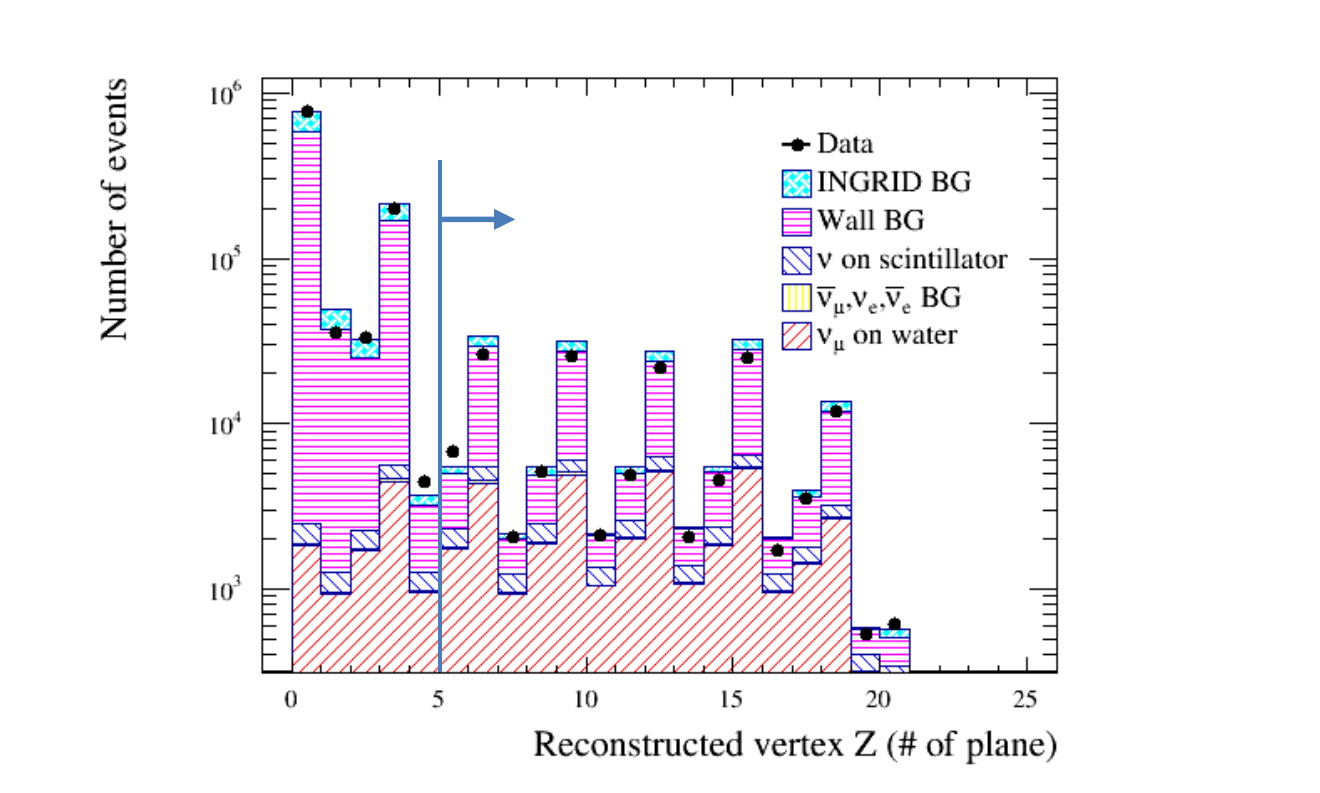}
\end{minipage}
\begin{minipage}{7.5cm}
        \centering
        \includegraphics[width=12cm, bb=0 0 503 338]{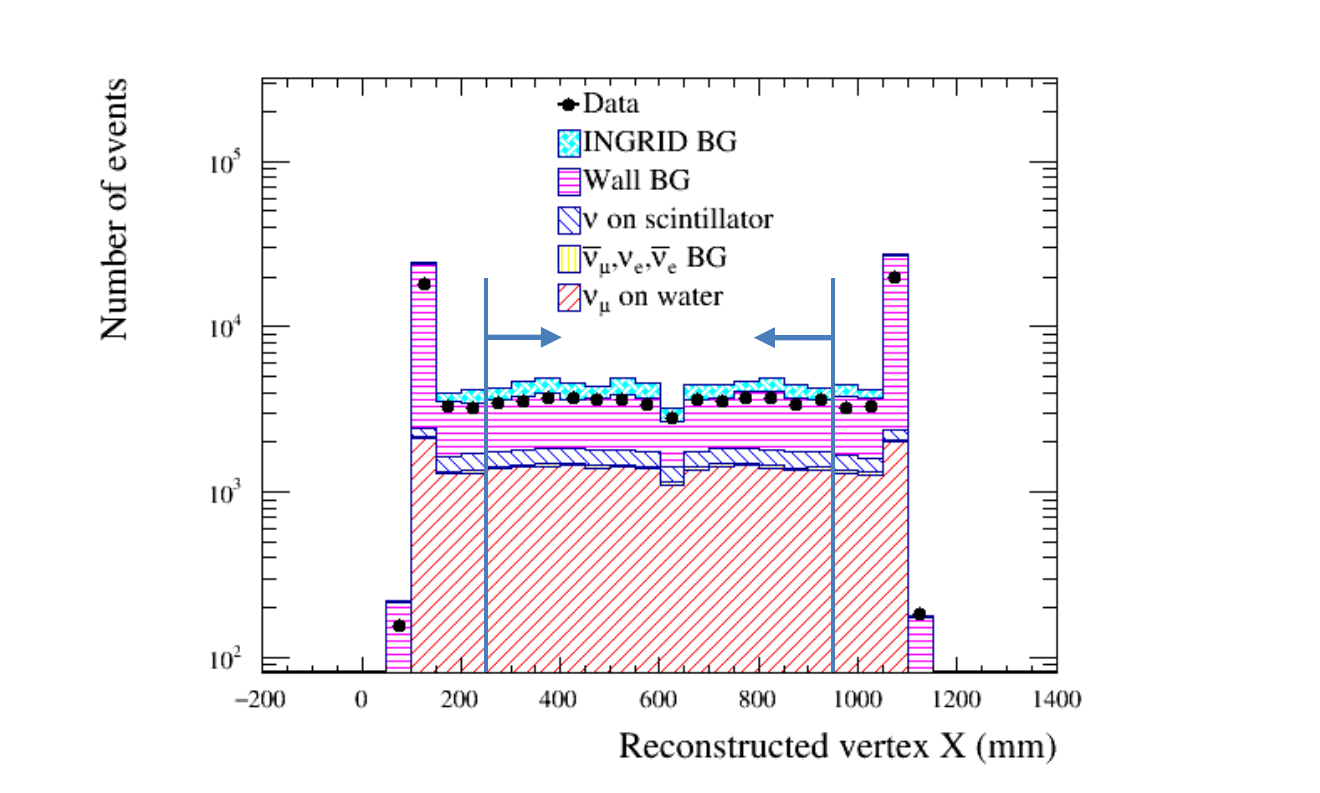}
\end{minipage}
\hspace*{30mm}
\begin{minipage}{7.5cm}
        \includegraphics[width=12cm, bb=0 0 503 238]{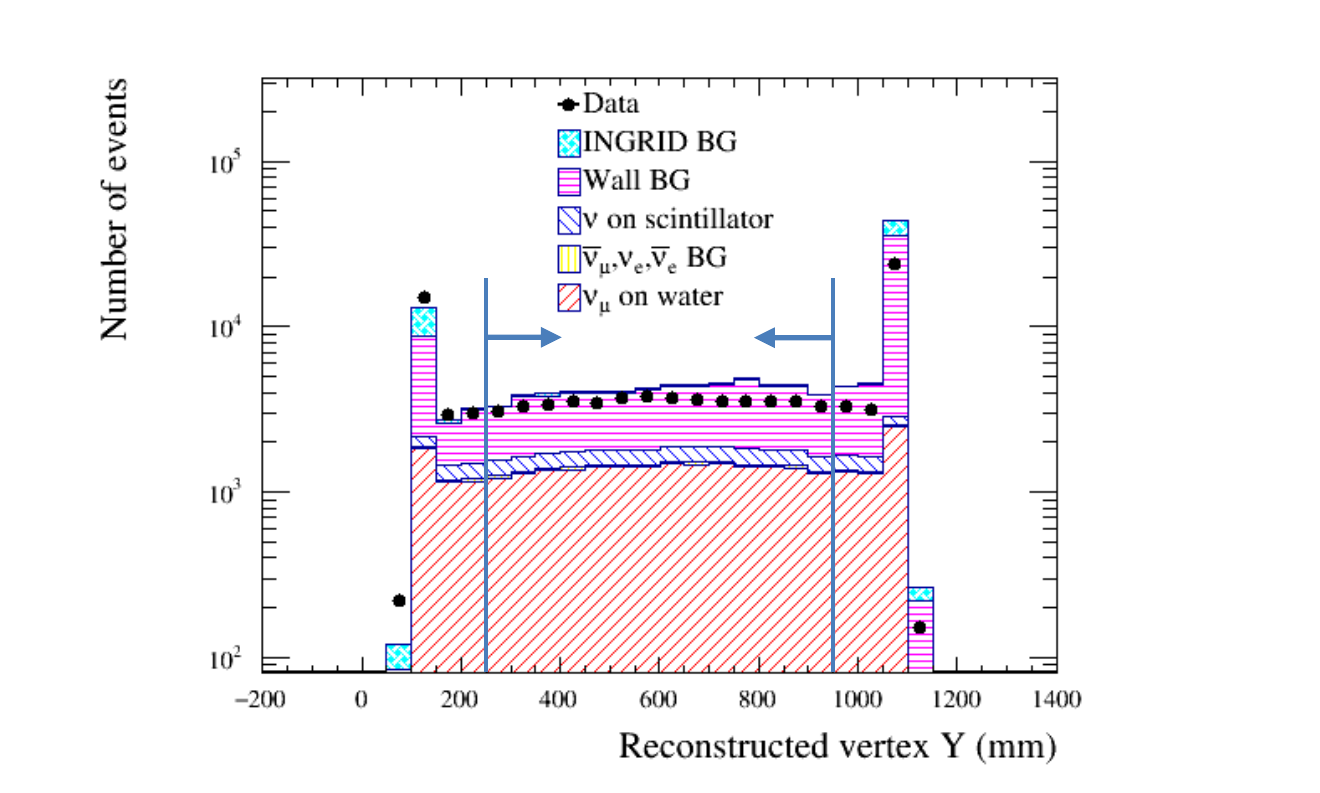}
\end{minipage}
\caption{ Reconstructed vertex {\it z} distribution in the {\it x}-{\it z} view before the front veto cut for the Water Module (upper left), reconstructed vertex {\it x} (upper right), and {\it y} (lower) distribution after the front veto cut for the Water Module. In the upper left plot, the $x$-axis shows the number of the plane and the most upstream plane is set to 0. The spikes for the plane numbers that are multiples of 3 are due to the parallel scintillators. In the upper right and lower plots, the center of the detector is set to 600~mm.  }
   \label{ fig:vertexxy_wm }
\end{figure}

The vertex is required to be within the fiducial volume for the neutrino event to be selected.
Figure~\ref{ fig:vertexxy_wm } shows distributions of the vertex used for these two cuts.

\subsubsection{Reconstructed angle cut}

The three-dimensional angle of the longest reconstructed track from a vertex is required to be smaller than 45$^{\circ}$ to reduce large-angle muons since the detection efficiency for such kind of events is less than 10\%, as described in Sec.~\ref{ sec:eff }.

\subsubsection{Event selections summary}
\begin{figure}[!htb]
\begin{minipage}{0.50\hsize}
\hspace*{-7mm}
        \centering
        \includegraphics[width=5.0cm,  bb=20 0 405 350]{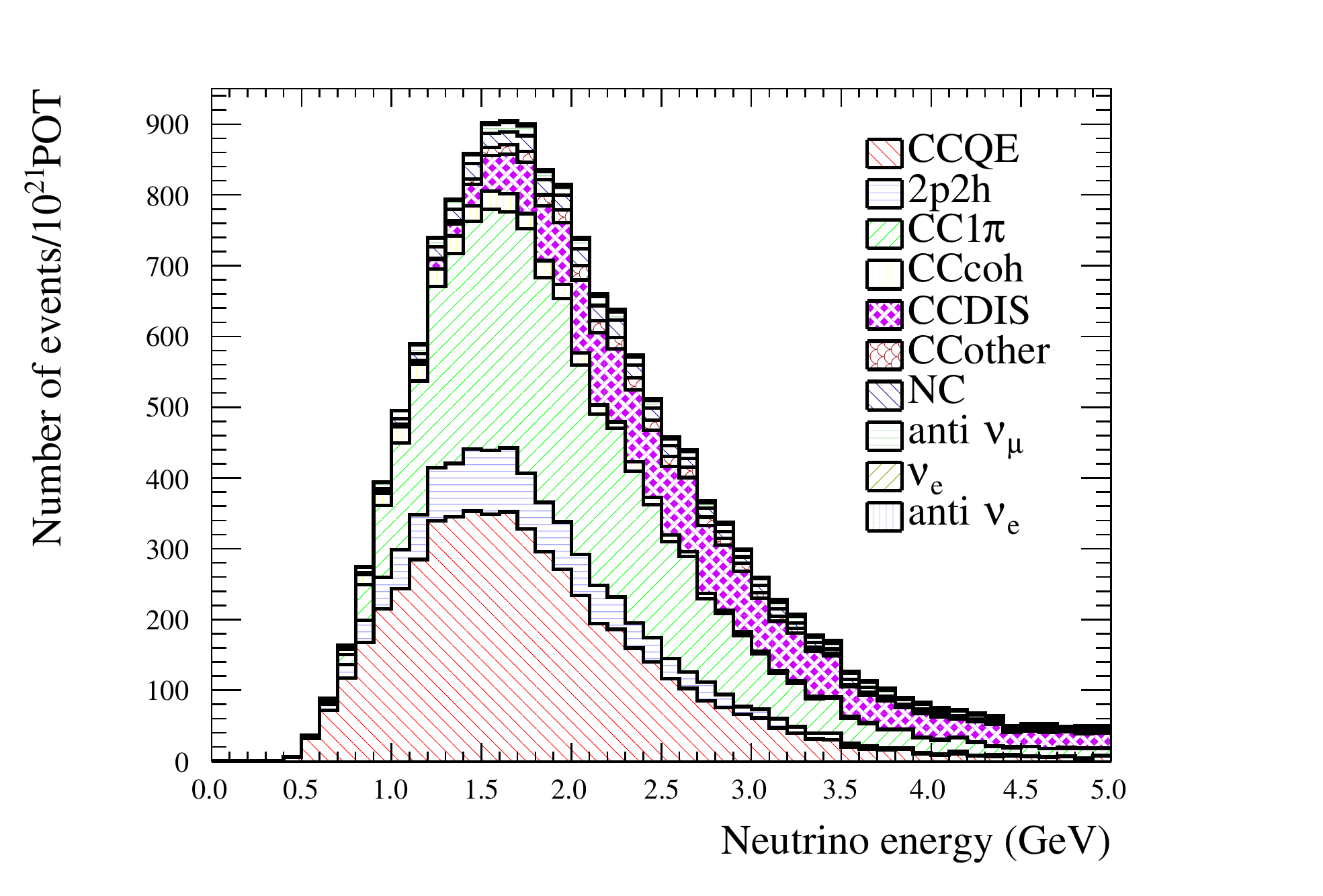}
\end{minipage}
\begin{minipage}{0.50\hsize}
\hspace*{-7mm}
        \centering
        \includegraphics[width=5.0cm,  bb=20 0 405 350]{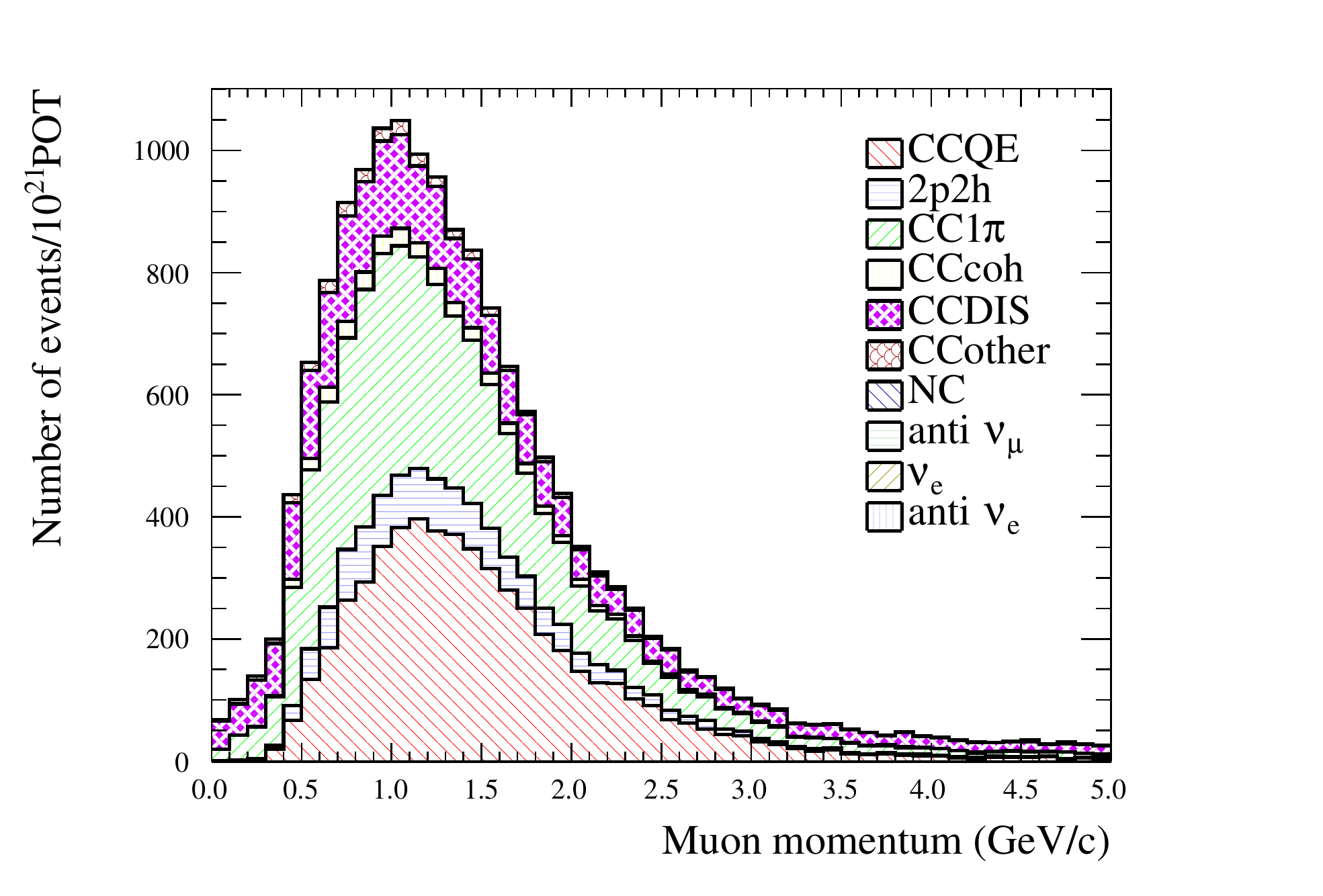}
\end{minipage}
\hspace*{30mm}
\begin{minipage}{0.50\hsize}
\vspace*{5mm}
        \centering
        \includegraphics[width=5.0cm,  bb=20 0 405 350]{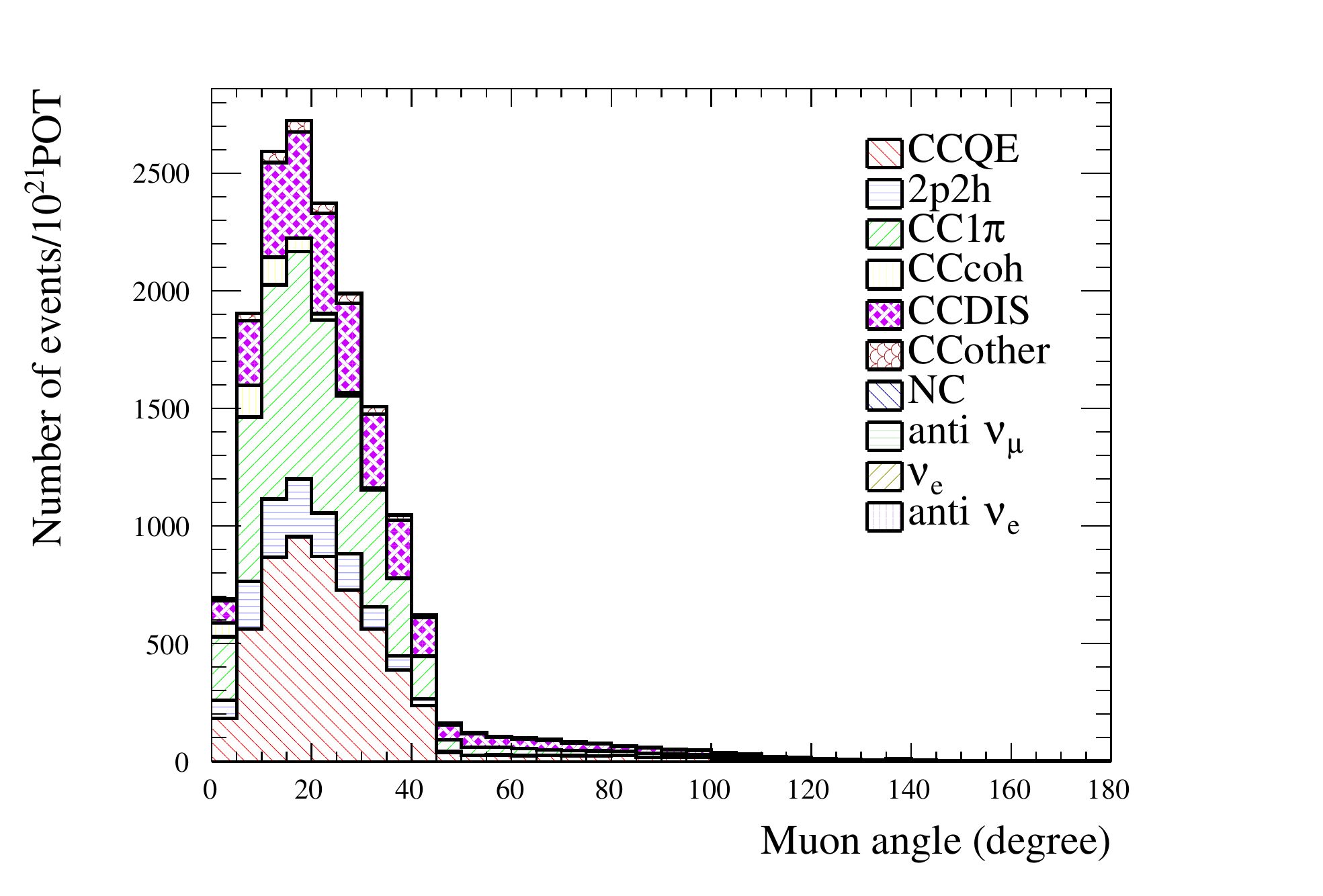}
\end{minipage}
\caption{ MC prediction of the true neutrino energy (upper left), muon momentum (upper right), and muon scattering angle (lower) of the selected events for the Water Module. }
   \label{ fig:enu_wm }
\end{figure}

\begin{figure}[!htb]
\begin{minipage}{7.5cm}
\hspace*{-7mm}
        \centering
        \includegraphics[width=12cm, bb=0 0 503 238]{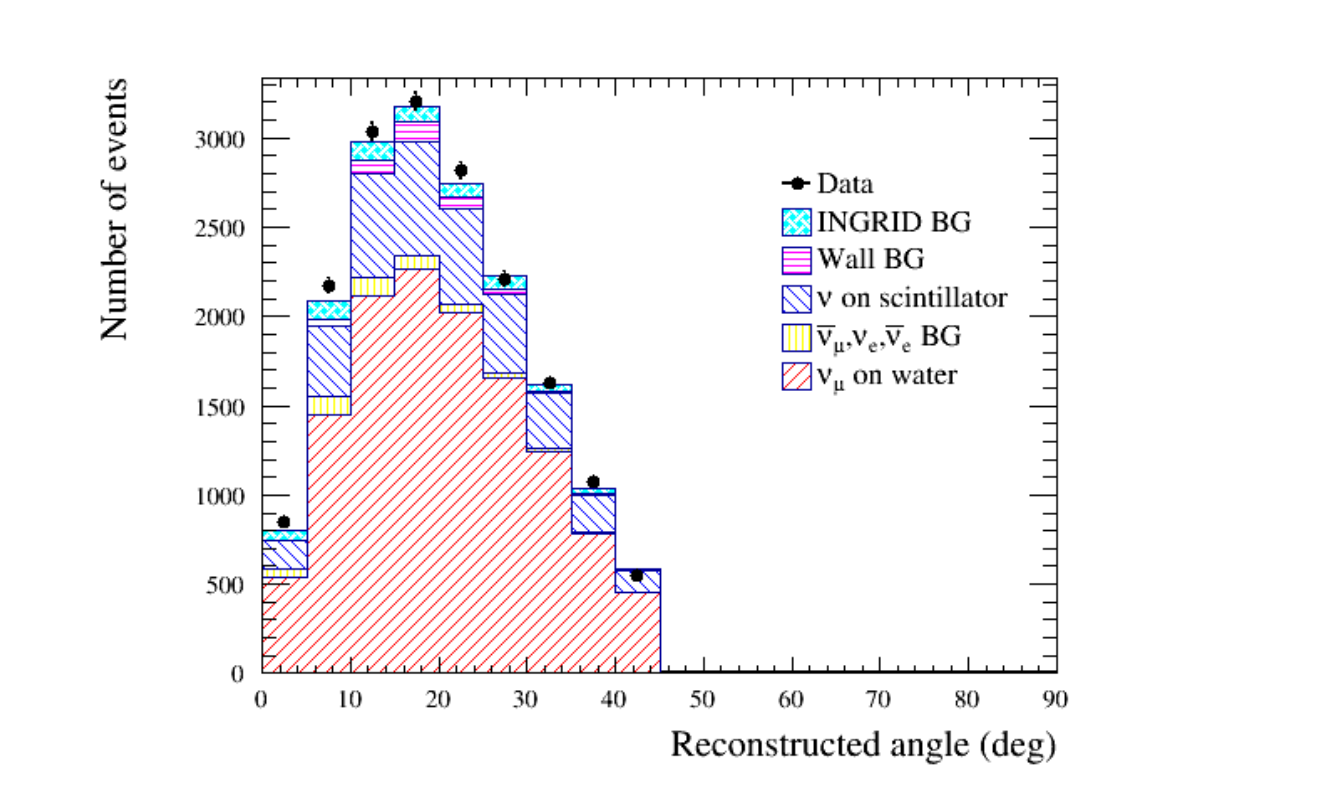}
	Water Module
\end{minipage}
\begin{minipage}{7.5cm}
\hspace*{-7mm}
        \centering
        \includegraphics[width=12cm, bb=0 0 503 238]{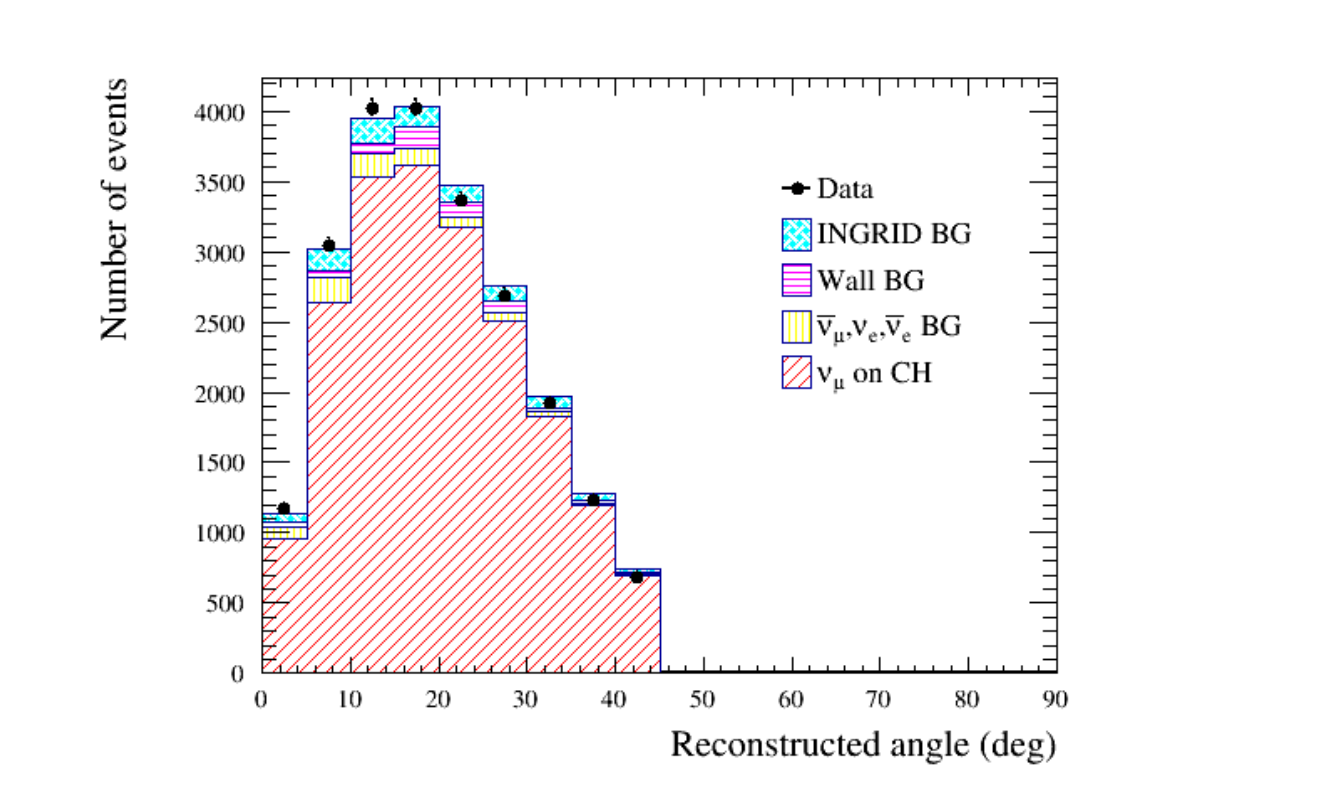}
	Proton Module
\end{minipage}
\hspace*{30mm}
\begin{minipage}{7.5cm}
        \centering
        \includegraphics[width=12cm, bb=0 0 503 238]{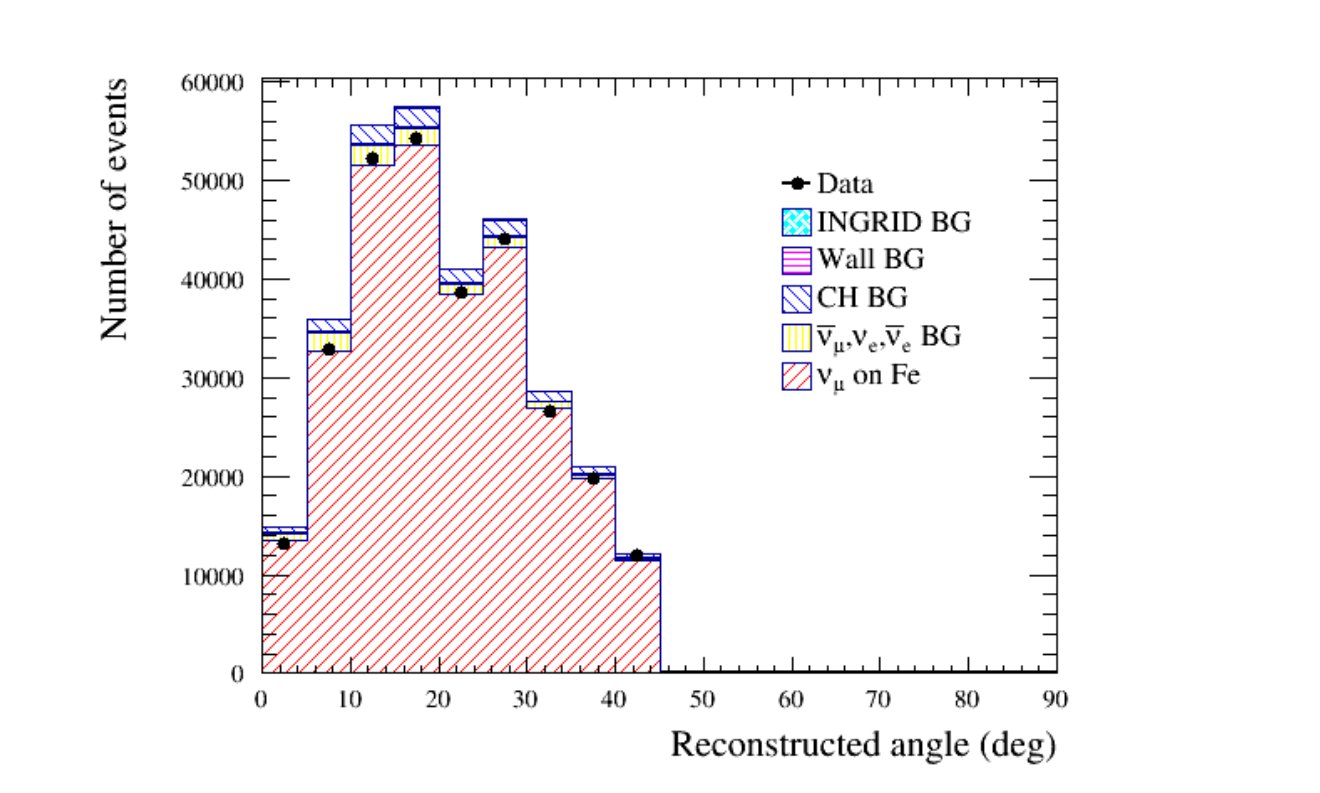}
	INGRID module
\end{minipage}
\caption{ Reconstructed angle of the longest track from a vertex after the event selection for the Water Module (upper left), Proton Module (upper right), and INGRID module (lower). }
   \label{ fig:angle_event_wm }
\end{figure}

Table~\ref{ tab:event_para } shows a summary of the parameters used for the event selection. The numbers of selected events and the backgrounds in the Water Module at each selection step are summarized in Table~\ref{ tab:selection_wm }. There are $1.73\times10^{4}$ events expected in the MC after the event selection. The purity of the $\nu_{\mu}$ CC interactions on $\rm H_{2}O$ is 69.0\% and the main background is from neutrino interactions on the scintillators (19.8\%). The remaining background sources are NC interactions (2.9\%) due to misidentification of pions, neutrino interactions of $\overline{\nu}_{\mu}$, $\nu_{e}$ and $\overline{\nu}_{e}$ (2.0\%), photons from $\pi^{0}$ produced by neutrino interactions on the walls of the detector hall (2.4\%), and backscattered production of neutrino interactions in the INGRID (3.1\%). The muon-like tracks, identified as the longest INGRID-matched track, have 87\% probability to be the true muons. Figure~\ref{ fig:enu_wm } shows the neutrino energy, muon momentum, and angle distributions of the selected events predicted by MC. The main interaction modes are CCQE, CC1$\pi$, CC multi-pion and DIS production. Figure~\ref{ fig:angle_event_wm } (upper left) shows the angle distribution of the reconstructed muon-like tracks for events which passed all event selection in the Water Module.

\begin{table}[!htb]
	\centering
	\caption{ Parameters used for the event selection criteria for the on-axis detectors.  }
		\label{ tab:event_para }
		\scalebox{0.85}{\begin{tabular}{ccccc}
			\hline
			\hline
			                  & Water Module & Proton Module  & INGRID module     \\
			\hline
                        Time clustering            & $\pm$50~nsec                & $\pm$50~nsec               & $\pm$50~nsec \\
			Track matching with INGRID & $\pm$35$^\circ$                 &  $\pm$35$^\circ$              & - \\
			                           & $\pm$150~mm                 &  $\pm$150~mm               & - \\
						   3D track matching          & $\leq$1~parallel plane        & $\leq$1~plane                &  $\leq$1~plane   \\
			Vertexing                  & $<$3~planes                 & $<$2~planes                & $<$2~planes \\
			                           & $<$150~mm                   & $<$150~mm                  & $<$150~mm   \\
			Beam timing                & $\pm$100~nsec               & $\pm$100~nsec              & $\pm$100~nsec \\
			Upstream veto              & $\geq$second~parallel plane   & $\geq$second~plane           &  $\geq$first~plane   \\
			Fiducial                   & 700~mm$\times$700~mm        & 700~mm$\times$700~mm       & 700~mm$\times$700~mm   \\
			Reconstructed angle        & $<$45$^\circ$               & $<$45$^\circ$              & $<$45$^\circ$   \\
			\hline    
			\hline                                                                                                         
		\end{tabular}}
\end{table}

\begin{table}[!htb]
	\begin{center}
		\caption{Summary of the event selection for the Water Module. The purities of CC interactions are shown in parentheses.}
		\label{ tab:selection_wm }
		\hspace*{-12mm}
		\scalebox{0.80}{
		\begin{tabular}{cccccccccc}
			\hline
			\hline
			Selection  & Data &    &    &                         &                                    &         MC  &           &          &     \\ \cline{3-10}
			&      & CC        & NC & $\overline{\nu}_{\mu}$, $\nu_{e}$, $\overline{\nu}_{e}$ & CH B.G. & Wall B.G. & INGRID B.G. & All \\
			\hline
			Vertexing cut  & 1175980    & $4.39\times10^{4}$ (4\%) &  1.66$\times10^{2}$  &               $1.12\times10^{3}$ &                $1.08\times10^{4}$   & $9.10\times10^{5}$ & $2.77\times10^{5}$ & $1.24\times10^{6}$ \\
			Front veto cut & 100790   & $2.77\times10^{4}$ (21\%) &  1.04$\times10^{3}$  &               $9.38\times10^{2}$ &                  $6.66\times10^{3}$   & $8.09\times10^{4}$ & $1.46\times10^{4}$ & $1.32\times10^{5}$ \\
			Fiducial cut  & 17992    & $1.25\times10^{4}$ (69\%)&  4.68$\times10^{2}$  &               $4.42\times10^{2}$ &                  $3.51\times10^{3}$   & $3.49\times10^{2}$ & $5.84\times10^{2}$ & $1.78\times10^{4}$ \\
			Track angle cut& 17528    & $1.20\times10^{4}$ (69\%)&  4.53$\times10^{2}$  &               $4.39\times10^{2}$ &                  $3.39\times10^{3}$   & $3.47\times10^{2}$ & $5.64\times10^{2}$ & $1.73\times10^{4}$ \\
			\hline
			\hline
		\end{tabular}}
	\end{center}
\end{table}

\subsection{Event selections for the Proton Module}

The event selections for the Proton Module and INGRID module are very similar to those for the Water Module. However, due to the difference in the scintillator layout, a few parameters for the cellular automaton algorithm and event selection have been optimized as listed in Table~\ref{ tab:event_para }. 

The numbers of selected events and the backgrounds in the Proton Module at each selection step are summarized in Table~\ref{ tab:selection_pm }. 
After the event selection, a total of $2.23\times10^{4}$ events are expected by MC. The purity of the CC interactions on CH is 85.4\%. Background sources are NC interactions (4.2\%), neutrino interactions of $\overline{\nu}_{\mu}$, $\nu_{e}$ and $\overline{\nu}_{e}$ (2.4\%), photons from $\pi^{0}$ produced by neutrino interactions on the walls of the detector hall (2.1\%) and backscattered events from neutrino interactions in the INGRID (5.2\%).
Figure~\ref{ fig:angle_event_wm } (upper right) shows the angle distribution of the reconstructed muon-like tracks for events which passed all event selection in the Proton Module.

\begin{table}[!htb]
	\begin{center}
		\caption{Summary of the event selection for the Proton Module. The purities of CC interactions are shown in parentheses.}
		\label{ tab:selection_pm }
		\scalebox{0.80}{
		\begin{tabular}{ccccccccc}
			\hline
			\hline
			Selection  & Data &           &    &                         &                                &         MC  &           &        \\ \cline{3-9} 
			&      & CC        & NC & $\overline{\nu}_{\mu}$, $\nu_{e}$, $\overline{\nu}_{e}$ &  Wall B.G. & INGRID B.G.  & All \\
			\hline
			Vertexing cut &  1321290    & $5.56\times10^{4}$ (4\%)&  2.66$\times10^{3}$  &               $2.00\times10^{3}$ &  $1.03\times10^{6}$ & $2.77\times10^{5}$ & $1.36\times10^{6}$ \\
			Front veto cut &  264550     & $4.69\times10^{4}$ (15\%)&  2.25$\times10^{3}$  &               $1.72\times10^{3}$ &  $2.17\times10^{5}$ & $3.63\times10^{4}$ & $3.04\times10^{5}$ \\
			Fiducial cut   &  22930      & $1.98\times10^{4}$ (85\%)&  9.52$\times10^{2}$  &               $7.31\times10^{2}$ &  $5.54\times10^{2}$ & $9.97\times10^{2}$ & $2.32\times10^{4}$ \\
			Track angle cut&  22165      & $1.92\times10^{4}$ (85\%)&  9.14$\times10^{2}$  &               $7.26\times10^{2}$ &  $5.51\times10^{2}$ & $9.50\times10^{2}$ & $2.23\times10^{4}$ \\
			\hline
			\hline
		\end{tabular}}
	\end{center}
\end{table}

\subsection{Event selections for the INGRID module}
\begin{figure}[!htb]
   \centering
   \vspace*{-30mm}
   \hspace*{20mm}
   \includegraphics[height=10cm, bb=0 0 620 340]{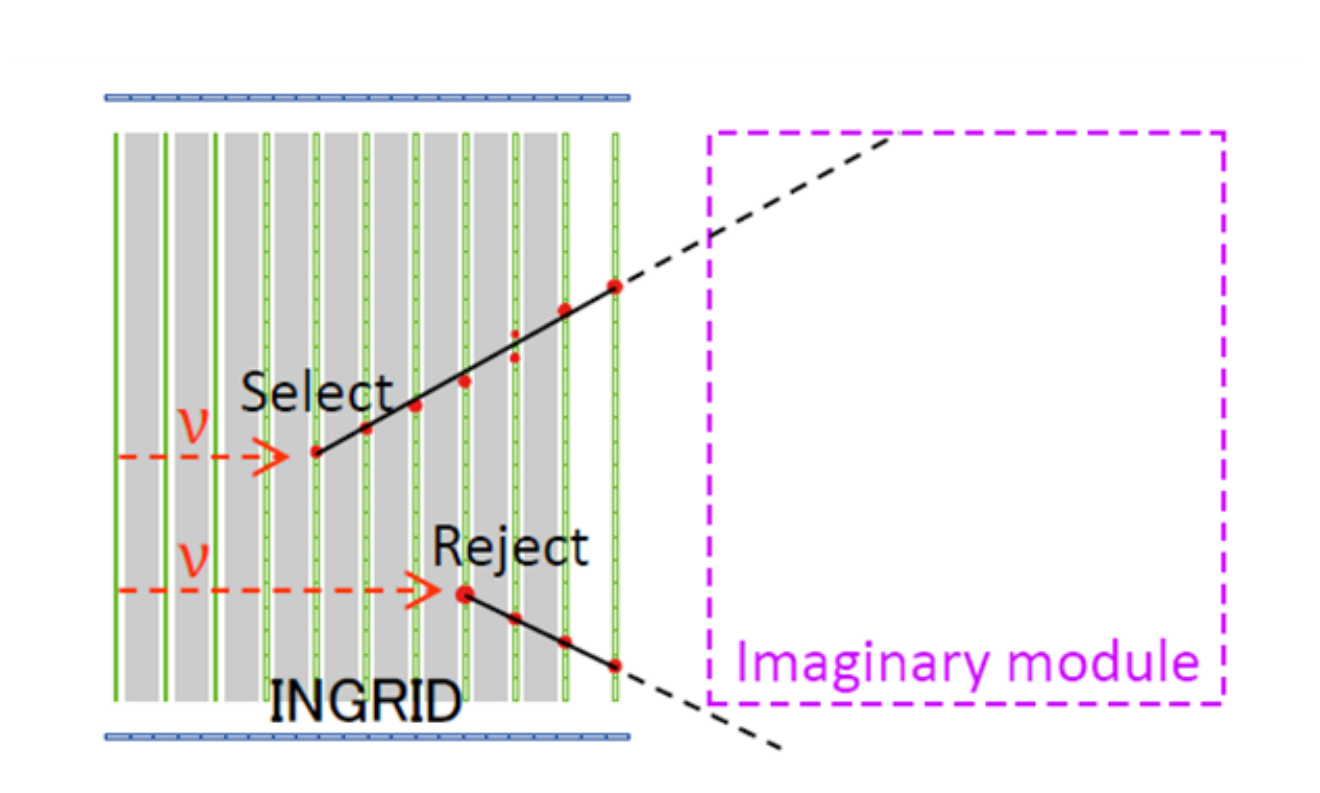}
   \caption{ An example of events selected and rejected by the $``$acceptance cut" for the INGRID module~\cite{kikawa}. If at least one extended reconstructed track from the vertex reaches the imaginary module, the event is selected. }
   \label{ fig:imagy_ing }
\end{figure}

The event selections are applied for the horizontal INGRID module located at the beam center with the parameters listed in Table~\ref{ tab:event_para }. In addition, an $``$acceptance cut" is applied only for the INGRID module in order to achieve a similar angular acceptance with the Water Module and Proton Module. An imaginary module located directly behind the INGRID module is defined, as shown in Fig.~\ref{ fig:imagy_ing }. The distance between the INGRID module and the imaginary module is the same as between the Water Module and the INGRID horizontal modules. The reconstructed tracks are then projected further downstream, even if the track has stopped in the INGRID module. If at least one reconstructed track from the vertex reaches the imaginary module, that event is selected.

The numbers of selected events and the backgrounds in the INGRID module at each selection step are summarized in Table~\ref{ tab:selection_ing }.
After the event selection, a total of $3.12\times10^{5}$ events are expected by MC. The purity of the $\nu_{\mu}$ CC interactions on Fe is 88.1\%. Background sources are NC interactions (5.2\%), neutrino interactions of $\overline{\nu}_{\mu}$, $\nu_{e}$ and $\overline{\nu}_{e}$ (2.9\%), neutrino interactions on scintillator (3.3\%), photons from $\pi^{0}$ produced by neutrino interactions on walls of the detector hall (0.3\%) and the other INGRID modules (0.2\%).
Figure~\ref{ fig:angle_event_wm } (lower) shows the angle distribution of the reconstructed muon-like tracks for events which passed all event selection in the INGRID module.

\begin{table}[!htb]
	\begin{center}
		\caption{Summary of the event selection for the INGRID module. The INGRID B.G.\ in the table represents backgrounds from the other INGRID modules. The purities of CC interactions are shown in parentheses. }
		\label{ tab:selection_ing }
		\hspace*{-12mm}
		\scalebox{0.80}{
		\begin{tabular}{cccccccccc}
			\hline
			\hline
			Selection  & Data &           &    &                                                             &         MC  &           &          &     \\ \cline{3-9}
			&      & CC        & NC & $\overline{\nu}_{\mu}$, $\nu_{e}$, $\overline{\nu}_{e}$ & CH B.G. & Wall B.G. & INGRID B.G. & All \\
			\hline
			Vertexing cut & 3019430    & $1.11\times10^{6}$ (44\%) &  $6.98\times10^{4}$  &               $3.20\times10^{4}$ &  $4.49\times10^{4}$   & $9.45\times10^{5}$ & $3.36\times10^{5}$ & $2.54\times10^{6}$ \\
			Front veto cut & 1468490    & $1.07\times10^{6}$ (74\%) &  $6.74\times10^{4}$  &               $3.07\times10^{4}$ &  $3.97\times10^{4}$   & $1.98\times10^{5}$ & $4.33\times10^{4}$ & $1.45\times10^{6}$ \\
			Fiducial cut  & 431211     & $4.10\times10^{5}$ (88\%) &  $2.58\times10^{4}$  &               $1.14\times10^{4}$ &  $1.49\times10^{4}$   & $1.52\times10^{3}$ & $1.06\times10^{2}$ & $4.65\times10^{5}$ \\
			Acceptance cut & 308971     & $2.88\times10^{5}$ (88\%) &  $1.81\times10^{4}$  &               $9.56\times10^{3}$ &  $1.07\times10^{4}$   & $9.26\times10^{2}$ & $6.73\times10^{2}$ & $3.28\times10^{5}$ \\
			Track angle cut& 293418     & $2.74\times10^{5}$ (88\%) &  $1.72\times10^{4}$  &               $9.31\times10^{3}$ &  $1.02\times10^{4}$   & $8.70\times10^{2}$ & $6.38\times10^{2}$ & $3.12\times10^{5}$ \\
			\hline
			\hline
		\end{tabular}}
	\end{center}
\end{table}


\subsection{Pileup correction for the INGRID module}
If more than one neutrino event occurs in the detector at the same bunch timing, we sometimes fail to count them. Therefore, a correction must be applied to account for this event pileup effect. For the INGRID module, this effect is estimated in each bin of the reconstructed track angle by merging multiple bunches to enrich the pileup rate artificially. Table~\ref{ tab:Closs } shows the number of selected events before and after the pileup correction. For the Water Module and Proton Module, the effect of the pileup is small due to the small target mass therefore no correction is applied.

\begin{table}[!htb]
	\begin{center}
		\caption{The number of selected events for the INGRID module before and after the pileup correction. }
		\label{ tab:Closs }
		\begin{tabular}{cccccccccc}
			\hline
			\hline
			reconstructed angle bin  &  $N_{\rm sel}$    & $N_{\rm corr}$ & $N_{\rm corr}/N_{\rm sel}$ \\ 
			\hline
                        0-5$^\circ$              &  13106            &  13582.0       & 1.036 \\ 
                        5-10$^\circ$             &  32928            &  33765.3       & 1.025 \\ 
                        10-15$^\circ$            &  52272            &  53671.3       & 1.027 \\ 
                        15-20$^\circ$            &  54205            &  55500.6       & 1.024 \\ 
                        20-25$^\circ$            &  38540            &  39119.4       & 1.015 \\ 
                        25-30$^\circ$            &  44097            &  45002.4       & 1.021 \\ 
                        30-35$^\circ$            &  26615            &  26984.1       & 1.014 \\ 
                        35-40$^\circ$            &  19709            &  20036.4       & 1.017 \\ 
                        40-45$^\circ$            &  11946            &  12094.0       & 1.012 \\ 
			\hline
                        Total                    &  293418           &  299755.5      & 1.022 \\ 
			\hline
			\hline
		\end{tabular}
	\end{center}
\end{table}

\subsection{Selection efficiencies} \label{ sec:eff }

\begin{figure}[!htb]
\begin{minipage}{0.50\hsize}
\hspace*{-20mm}
        \centering
        \includegraphics[width=5.0cm,  bb=20 0 405 350]{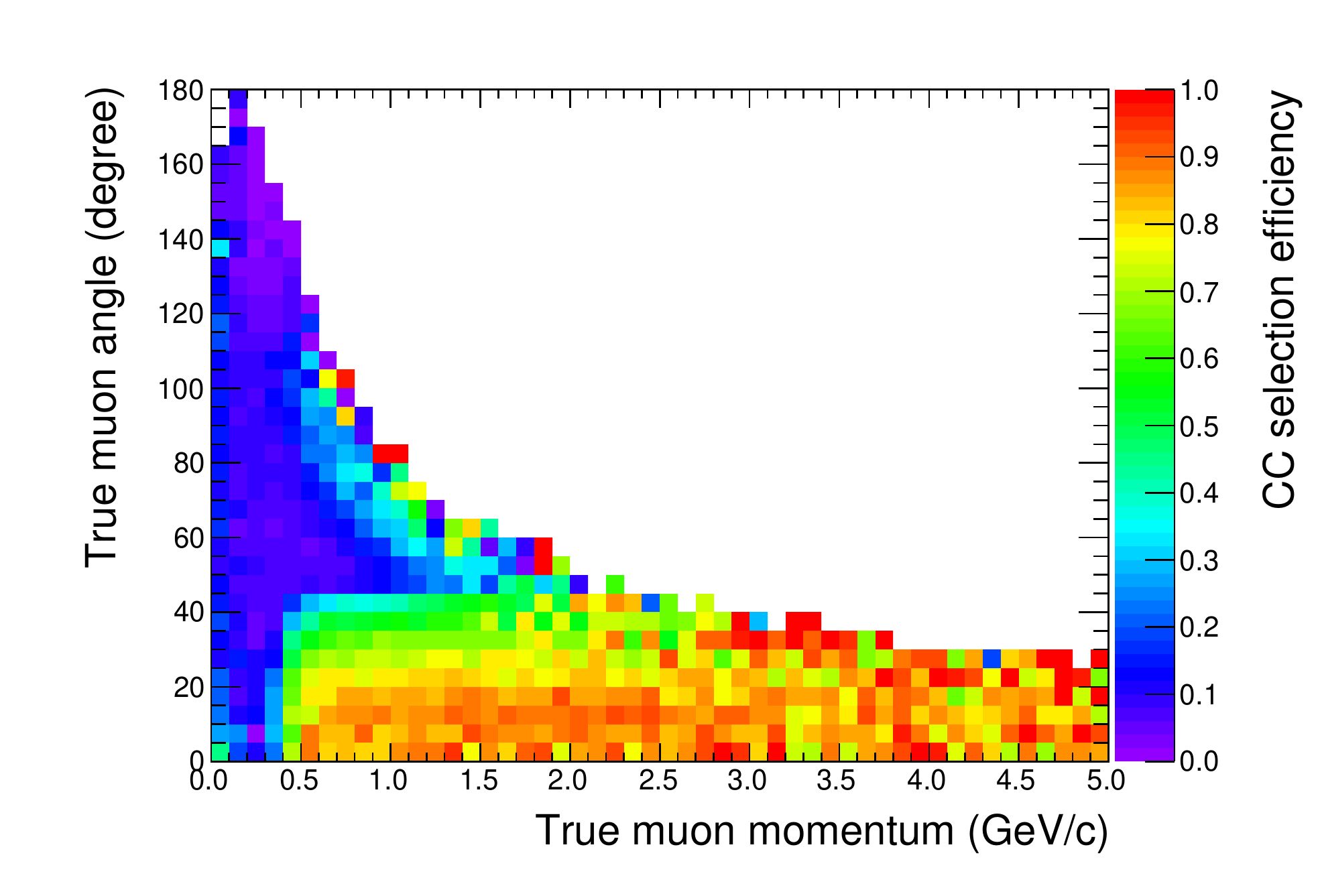}
	\\Water Module
\end{minipage}
\begin{minipage}{0.50\hsize}
\hspace*{-20mm}
        \centering
        \includegraphics[width=5.0cm,  bb=20 0 405 350]{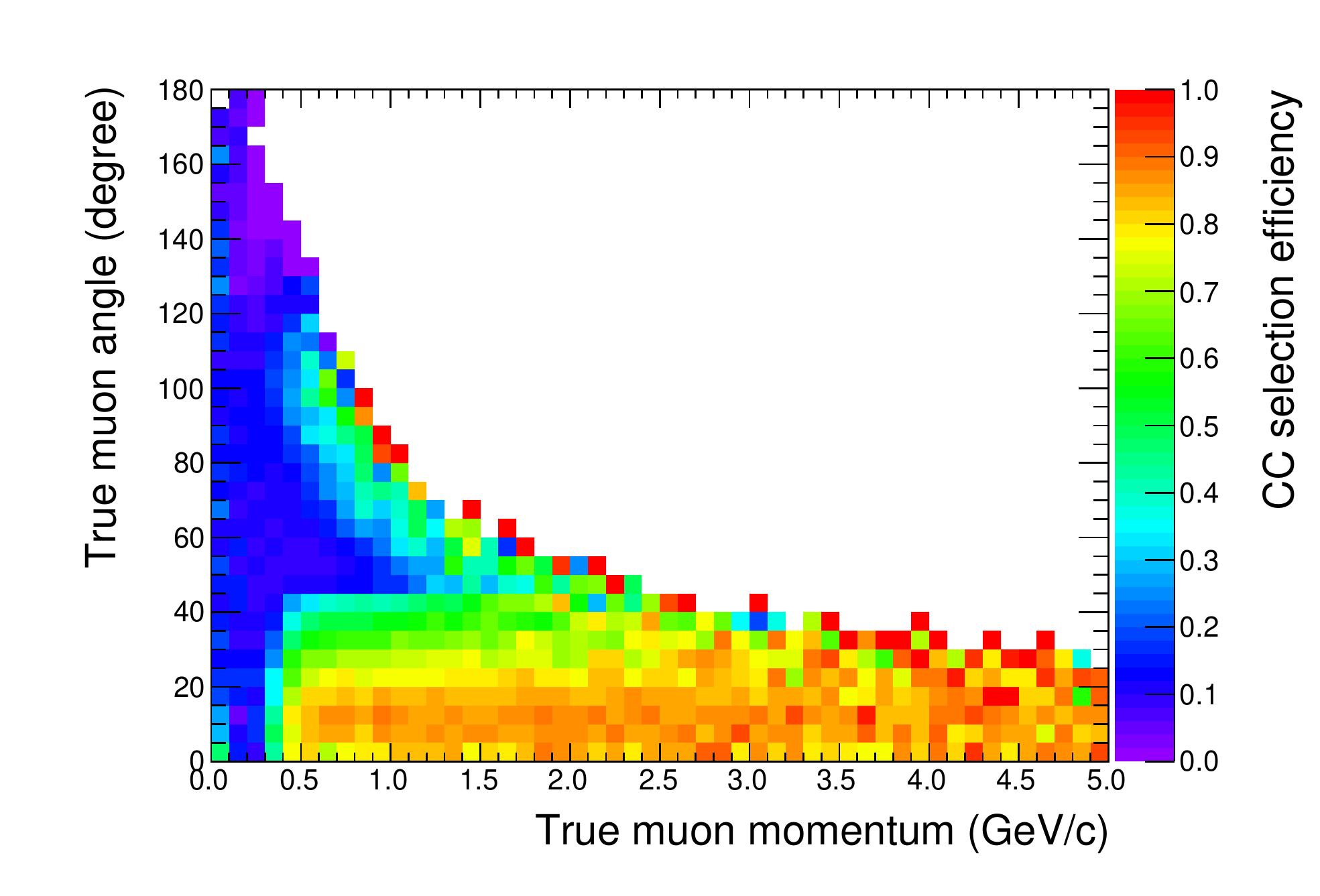}
	\\Proton Module
\end{minipage}
\hspace*{35mm}
\begin{minipage}{0.50\hsize}
\vspace*{10mm}
\hspace*{-15mm}
        \centering
        \includegraphics[width=5.0cm,  bb=20 0 405 350]{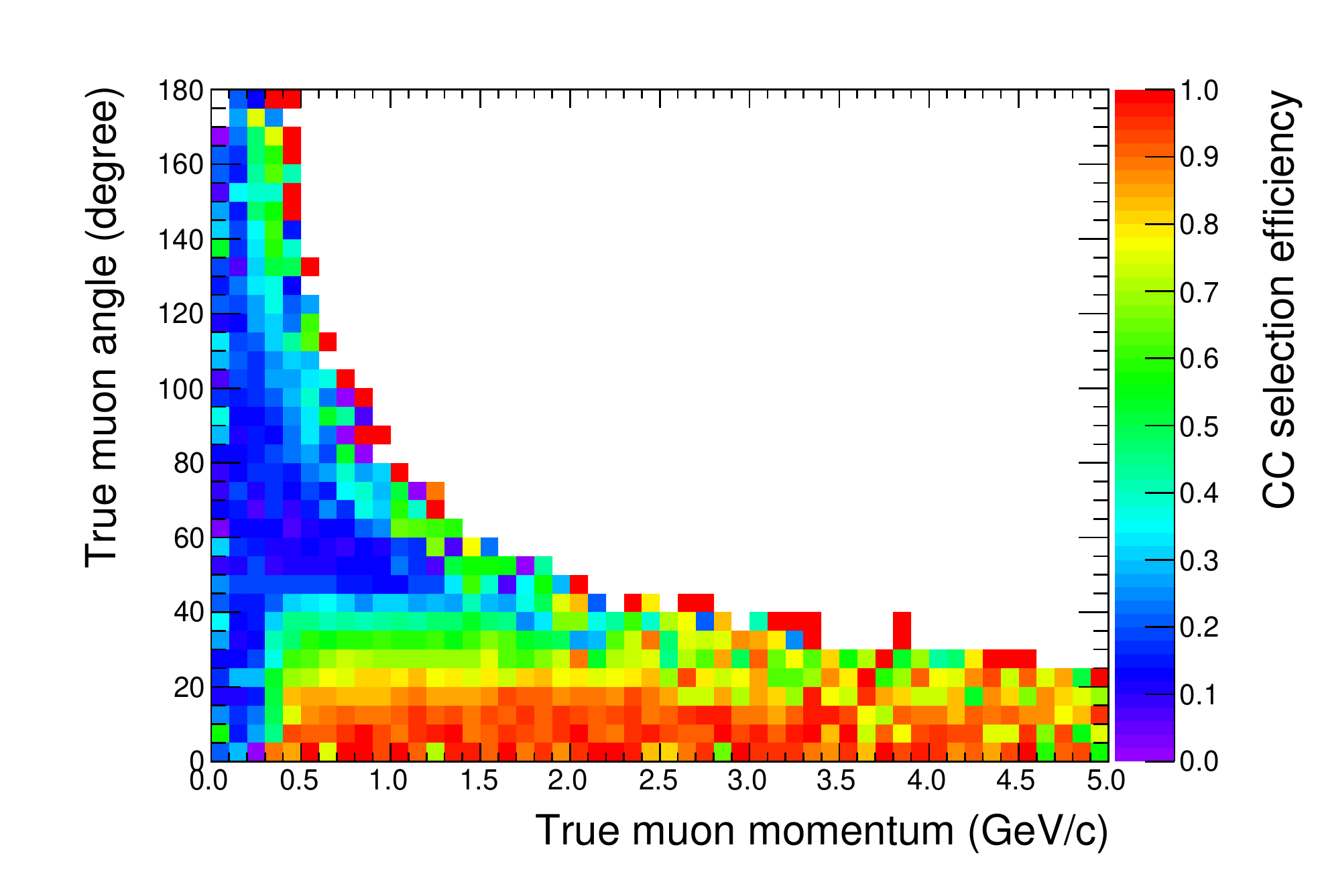}
	\\INGRID module
\end{minipage}
\caption{ Neutrino selection efficiency for CC interactions as a function of true muon scattering angle and momentum for the Water Module (upper left), Proton Module (upper right), and the INGRID module (lower). }
   \label{ fig:eff_2d }
\end{figure}

\begin{figure}[!htb]
\begin{minipage}{0.50\hsize}
\hspace*{-10mm}
        \centering
        \includegraphics[width=6.0cm,  bb=20 0 405 350]{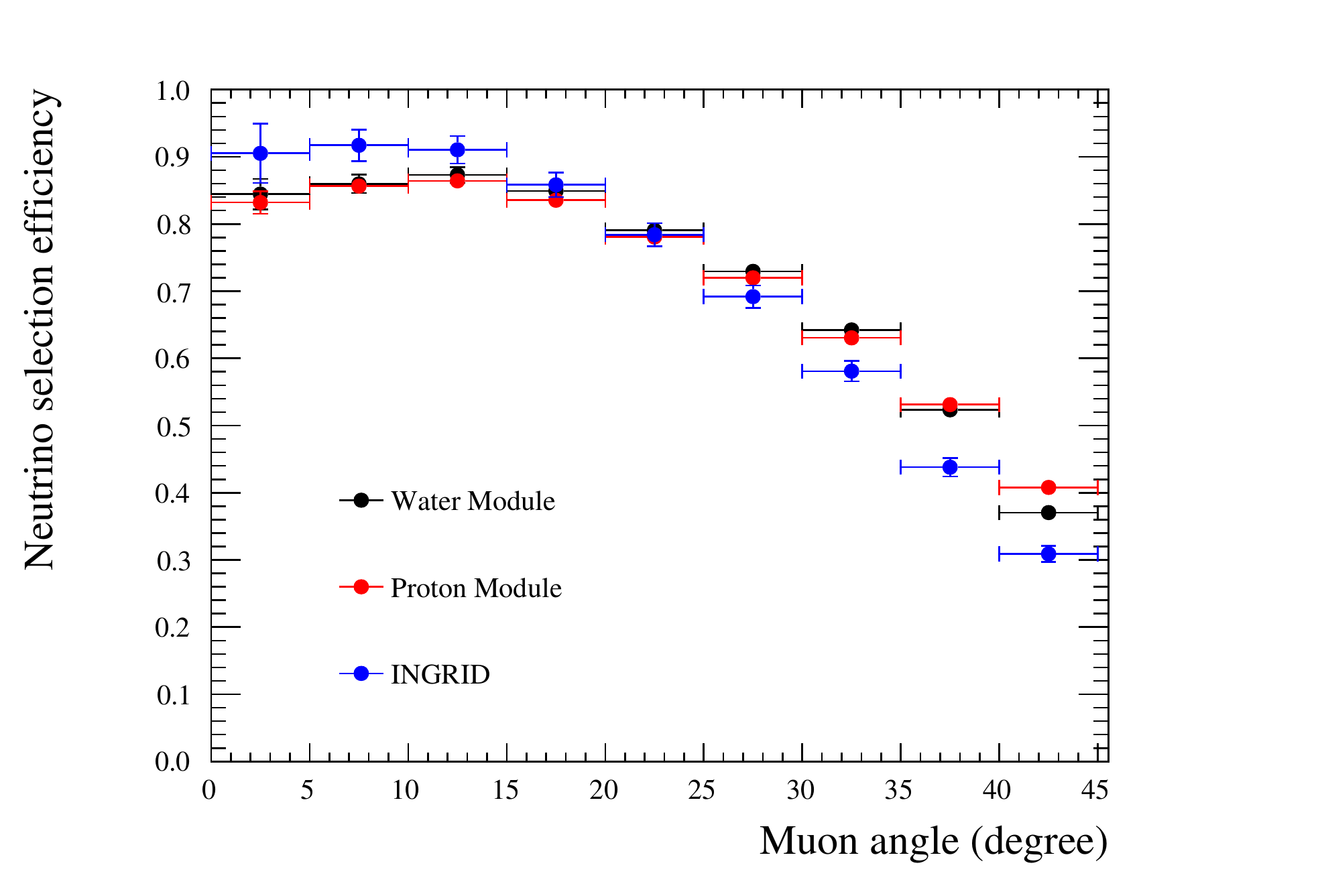}
\end{minipage}
\begin{minipage}{0.50\hsize}
\hspace*{-10mm}
        \centering
        \includegraphics[width=6.0cm,  bb=20 0 405 350]{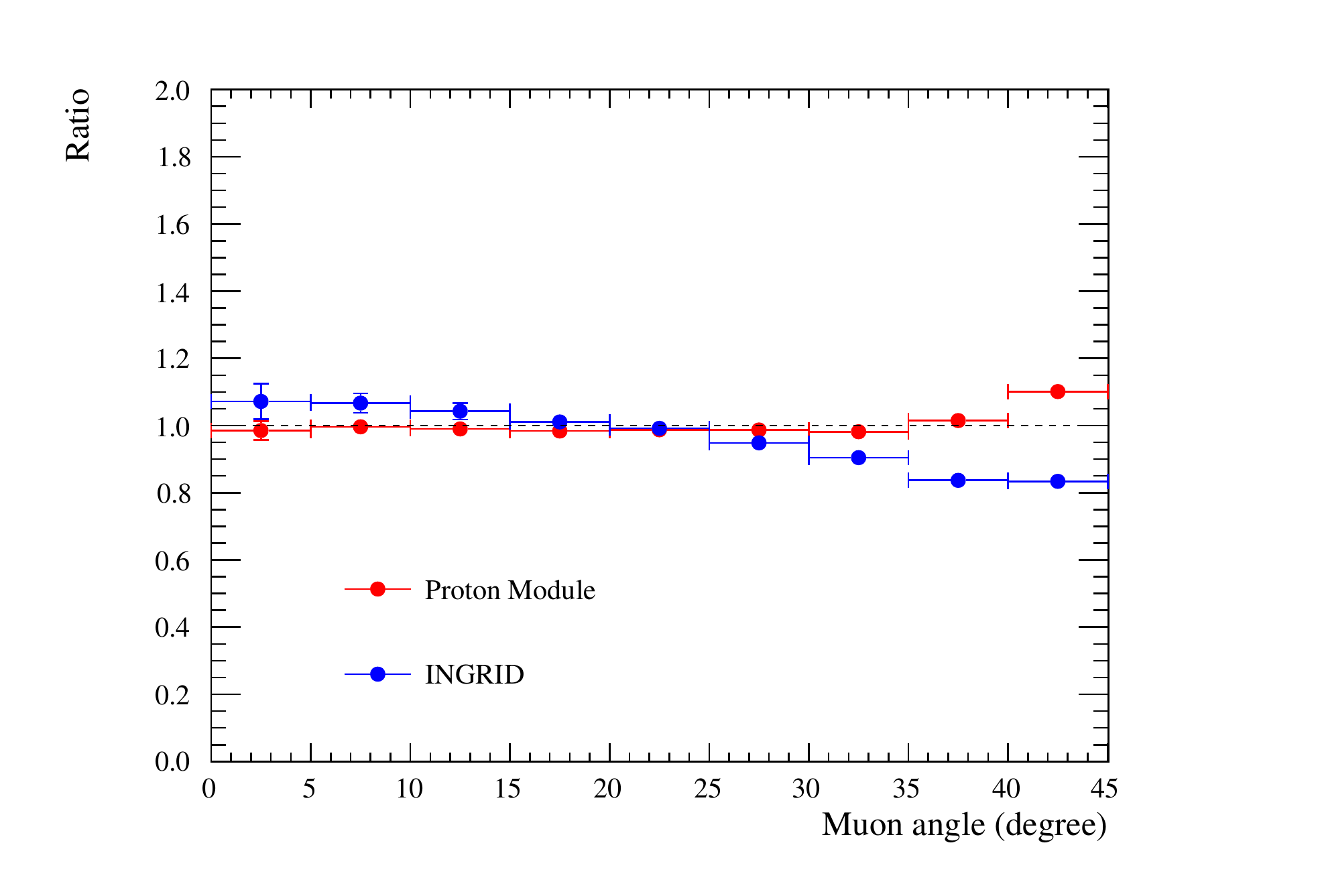}
\end{minipage}
\caption{ Selection efficiency of the signal as a function of the muon scattering angle for the three detectors (left) and their ratio with respect to the Water Module (right). }
   \label{ fig:eff_angle }
\end{figure}

Figure~\ref{ fig:eff_2d } shows the selection efficiency of CC interactions for the Water Module, Proton Module, and the INGRID module as a function of true muon scattering angle and momentum.
Because the selection efficiencies for the CC interactions with $\theta_{\mu}>$45$^{\circ}$ or $p_{\mu}<$400~MeV, are less than 10\%, these events are excluded from the signal sample defined in this analysis.
Figure~\ref{ fig:eff_angle } shows the efficiency of the signal for the three detectors and their ratios as a function of the muon scattering angle. 
The signal efficiency is almost constant as a function of muon momentum, while it depends on the muon scattering angle. 
In this analysis, the cross section is calculated by a sum of the differential cross sections as a function of the muon scattering angle, as described in Sec.~\ref{ sec:analysis }. In this method, the efficiency is calculated for each bin of the scattering angle and the dependence of the signal efficiency on the MC models used in this analysis is reduced.

\section{Cross section analysis}\label{ sec:analysis }

\subsection{Analysis method}

The flux-integrated $\nu_{\mu}$ cross sections of CC interactions on water ($\sigma_{\rm{H_{2}O}}$), hydrocarbon ($\sigma_{\rm{CH}}$), and iron ($\sigma_{\rm{Fe}}$) defined in a restricted phase space of the induced muon, $\theta_{\mu}<45^{\circ}$ and $p_{\mu}>$0.4~GeV/$c$, are measured as a sum of the differential cross sections as a function of the muon scattering angle. They are calculated as follows:
\begin{eqnarray}
\label{ math:eq_xsec_h2o }
\sigma_{A} &=&  \sum_{ij}\frac{ U_{ij {~D}} (N^{\rm{sel}}_{j{~D}}-N^{\rm{BG}}_{j{~D}}) }{\Phi^{A}_{{D}}T^{A}_{{D}} \varepsilon^{A}_{i{~D}}},
\end{eqnarray}
where ${A}$ represents the type of the target material ($\rm{H_{2}O}$, CH, and Fe) and $D$ is the corresponding detector (Water Module, Proton Module, and INGRID). $N^{\rm{sel}}$ is the number of selected events, $N^{\rm{BG}}$ is the number of expected backgrounds, $\Phi$ is the integrated $\nu_{\mu}$ flux, $T$ is the number of target nucleons, and $\varepsilon$ is the detection efficiency of the signal. Subscript $i$ is a bin index of the true muon scattering angle and subscript $j$ is a bin index of the reconstructed angle of the muon-like track. The true and reconstructed muon scattering angle bins are defined as 9 bins from 0$^{\circ}$ to 45$^{\circ}$ with a bin width of 5$^{\circ}$, which are optimized based on the detector resolution.
$U_{ij}$ is a probability that events in the reconstructed angle bin $j$ are in the true muon scattering angle bin $i$. 
The CC cross section ratios are estimated by taking the ratios of the $\sigma_{\rm{H_{2}O}}$, $\sigma_{\rm{CH}}$ and $\sigma_{\rm{Fe}}$.

The $N^{\rm sel}$ is estimated based on data as shown in Fig.~\ref{ fig:angle_event_wm } for the Water Module and Proton Module, and Table~\ref{ tab:Closs } for the INGRID module with the pileup correction. 
Except for the $\sigma_{\rm H_{2}{O}}$ measurement with the Water Module, in which the backgrounds from CC interactions on plastic scintillator ($N^{\rm{CH~BG}}_{\rm WM}$) are estimated with data from the Proton Module, other backgrounds $N^{\rm BG}$ are estimated by MC simulation.
The $N^{\rm{CH~BG}}_{\rm WM}$ is estimated as follows:
\begin{eqnarray}
N^{\rm{CH~BG}}_{\rm{WM}} &=& \sum_{i} \sigma_{i~\rm{CH}}\Phi^{\rm{CH}}_{\rm{WM}}T^{\rm{CH}}_{\rm{WM}}\varepsilon^{\rm{CH}}_{i~\rm{WM}}  \nonumber \\
\label{ math:eq_xsec_bg_calc2 }
&=& \sum_{ij} U_{ij \rm{~PM}}(N^{\rm{sel}}_{j\rm{~PM}}-N^{\rm{BG}}_{j\rm{~PM}})\frac{ \Phi^{\rm{CH}}_{\rm{WM}}T^{\rm{CH}}_{\rm{WM}}\varepsilon^{\rm{CH}}_{i~\rm{WM}} }{ \Phi^{\rm{CH}}_{\rm{PM}}T^{\rm{CH}}_{\rm{PM}} \varepsilon^{\rm{CH}}_{i~\rm{PM}} }~ ~ ~.
\end{eqnarray}
where 
$\sigma_{i~\rm{CH}}$ is the differential cross section on the CH target with $i$-th muon scattering angle bin. The other backgrounds are estimated by MC as summarized in Table~\ref{ tab:bgsummary } in detail.
The integrated $\nu_{\mu}$ fluxes $\Phi$ are estimated to be $\Phi^{\rm H_{2}O}_{\rm WM}=3.72\times 10^{13}$~/cm$^{2}$ with 7.25$\times$10$^{21}$ POT, $\Phi^{\rm CH}_{\rm PM}=3.02\times 10^{13}$~/cm$^{2}$ with 5.89$\times$10$^{21}$ POT, and $\Phi^{\rm Fe}_{\rm INGRID}=2.99\times 10^{13}$~/cm$^{2}$ with 5.89$\times$10$^{21}$ POT by MC, as shown in Table~\ref{ tab:flux_integ }. Although data samples used for the Proton Module and INGRID module are at the same delivered POT, the fact that the Proton Module is 1.2~m closer to the production target than the INGRID module, leads to a small difference in the integrated flux between them. 
The number of target nucleons, $T$, is calculated based on measurements performed during the detector construction as shown in Table~\ref{ tab:mass }. 
The detection efficiency of the signal, $\varepsilon$, is estimated by MC as shown in Fig.~\ref{ fig:eff_angle } in each true muon scattering angle bin.

\begin{table}[!htb]
	\begin{center}
		\caption{Summary of the fraction of the backgrounds after the event selection. Non-target element backgrounds are neutrino interactions on neither CH nor $\rm H_{2}O$ for the Water Module, on O, N, and Ti for the Proton Module, and on scintillators for the INGRID module.}
		\label{ tab:bgsummary }
		\scalebox{0.90}{
		\begin{tabular}{cccccccccc}
			\hline
			\hline
			Detector   & angle bin    &  CC out of      & Non-target            & NC           & $\overline{\nu}_{\mu}$, $\nu_{e}$, $\overline{\nu}_{e}$ &  Wall & INGRID & All BG \\
			           &              &  phase space    & element           &              & \\
			\hline
			Water      & 0-5$^\circ$       &  44.5 & 26.1 & 28.6  &  43.5   &    4.90   & 55.2 &  216  \\
			Module     & 5-10$^\circ$      &  98.2 & 55.1 & 67.4  &  99.2   &    36.7   & 96.6 &  477  \\
			           & 10-15$^\circ$     &  145  & 72.0 & 83.7  &  103    &    73.8   & 10.3 &  615  \\
			           & 15-20$^\circ$     &  171  & 76.3 & 86.4  &  75.6   &    113    & 90.0 &  654  \\
			           & 20-25$^\circ$     &  165  & 58.2 & 76.7  &  51.7   &    58.9   & 77.2 &  527  \\
			           & 25-30$^\circ$     &  113  & 43.6 & 54.8  &  30.9   &    32.7   & 72.9 &  377  \\
			           & 30-35$^\circ$     &  84.4 & 27.6 & 32.4  &  19.6   &    13.2   & 33.4 &  229  \\
			           & 35-40$^\circ$     &  35.0 & 15.6 & 16.3  &  10.9   &    12.4   & 25.8 &  126  \\
			           & 40-45$^\circ$     &  40.2 & 7.70 & 6.99  &  4.72   &    1.24   & 9.74 &  82.4 \\
			           & Total        &  896  & 382  & 453   &  439    &    3.47   & 564  &  3300 \\
			\hline
			Proton     &  0-5$^\circ$      &  99.0 & 12.9 & 60.3  &  79.7   &    38.2   & 57.2 &  346 \\
			Module     &  5-10$^\circ$     &  255  & 35.7 & 145   &  172    &    47.4   & 154  &  905 \\
			           &  10-15$^\circ$    &  338  & 48.0 & 174   &  162    &    75.2   & 183  &  975 \\
			           &  15-20$^\circ$    &  352  & 49.1 & 177   &  129    &    145    & 150  &  997 \\
			           &  20-25$^\circ$    &  313  & 43.3 & 144   &  78.8   &    104    & 124  &  803 \\
			           &  25-30$^\circ$    &  243  & 34.5 & 101   &  50.6   &    83.7   & 107  &  616 \\
			           &  30-35$^\circ$    &  148  & 25.3 & 63.4  &  30.1   &    23.9   & 90.4 &  379 \\
			           &  35-40$^\circ$    &  67.6 & 16.6 & 32.4  &  15.2   &    20.5   & 56.1 &  207 \\
			           &  40-45$^\circ$    &  83.5 & 9.69 & 17.3  &  8.96   &    12.4   & 28.3 &  159 \\
			           &  Total       &  1870 & 275  & 914   &  726    &    551    & 950  &  5290\\
			\hline
			INGRID     &  0-5$^\circ$      &  1370  & 507  & 769   &  766   &    95.7   & 7.96 &  3540  \\
			module     &  5-10$^\circ$     &  2910  & 1310 & 1690  &  1740  &    145    & 101  &  7900  \\
			           &  10-15$^\circ$    &  4990  & 1990 & 2680  &  2020  &    147    & 122  &  11900 \\
			           &  15-20$^\circ$    &  5630  & 2020 & 3280  &  1720  &    114    & 216  &  13000 \\
			           &  20-25$^\circ$    &  3990  & 1440 & 2100  &  1010  &    109    & 49.0 &  8690  \\
			           &  25-30$^\circ$    &  5520  & 1680 & 3070  &  993   &    126    & 88.8 &  11500 \\
			           &  30-35$^\circ$    &  3320  & 997  & 1660  &  588   &    58.0   & 19.7 &  6650  \\
			           &  35-40$^\circ$    &  3650  & 702  & 1170  &  338   &    34.9   & 19.2 &  5920  \\
			           &  40-45$^\circ$    &  3080  & 456  & 801   &  144   &    40.2   & 13.2 &  4530  \\
			           &  Total       &  34500 & 11100& 17200 &  9310  &    870    & 638  &  73600 \\
			\hline
			\hline
		\end{tabular}}
	\end{center}
\end{table}

\begin{table}[!htb]
	\centering
		\caption{ Integrated $\nu_{\mu}$ flux in the fiducial volume of each detector.   }
		\label{ tab:flux_integ }
		\begin{tabular}{ccccccc}
			\hline
			\hline
			                               &   Water Module      &  Proton Module      & INGRID module  \\
			\hline
			Integrated $\nu_{\mu}$ flux per $10^{21}$ POT ($\rm /cm^{2}$)  &  $5.13\times10^{13}$ & $5.13\times10^{13}$ & $5.08\times10^{13}$ \\
			Used POT in this analysis                      &  $7.25\times10^{20}$ & $5.89\times10^{20}$ & $5.89\times10^{20}$ \\
			Integrated $\nu_{\mu}$ flux per used POT ~($\rm /cm^{2}$)      &  $3.72\times10^{13}$ & $3.02\times10^{13}$ & $2.99\times10^{13}$ \\
			\hline                                                                                                         
			\hline                                                                                                         
		\end{tabular}
\end{table}


\begin{table}[!htb]
	\begin{center}
		\caption{Summary of the number of target nucleons.}
		\label{ tab:mass }
		\begin{tabular}{ccc}
			\hline
			                        &  Number of target nucleons         \\
			\hline
			$T^{\rm{H_{2}O}}_{\rm{WM}}$             &    $4.939\times10^{28}$      \\
			$T^{\rm{CH}}_{\rm{WM}}$                 &    $1.090\times10^{28}$      \\
			$T^{\rm{CH}}_{\rm{PM}}$                 &    $9.230\times10^{28}$      \\
			$T^{\rm{Fe}}_{\rm{ING}}$                &    $1.206\times10^{30}$      \\
			\hline
		\end{tabular}
	\end{center}
\end{table}

\begin{figure}[!htb]
\begin{minipage}{7.5cm}
\hspace*{-20mm}
   \centering
   \includegraphics[width=5.0cm,  bb=20 0 405 350]{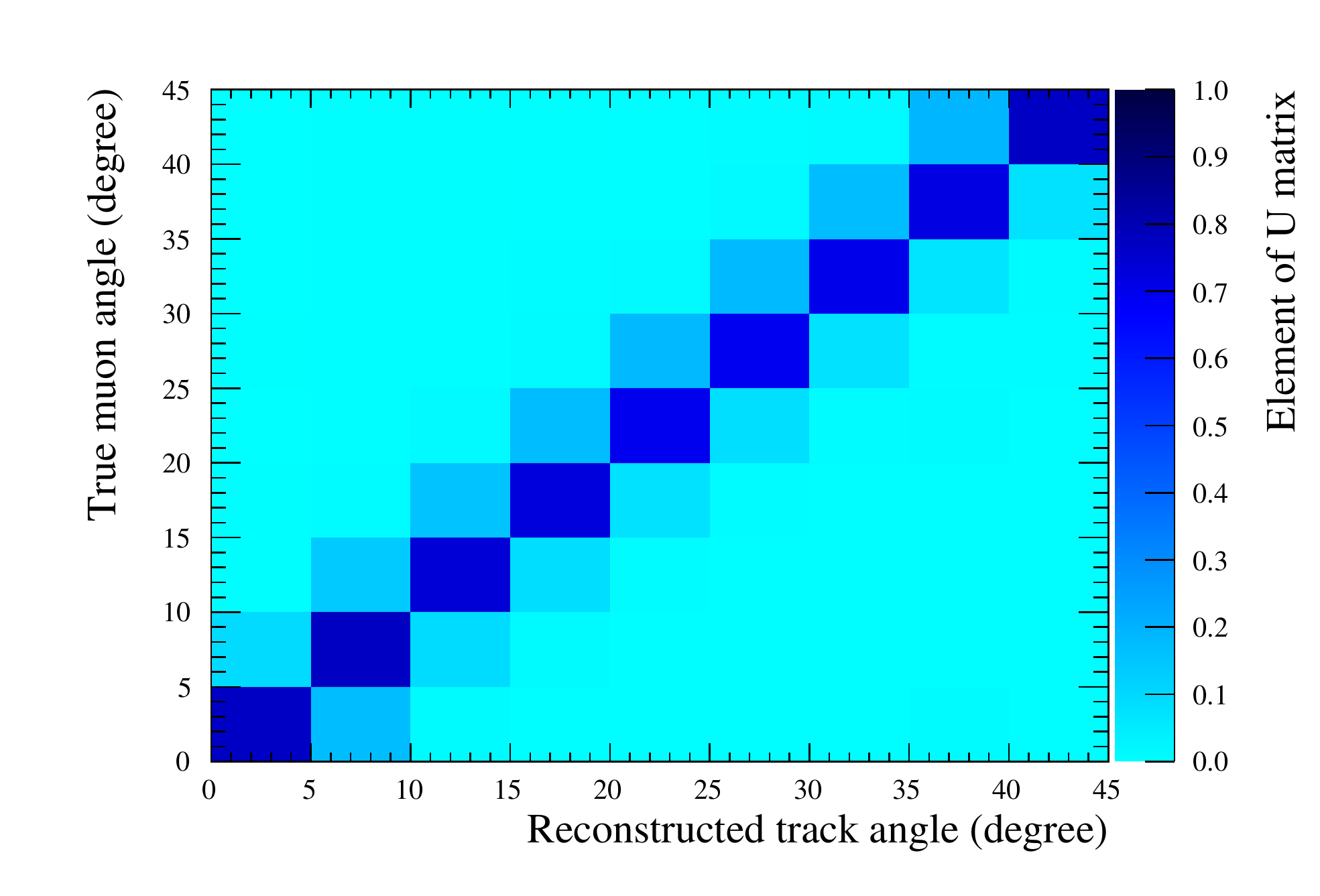}
   \\Water Module
\end{minipage}
\begin{minipage}{7.5cm}
\hspace*{-20mm}
   \centering
   \includegraphics[width=5.0cm,  bb=20 0 405 350]{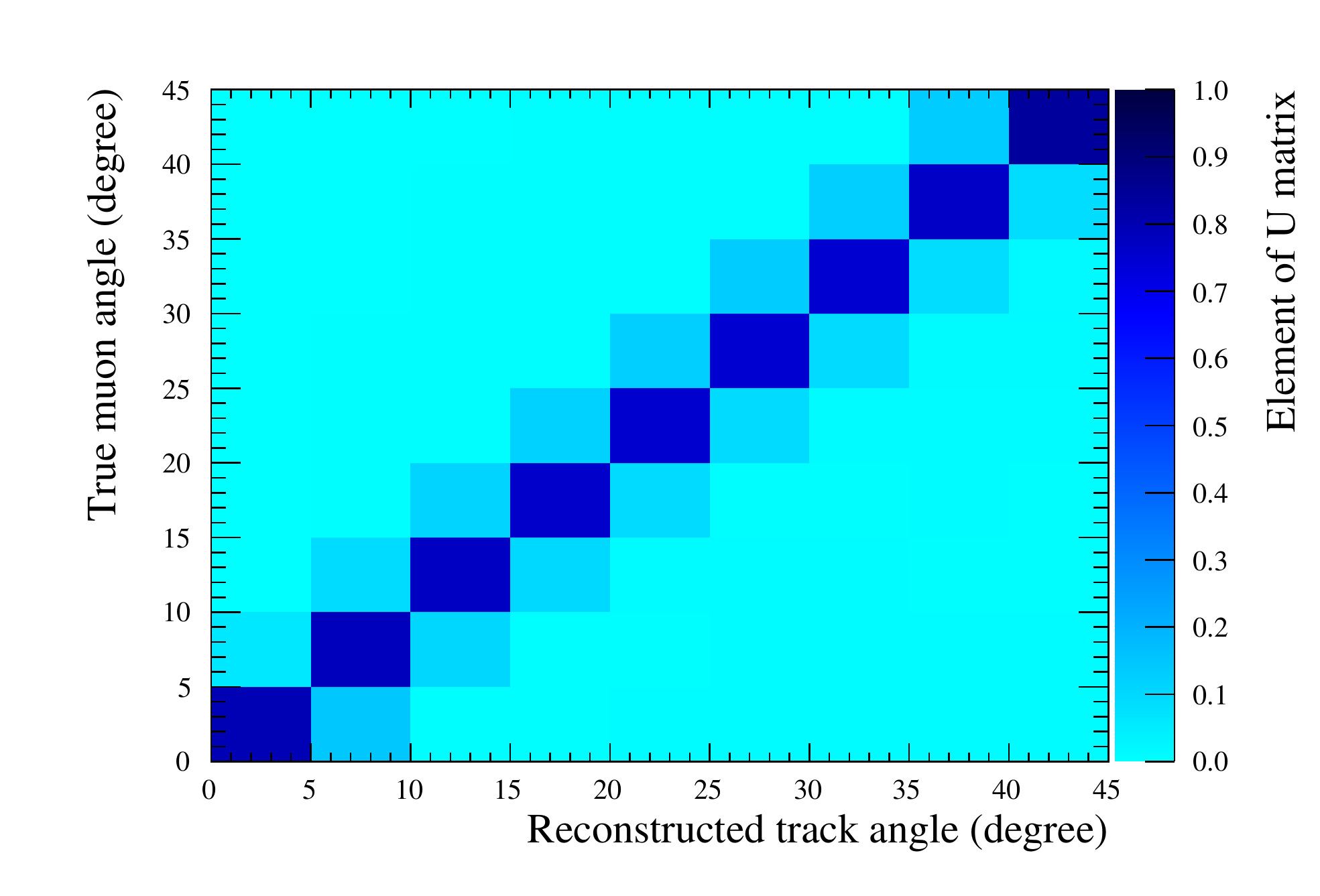}
   \\Proton Module
\end{minipage}
\hspace*{35mm}
\begin{minipage}{7.5cm}
\vspace*{10mm}
\hspace*{-15mm}
   \centering
   \includegraphics[width=5.0cm,  bb=20 0 405 350]{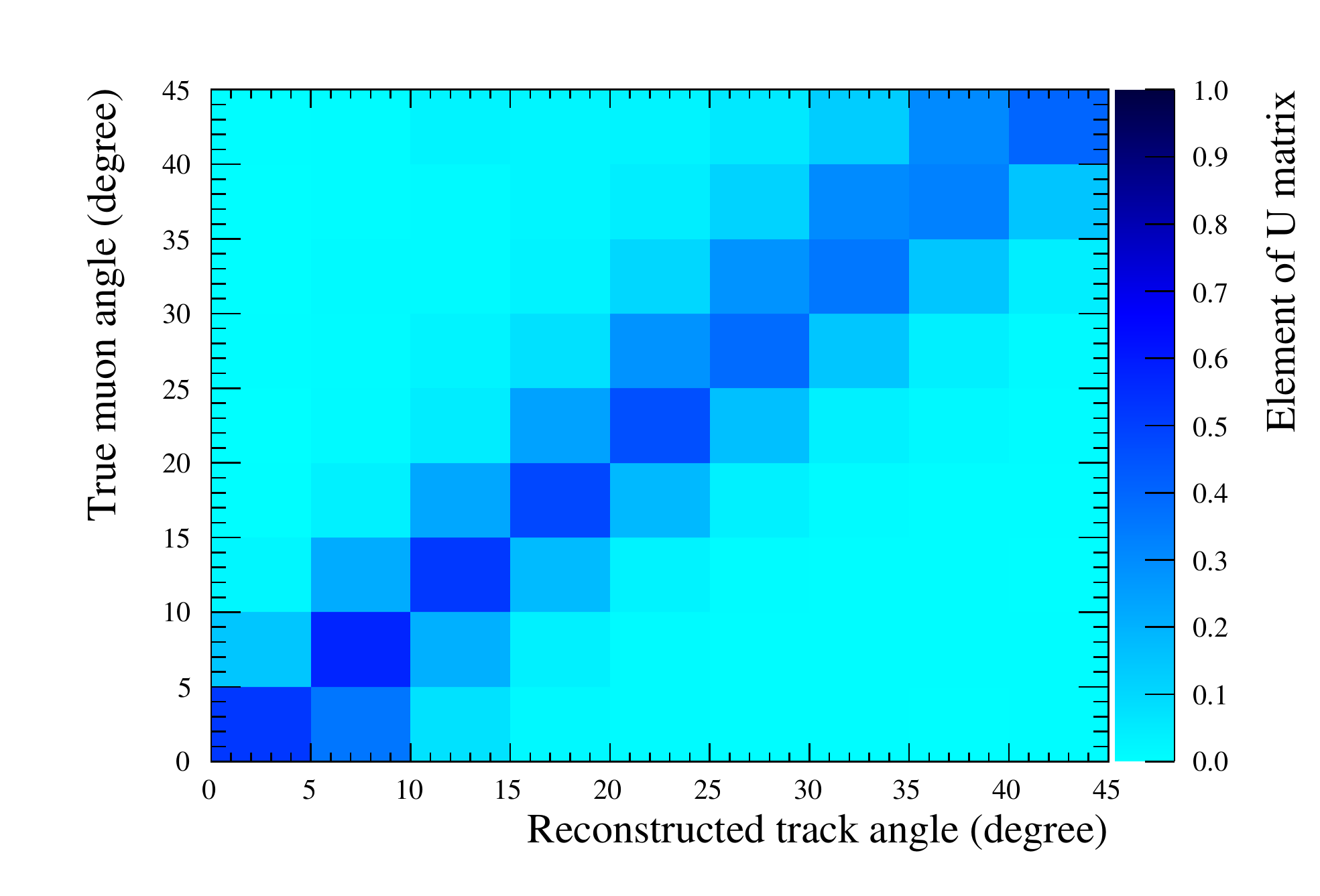}
   \\INGRID module
\end{minipage}
\caption{ Calculated reconstructed-to-true transfer matrix for the Water Module (upper left), Proton Module (upper right), and INGRID module (lower). The angle resolution for the INGRID module is worse than that for the Water Module and Proton Module due to differences of the scintillator width. }
   \label{ fig:pmat }
\end{figure}

The $U_{ij}$, probability that events in the reconstructed angle bin $j$ are in the true muon scattering angle bin $i$, is calculated as follows based on Bayes's theorem:
\begin{eqnarray}
	U_{ij} &=& P(\theta^{\rm{true}}_{i}|\theta^{\rm{recon}}_{j}) \nonumber \\
	&=& P(\theta^{\rm{recon}}_{j}|\theta^{\rm{true}}_{i})\times P(\theta^{\rm{true}}_{i})/P(\theta^{\rm{recon}}_{j}) \nonumber  \\
	&=& P(\theta^{\rm{recon}}_{j}|\theta^{\rm{true}}_{i})\times P(\theta^{\rm{true}}_{i})/\sum_{k}P(\theta^{\rm{recon}}_{j}|\theta^{\rm{true}}_{k})P(\theta^{\rm{true}}_{k}),
\end{eqnarray}
where $P(\theta^{\rm{recon}}_{j}|\theta^{\rm{true}}_{i})$ is calculated by MC as shown in Fig.~\ref{ fig:pmat }. The $P$($\theta^{\rm{true}}_{i}$) is calculated by an iterative unfolding method~\cite{DAgostini:1994fjx}, which is briefly described as follows:
\begin{enumerate}
\item $P(\theta^{\rm{true}}_{i})$ is set to a flat prior,
\item calculate $U_{ij}$,
\item $P(\theta^{\rm{true}}_{i})$ is set to $\sum_{j}U_{ij}(N^{\rm{sel}}_{j}-N^{\rm{BG}}_{j}) / \sum_{ij}U_{ij}(N^{\rm{sel}}_{j}-N^{\rm{BG}}_{j})$,
\item repeat (2)--(3).
\end{enumerate}
The number of required iterations is set to 10 as described in Sec.~\ref{ sec:closure }. 

\subsection{Consistency test}\label{ sec:closure }
\begin{figure}[!htb]
\hspace*{-20mm}
   \centering
   \includegraphics[width=6.0cm,  bb=20 0 405 350]{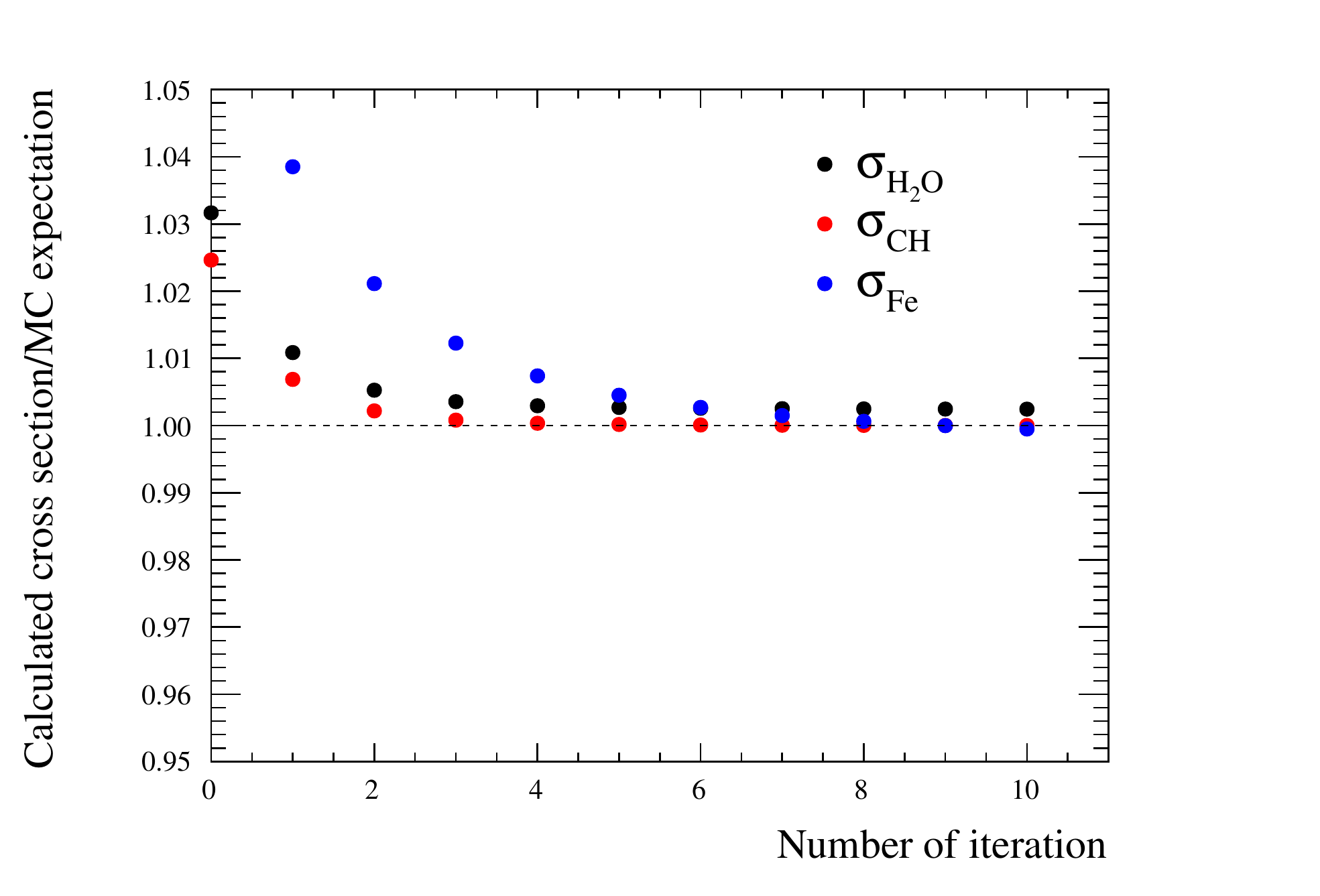}
   \caption{ Relation between the number of iterations and deviations of calculated cross sections from MC expectation. }
   \label{ fig:iteration }
\end{figure}

From the number of selected events and the quantities described earlier in this section, the flux-integrated CC cross sections on $\rm H_{2}O$, CH, Fe, and their ratios are calculated based on Eq.~\ref{ math:eq_xsec_h2o }. In this section, a consistency test is performed by replacing the number of selected events of data with that of the MC expectation, in order to check the consistency between the calculated cross section and MC expectation. Figure~\ref{ fig:iteration } shows the relation between the number of iterations and deviations of the calculated cross sections from MC expectation and the number of iterations when it is set to 10. Table~\ref{ tab:xsec_closure } shows the calculated cross sections and their consistency with the MC expectation. The consistency test is performed with not only the nominal cross section model but also a few alternative models. 


\begin{table}[!htb]
	\centering
	\caption{ Calculated cross sections using numbers of events expected by MC and their true values with the nominal model. }
		\label{ tab:xsec_closure }
		\begin{tabular}{ccc}
			\hline
			\hline
			Target       &  Calculated cross sections & Expected cross sections      \\
			\hline                                                                                                         
			$\rm H_{2}O$ &  $ 0.821 \times 10^{-38} \rm{ cm^{2}}$   &  $ 0.819 \times 10^{-38} \rm{ cm^{2}}$          \\
			CH           &  $ 0.832 \times 10^{-38} \rm{ cm^{2}}$   &  $ 0.832 \times 10^{-38} \rm{ cm^{2}}$          \\
			Fe           &  $ 0.904 \times 10^{-38} \rm{ cm^{2}}$   &  $ 0.904 \times 10^{-38} \rm{ cm^{2}}$          \\
			\hline                                                                                                         
			\hline                                                                                                         
		\end{tabular}
\end{table}

\section{Systematic uncertainties}

%
%
%
%

There are three main sources of systematic uncertainties for the cross section measurements: neutrino flux, neutrino interaction models, and detector response. The uncertainty evaluation for each source is detailed in this section.

\subsection{Systematic uncertainties from the neutrino flux }\label{ cross:sys_flux }

\begin{figure}[!htb]
   \hspace*{-10mm}
        \centering
        \includegraphics[width=10.0cm, bb=0 0 328 244]{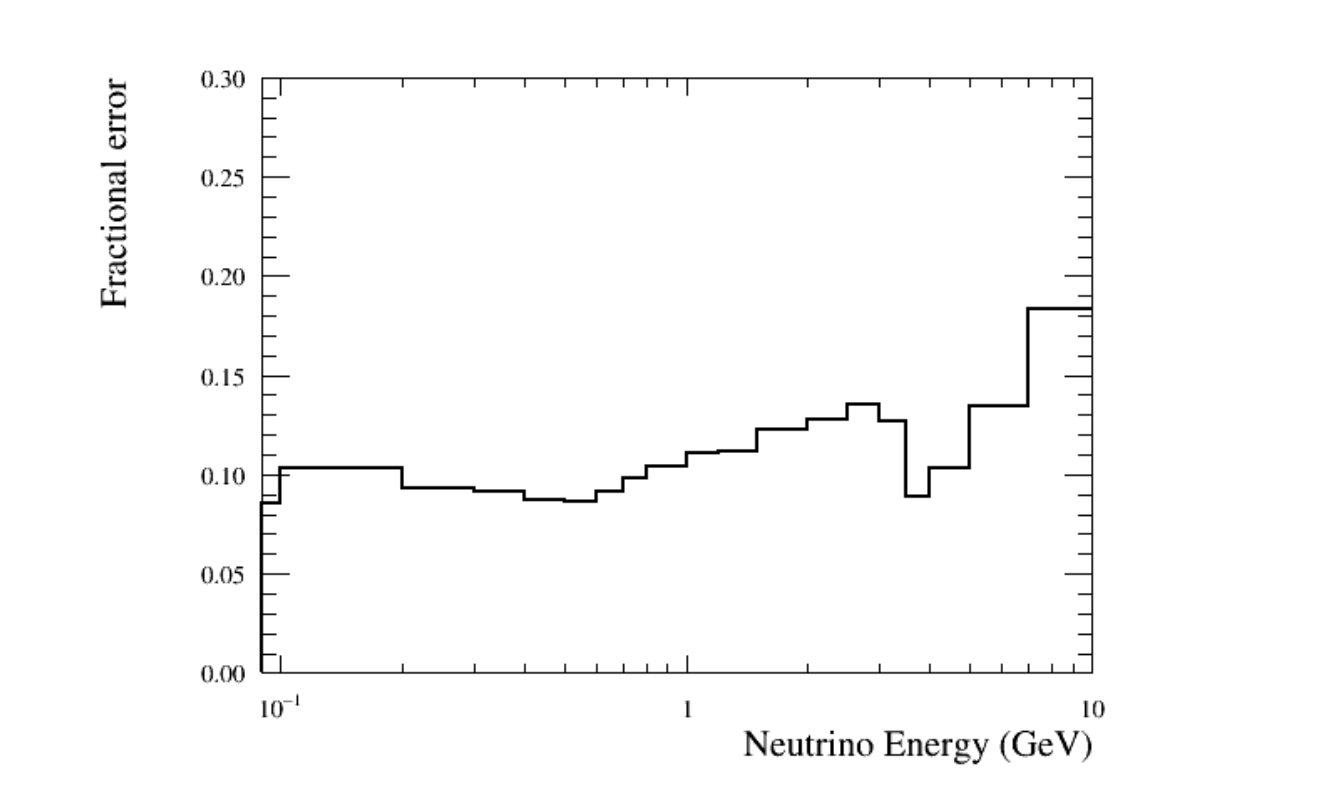}
\caption{ Total uncertainty of the muon neutrino on-axis flux in each true neutrino energy bin. }
  \label{ fig:flux_error }
\end{figure}

The T2K neutrino flux simulation, based on JNUBEAM mentioned in Sec.~\ref{ sec:mc }, relies on several measurements as inputs, including the hadron production measurements and information from the J-PARC beam line monitors. 
The uncertainty on the flux prediction takes into account the uncertainties in the measurements of the hadron scattering experiments, the hadronic interaction models and the uncertainties in the beam profile measurements with the beam line monitors.
Details about the sources of the flux uncertainty can be found in \cite{jnubeam}. Figure~\ref{ fig:flux_error } shows the calculated total on-axis flux uncertainty as a function of neutrino energy. 


The uncertainty of the neutrino flux is related to the systematic uncertainties on the number of expected backgrounds ($N^{\rm{BG}}$), integrated flux ($\Phi$), detection efficiency ($\varepsilon$), and reconstructed-to-true transfer matrix ($U$). To evaluate the systematic effects on the cross section measurement, the number of produced and selected neutrino events in each bin of the reconstructed track angle and true muon scattering angle is varied by using the calculated flux uncertainty, including correlations between the true neutrino energy bins. Therefore the variations of $N^{\rm{BG}}$, $\Phi$, $\varepsilon$, and $U$ are calculated and the variation of the cross section result is determined. This is repeated for many toy data sets and the 68\% range of the distribution of the cross section variation around the central value is taken as the size of the flux-related systematic uncertainty. The first row in Table~\ref{ tab:neut_param_xsec } shows the calculated flux uncertainties. They are approximately 10\% for the absolute cross section measurement and 1--2\% for the cross section ratios.

In addition, uncertainties due to difference in position of the INGRID module compared with the Water Module and Proton Module, and difference of the running periods between the Water Module, Proton Module, and INGRID module are estimated separately.
The former is estimated to be 0.31\% based on measurement of the detector location. The latter is estimated to be 1.03\% based on the beam stability measurements of the INGRID module between the different running periods. Their quadratic sums are summarized in the second row of Table~\ref{ tab:neut_param_xsec }.

\begin{table}[!htb]
	\centering
		\caption{ Summary of the systematic uncertainties for the cross section measurements (\%). }
		\label{ tab:neut_param_xsec }
		\scalebox{0.9}{\begin{tabular}{lccccccc}
			\hline
			\hline
			Systematic uncertainty                         &  $\sigma_{\rm H_{2}O}$  &  $\sigma_{\rm CH}$  &  $\sigma_{\rm Fe}$  &  $\sigma_{\rm H_{2}O}/\sigma_{\rm CH}$ & $\sigma_{\rm Fe}/\sigma_{\rm H_{2}O}$   &  $\sigma_{\rm Fe}/\sigma_{\rm CH}$      \\
			\hline
			\hline
			Flux-related          &  ${+10.8}$   & ${+11.5}$    & ${+13}$   & ${+0.6}$  & ${+1.8} $  & ${+1.1}$          \\
			(hadron production and beam line)                                               &  ${-8.9}$  & ${-9.6}$   & ${-11}$ & ${-0.6}$  & ${-1.8} $  & ${-1.2}$          \\
			\hline
			Flux-related    &  ${+0.3}$   & --    & --   & ${+1.3}$  & ${+1.1} $  & ${+0.3}$          \\
			(difference of running periods and location)                                               &  ${-0.3}$   & --    & --   & ${-1.3}$  & ${-1.1} $  & ${-0.3}$          \\
			\hline
			Interaction model-related                                    &  ${+2.6}$ & ${+3.1}$  & ${+5.2}$ & ${+2.3}$ & ${+4.0}$  & ${+2.7}$          \\
			                                               &  ${-2.6}$ & ${-3.1}$  & ${-5.2}$ & ${-2.3}$ & ${-4.0}$  & ${-2.7}$          \\
			\hline
			Detector response-related                              &  ${+2.9}$ & ${+2.5}$  & ${+1.5}$ & ${+4.5}$ & ${+3.4}$  & ${+2.8}$          \\
			                                               &  ${-2.9}$ & ${-2.5}$  & ${-1.5}$ & ${-4.5}$ & ${-3.4}$  & ${-2.8}$          \\
			\hline
			\hline
			Total                                          &  ${+11.5}$  & ${+13}$  & ${+14}$ & ${+5.2}$ & ${+5.7}$  & ${+4.1}$          \\
			&  ${-9.7}$   & ${-10}$  & ${-12}$ & ${-5.2}$ & ${-5.7}$  & ${-4.1}$          \\
			\hline                                                                                                         
			\hline                                                                                                         
		\end{tabular}}
\end{table}

\subsection{Systematic uncertainties from the neutrino interaction models } \label{ sec:sys_neut }

The NEUT neutrino interaction model has a number of uncertainties that can affect the detection efficiency ($\varepsilon$), background contamination ($N^{\rm{BG}}$), and reconstructed-to-true transfer matrix ($U$). 
To evaluate the model-related effect on the cross section measurement, for each $\pm$ 1$\sigma$ variation of a given interaction model parameter, a deviation of the cross section from the nominal value calculated based on the induced variation of $\varepsilon$, $N^{\rm{BG}}$, and $U$ are set as a systematic uncertainty.
Table~\ref{ tab:neut_param } shows the nominal values and the uncertainties of the neutrino interaction parameters. More details about the simulation models used can be found in \cite{Abe:2017vif}. In addition, uncertainties from pion final state interactions inside nuclei are taken into account: for each type of interaction, the uncertainties are assigned as normalization, as shown in Table~\ref{ tab:neut_param }.

When the uncertainty is calculated, no correlation amongst the different target nuclei for the Fermi momentum ($p_{F}$), binding energy ($E_{b}$), 2p2h, and CC coherent normalizations is assumed. Full correlation amongst the different nuclei is assumed for the other parameters. 
Table~\ref{ tab:neut_param_xsec2 } shows the calculated uncertainties and they are in the range between 2.6\% and 5.2\%. 
The dominant ones come from the uncertainties of the axial vector mass of the CCQE, CC1$\pi$, and the energy-dependent normalization of the CC multi-pion and DIS production. The uncertainty of the beam-induced backgrounds coming from outside of the detector is not included here, although it affects the $N^{BG}$. It is calculated as one of the detector systematics, as described in Sec.~\ref{ sec:sys_det }.

In addition to the systematic effects estimated by NEUT, the uncertainties of backscattered protons and pions produced by neutrino interactions with nuclei, which mainly affects the position of the reconstructed vertex, are estimated independently. A fraction of the events generated inside the fiducial volume have reconstructed vertices outside the fiducial volume due to backscattered secondary protons or pions. The fraction of such events is 3.0\% for the Water Module, 1.6\% for the Proton Module, and 2.0\% for the INGRID module with respect to the total number of selected events. The number and the uncertainty of such backscattered secondary particles may not be simulated well by NEUT, so a 50\% conservative uncertainty is assumed, which leads to 1.5\%, 0.8\%, and 1.0\% uncertainties for the Water Module, Proton Module, and INGRID module respectively in the total number of selected events. This is taken to be the 1$\sigma$ uncertainty for all reconstructed angle bins. In addition, no correlations between the target materials are assumed for this error.

\begin{table}[!htb]
	\centering
		\caption{ List of the interaction model parameters and uncertainties used in the analysis. }
		\label{ tab:neut_param }
		\begin{tabular}{llccc}
			\hline
			\hline
			Parameter           & Nominal value  & Uncertainties (1$\sigma$)       \\
			\hline
			\hline
			\multicolumn{3}{l}{CCQE-like} \\
			\hline
			$\rm M^{QE}_{A}$                          &  1.15~GeV/$c^{2}$           & 0.18~GeV/$c^{2}$                 \\
			$\rm p_{F}~^{12}C             $              &  217~MeV/c        & 31~MeV/c               \\
			$\rm p_{F}~^{16}O             $              &  225~MeV/c        & 31~MeV/c               \\
			$\rm p_{F}~^{56}Fe            $              &  250~MeV/c        & 35~MeV/c               \\
			$\rm E_{b}~^{12}C             $              &  25~MeV/c         & 9~MeV/c                \\
			$\rm E_{b}~^{16}O             $              &  27~MeV/c         & 9~MeV/c                \\
			$\rm E_{b}~^{56}Fe            $              &  33~MeV/c         & 11~MeV/c               \\
			$\rm 2p2h~normalization ^{12}C      $     &  100\%           & 100\%                 \\
			$\rm 2p2h~normalization ^{16}O      $     &  100\%           & 100\%                 \\
			$\rm 2p2h~normalization ^{56}Fe     $     &  100\%           & 100\%                 \\
			\hline
			\multicolumn{3}{l}{1$\pi$} \\
			\hline
			$\rm C_{A5}$                              &  1.01           & 0.12                 \\
			$\rm M^{Res}_{A}$                         &  0.95~GeV/$c^{2}$           & 0.15~GeV/$c^{2}$                 \\
			$\rm Isospin \frac{1}{2} bg$              &  1.30           & 0.20                 \\
			\hline
			\multicolumn{3}{l}{CC multi-pion and DIS production}           \\
			\hline
			\multicolumn{3}{l}{Normalization uncertainty is applied depending on neutrino energy by 0.4/$E_{\nu}$ (GeV) }           \\
			\hline
			\multicolumn{3}{l}{CC coherent }           \\
			\hline
			
			$\rm CC~coherent~normalization~^{12}C          $              &  100\%           & 30\%                 \\
			$\rm CC~coherent~normalization~^{16}O          $              &  100\%           & 30\%                 \\
			\hline
			\multicolumn{3}{l}{Normalization of NC interactions}                                                \\
			\hline
			$\rm NC~coherent~normalization     $              &  100\%           & 30\%                 \\
			NC multi-pion and DIS production normalization                 &  100\%           & 30\%                 \\
			\hline
			\multicolumn{3}{l}{Secondary interaction of pions}                                                \\
			\hline
			Pion absorption normalization                  &  100\%            & 50\%                \\
			Pion charge exchange normalization ($p_{\pi}<$500~MeV/$c$)            &  100\%            & 50\%                \\
			Pion charge exchange normalization ($p_{\pi}>$500~MeV/$c$)    &  100\%            & 30\%                \\
			Pion quasi elastic normalization ($p_{\pi}<$500~MeV/$c$)                 &  100\%            & 50\%                \\
			Pion quasi elastic normalization ($p_{\pi}>$500~MeV/$c$)                 &  100\%            & 30\%                \\
			Pion inelastic normalization                  &  100\%            & 50\%                \\
			\hline                                                                                                         
			\hline                                                                                                         
		\end{tabular}
\end{table}

\begin{table}[!htb]
	\centering
	\caption{ Summary of the neutrino interaction model-related uncertainties for each cross section measurement (\%). Only the dominant systematic parameters are shown. }
		\label{ tab:neut_param_xsec2 }
		\scalebox{0.85}{
		\begin{tabular}{lcccccc}
			\hline
			\hline
			Parameter                    &  $\sigma_{\rm H_{2}O}$  &  $\sigma_{\rm CH}$  &  $\sigma_{\rm Fe}$  &  $\sigma_{\rm H_{2}O}/\sigma_{\rm CH}$ & $\sigma_{\rm Fe}/\sigma_{\rm H_{2}O}$   &  $\sigma_{\rm Fe}/\sigma_{\rm CH}$      \\
			\hline
			$\rm M^{QE}_{A}$             &  1.1 & 1.2  & 1.6 & 0.1 & 0.5 & 0.5         \\
			$\rm M^{Res}_{A}$            &  0.6 & 1.2  & 2.0 & 0.5 & 1.5 & 1.0         \\
			$\rm C_{A5}$                 &  0.1 & 0.3  & 1.0 & 0.3 & 1.0 & 0.7         \\
			$\rm Isospin \frac{1}{2} bg$ &  0.1 & 0.2  & 0.6 & 0.2 & 0.5 & 0.3         \\
			CC multi-pion and DIS production       &  0.2 & 0.7  & 1.3 & 0.5 & 1.1 & 0.6         \\
			NC multi-pion and DIS production       &  0.8 & 1.1  & 1.7 & 0.3 & 0.9 & 0.6         \\
			$\rm 2p2h~normalization ^{12}C  $ &  0.1 & 0.4  & 0.3 & 0.4 & 0.3 & 0.4         \\
			$\rm 2p2h~normalization ^{16}O  $ &  0.2 & 0    & 0   & 0.2 & 0.2 & 0           \\
			$\rm 2p2h~normalization ^{56}Fe $ &  0   & 0    & 0.4 & 0   & 0.4 & 0.4         \\
			Pion absorption normalization&  0.2 & 0.1  & 0.2 & 0.2 & 0.4 & 0.2         \\
			Pion quasi elastic normalization ($p_{\pi}<$500~MeV/$c$)&  0.2 & 0.1  & 0.1 & 0.2 & 0.4 & 0.4         \\
			Pion quasi elastic normalization ($p_{\pi}>$500~MeV/$c$)&  0.2 & 0.1  & 0.3 & 0.1 & 0.1 & 0.1         \\
			Pion inelastic normalization    &  0.2 & 0.2  & 0.3 & 0.1 & 0.1 & 0.2         \\
			Backscattered protons and pions &  1.5 & 0.8  & 1.0 & 1.8 & 1.8 & 1.3         \\
			\hline                                                                                                                                                        
			$\rm Total                 $ &  2.6 & 3.1  & 5.2 & 2.3 & 4.0  & 2.7          \\
			\hline                                                                                                         
			\hline                                                                                                         
		\end{tabular}}
\end{table}

\subsection{Systematic uncertainties from the detector responses } \label{ sec:sys_det }

Uncertainties of the detector response are estimated based on the difference between data and MC for the cosmic rays and beam-induced muons coming from outside of the detectors. We take into account the following errors: target mass, MPPC noise, scintillator crosstalk, hit efficiency of the scintillator, event pileup, beam-induced backgrounds from outside of the detector, two dimensional tracking efficiency, and three-dimensional tracking efficiency. In addition, the uncertainties of the reconstructed variables used for the event selections are taken into account as follows: two-dimensional track matching with the INGRID modules, three-dimensional track matching, vertexing, beam timing cut, veto and fiducial volume cut, and reconstructed angle cut. The effect from non beam-induced backgrounds is estimated to be less than 0.1\% with beam-off data and is not included in the systematic uncertainties.

In order to evaluate these uncertainties on the cross section measurement, MC simulations are produced by varying detector parameters independently within their uncertainties by 1$\sigma$. The difference in the number of selected events in each bin of reconstructed track angle with respect to varying their uncertainty by 1$\sigma$ defines the 1$\sigma$ standard deviation systematic uncertainty in the number of events. 
Table~\ref{ tab:detsys_sum_xsec_abs } shows a summary of the uncertainties from the detector response for the absolute cross sections.
For the measurements of the cross section ratios, no correlation is assumed between the three detectors except for the beam-induced backgrounds from the outside the detector, which is treated as a common uncertainty. The fourth row in Table~\ref{ tab:neut_param_xsec } shows the total uncertainty from the detector response. They are approximately 2\% for the absolute cross section measurement and 4\% for the cross section ratios because most of the systematics do not cancel between the detectors.

The total systematic uncertainties of the cross section measurements are estimated as a quadratic sum of the uncertainties of the neutrino flux, neutrino interaction, and detector response. Table~\ref{ tab:neut_param_xsec } shows the total systematic uncertainties and they are between 10\% and 14\% for the absolute cross section measurements and approximately 5\% for the cross section ratios.


\begin{table}[!htb]
	\centering
	\caption{ Summary of the detector systematic uncertainties for the absolute cross section measurements (\%). }
		\label{ tab:detsys_sum_xsec_abs }
		\scalebox{0.95}{
		\begin{tabular}{l|cc|c|c}
			\hline
			\hline
			cross section          & \multicolumn{2}{c|}{$\sigma_{\rm H_{2}O}$}  & $\sigma_{\rm CH}$ & $\sigma_{\rm Fe}$             \\
			\hline
			Detector               & Water Module                & Proton Module            & Proton Module         & INGRID           \\
			\hline
			Target mass            & 0.68            & 0.05        & 0.27     & 0.14           \\
			MPPC noise             & 0.01            & 0.09        & 0.39     & 0.09           \\
			Scintillator crosstalk & 0.30            & --          & --       & --             \\
			Hit efficiency         & 0.27            & 0.02        & 0.50     & 0.94           \\
			Event pileup           & 0.72            & 0.15        & 0.64     & 0.09           \\
			Beam-related background       & 1.09            & 0.31      & 1.31     & 0.38           \\
			Non-beam-related background   & 0.04            & 0.02      & 0.02     & 0.01           \\
	  	     2D Track reconstruction   & 0.60            & 0.28        & 1.18     & 0.43           \\
		  Track matching with INGRID   & 1.42            & 0.20        & 0.84     & --             \\
			3D track matching      & 0.89            & 0.13        & 0.56     & 0.35           \\
			Vertexing              & 0.44            & 0.05        & 0.20     & 0.28           \\
			Beam timing cut        & 0.06            & 0.01        & 0.01     & 0.01           \\
			VETO and FV cut        & 1.19            & 0.18        & 0.72     & 0.52           \\
			Acceptance cut         & --                & --            & --         & 0.61           \\
			\hline
			Total                  & \multicolumn{2}{c|}{2.88}       & 2.52     & 1.54           \\
			\hline
			\hline
		\end{tabular}}
\end{table}

\section{Results}

The measured flux-integrated cross sections of $\nu_\mu$ CC interactions per nucleon at a mean neutrino energy of 1.5~GeV, defined in a restricted phase space of induced muon, $\theta_{\mu}<45^{\circ}$ and $p_{\mu}>$0.4~GeV/$c$, on $\rm H_{2}O$, CH, and Fe are 
\begin{equation}
	\sigma^{\rm{H_{2}O}}_{\rm{CC}} = ( 0.840\pm 0.010 \rm{(stat.)} ^{+0.10}_{-0.08} \rm{(syst.)} )  \times 10^{-38} ~\rm{cm^2/nucleon}, 
\end{equation}

\begin{equation}
	\sigma^{\rm{CH}}_{\rm{CC}}     = ( 0.817\pm 0.007 \rm{(stat.)} ^{+0.11}_{-0.08} \rm{(syst.)} )  \times 10^{-38} ~\rm{cm^2/nucleon},
\end{equation}

\begin{equation}
	\sigma^{\rm{Fe}}_{\rm{CC}}     = ( 0.859\pm 0.003 \rm{(stat.)} ^{+0.12}_{-0.10} \rm{(syst.)} )  \times 10^{-38} ~\rm{cm^2/nucleon}. 
\end{equation}
The cross section ratios are 
\begin{equation}
	\frac{\sigma^{\rm{H_{2}O}}_{\rm{CC}}}{\sigma^{\rm{CH}}_{\rm{CC}}} = 1.028\pm 0.016 \rm{(stat.)} \pm 0.053 \rm{(syst.)},  
\end{equation}

\begin{equation}
	\frac{\sigma^{\rm{Fe}}_{\rm{CC}}}{\sigma^{\rm{H_{2}O}}_{\rm{CC}}} = 1.023\pm 0.012 \rm{(stat.)} \pm 0.058 \rm{(syst.)},  
\end{equation}

\begin{equation}
	 \frac{\sigma^{\rm{Fe}}_{\rm{CC}}}{\sigma^{\rm{CH}}_{\rm{CC}}} = 1.049 \pm0.010 \rm{(stat.)} \pm 0.043 \rm{(syst.)}. 
\end{equation}

\begin{figure}[!htb]
\begin{minipage}{7.5cm}
     	\hspace*{-10mm}
        \centering
        \includegraphics[width=6.0cm,  bb=20 0 405 350]{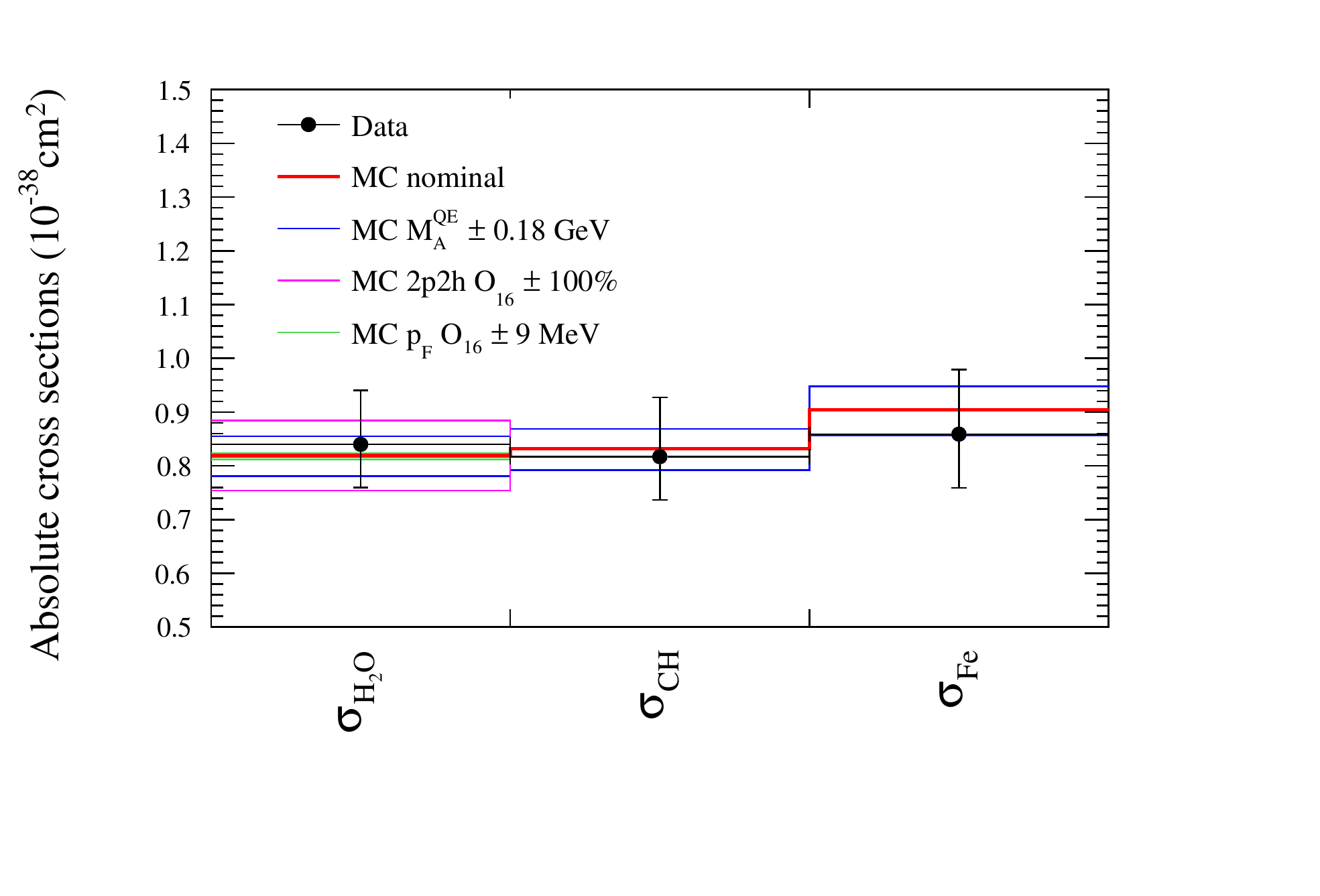}
\end{minipage}
\begin{minipage}{7.5cm}
        \centering
        \includegraphics[width=6.0cm, bb=20 0 405 350]{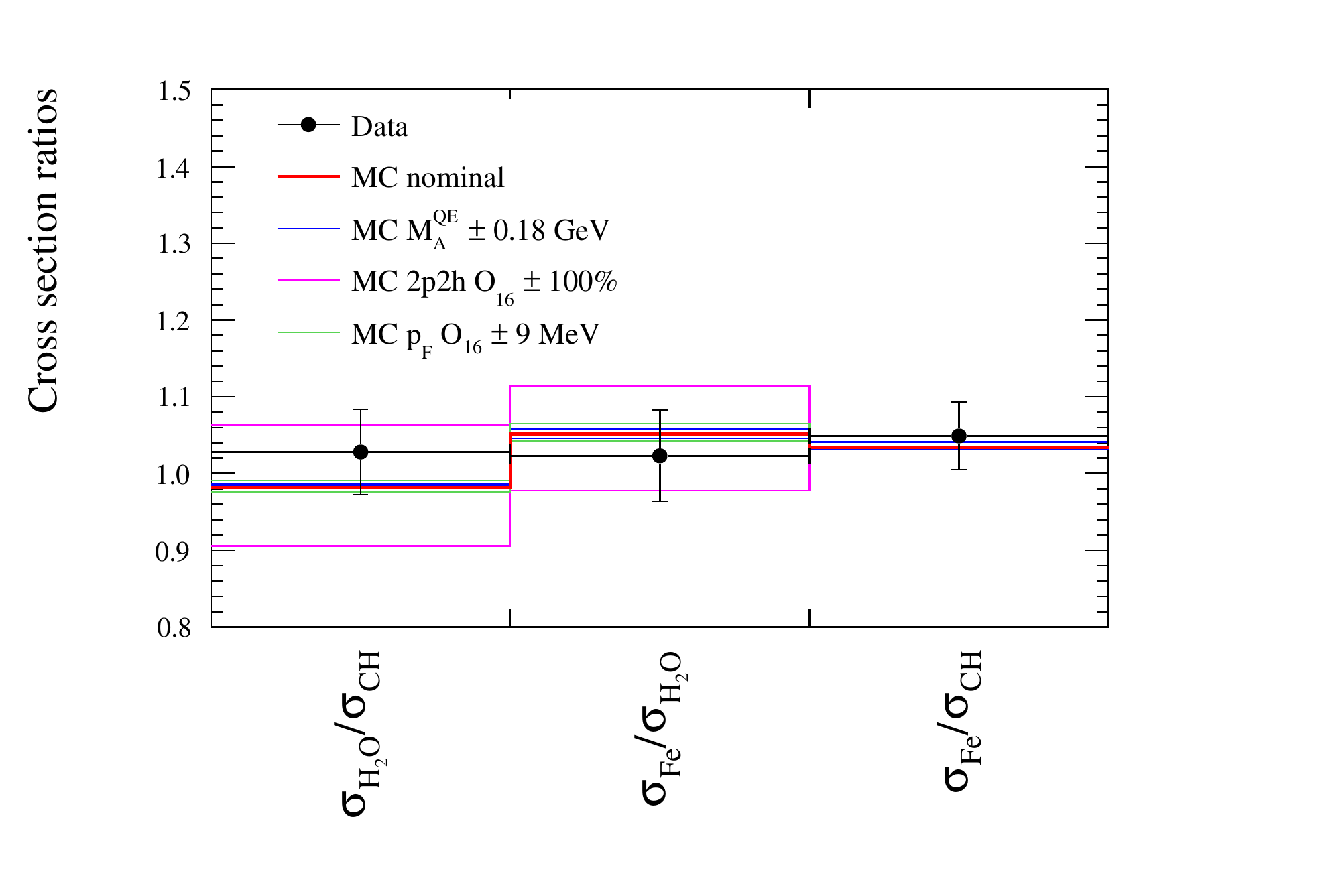}
\end{minipage}
\caption{ Results of the absolute cross sections (left) and cross section ratios (right) measurements with total uncertainties and theoretical predictions by NEUT. }
  \label{ fig:result }
\end{figure}

\begin{figure}[!htb]
\hspace*{-7mm}
\begin{minipage}{0.50\hsize}
        \centering
        \includegraphics[width=5.0cm, bb=20 0 405 350]{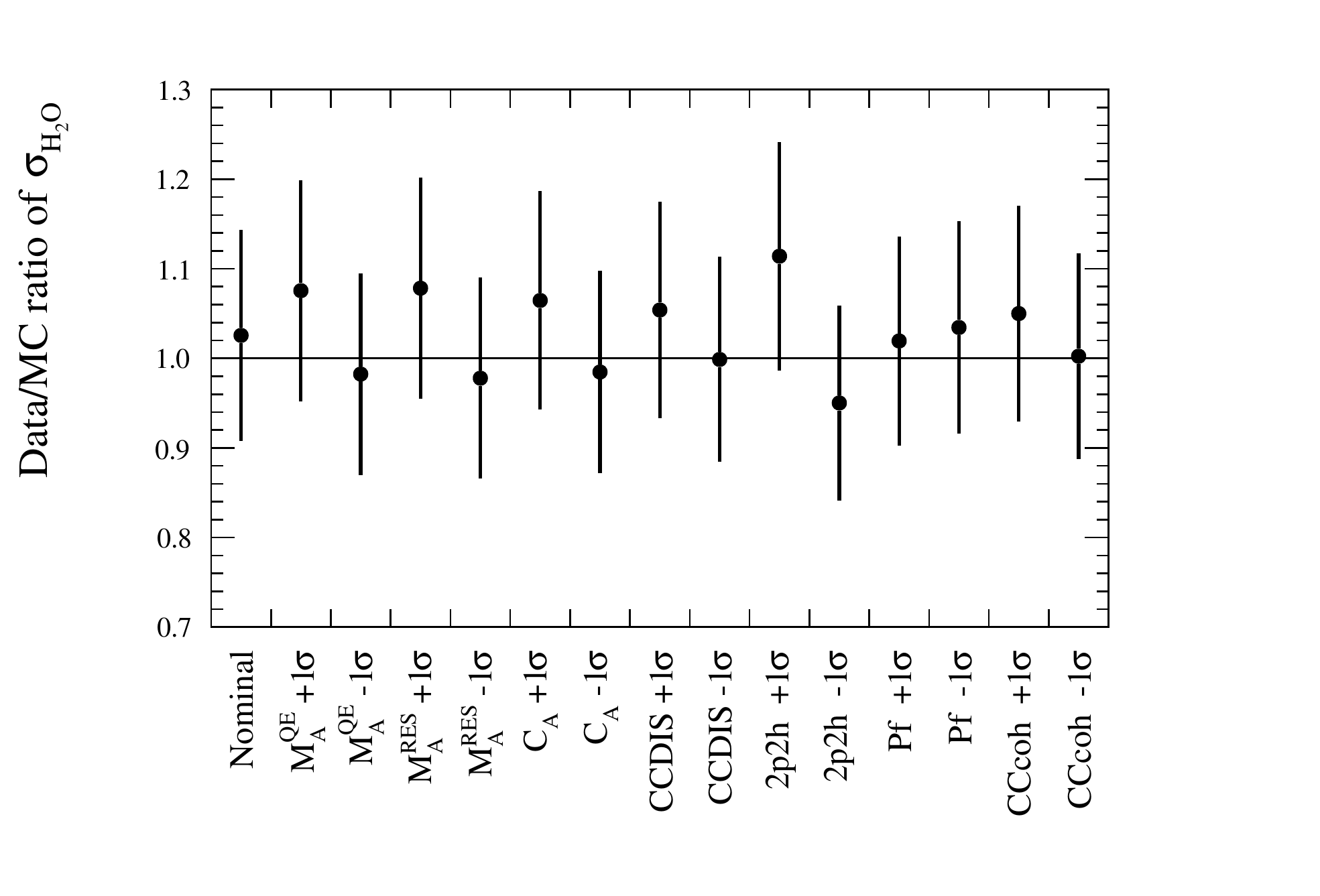}
\end{minipage}
\hspace*{-7mm}
\begin{minipage}{0.50\hsize}
        \centering
        \includegraphics[width=5.0cm, bb=20 0 405 350]{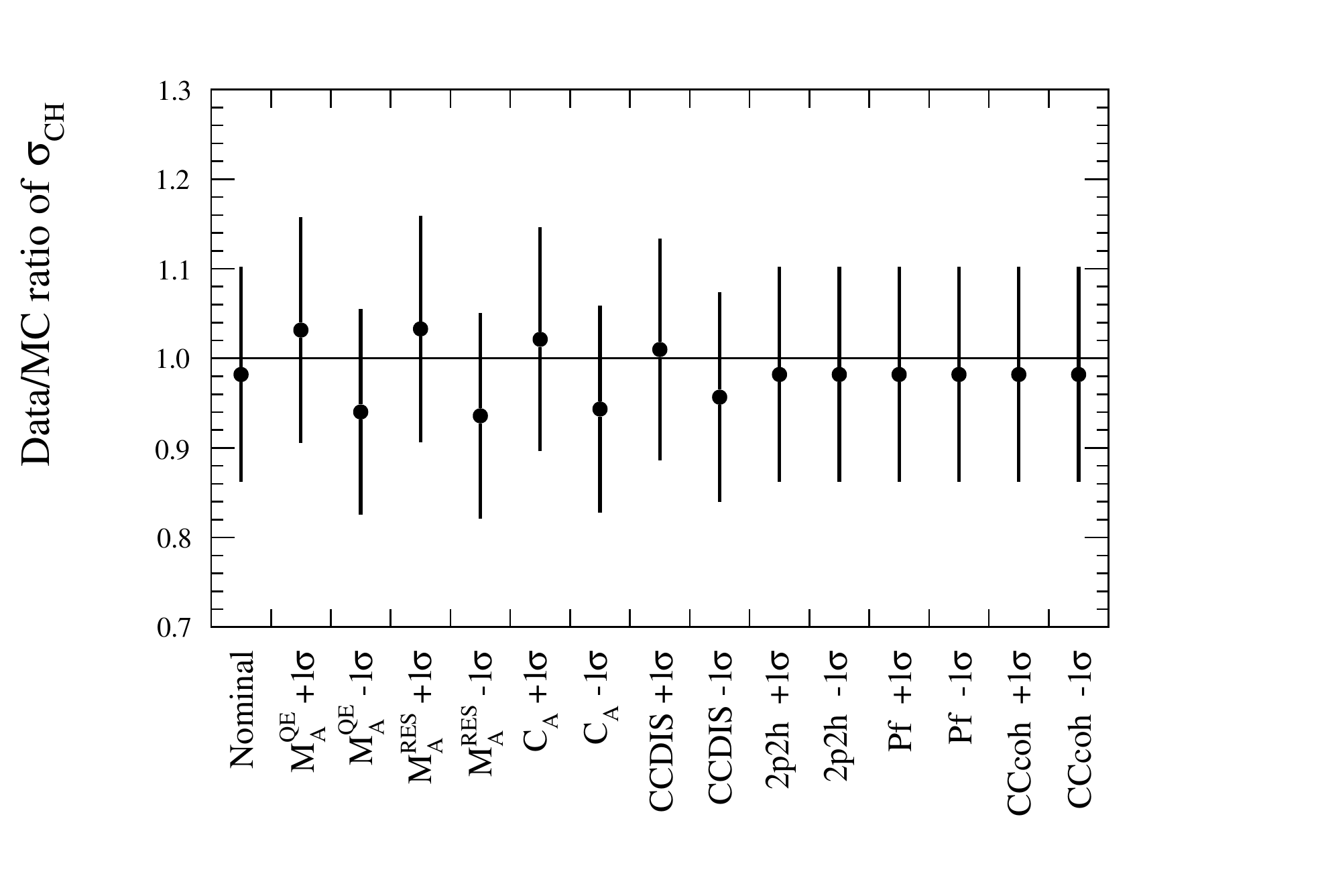}
\end{minipage}\\
\hspace*{-7mm}
\begin{minipage}{0.50\hsize}
        \centering
        \includegraphics[width=5.0cm, bb=20 0 405 350]{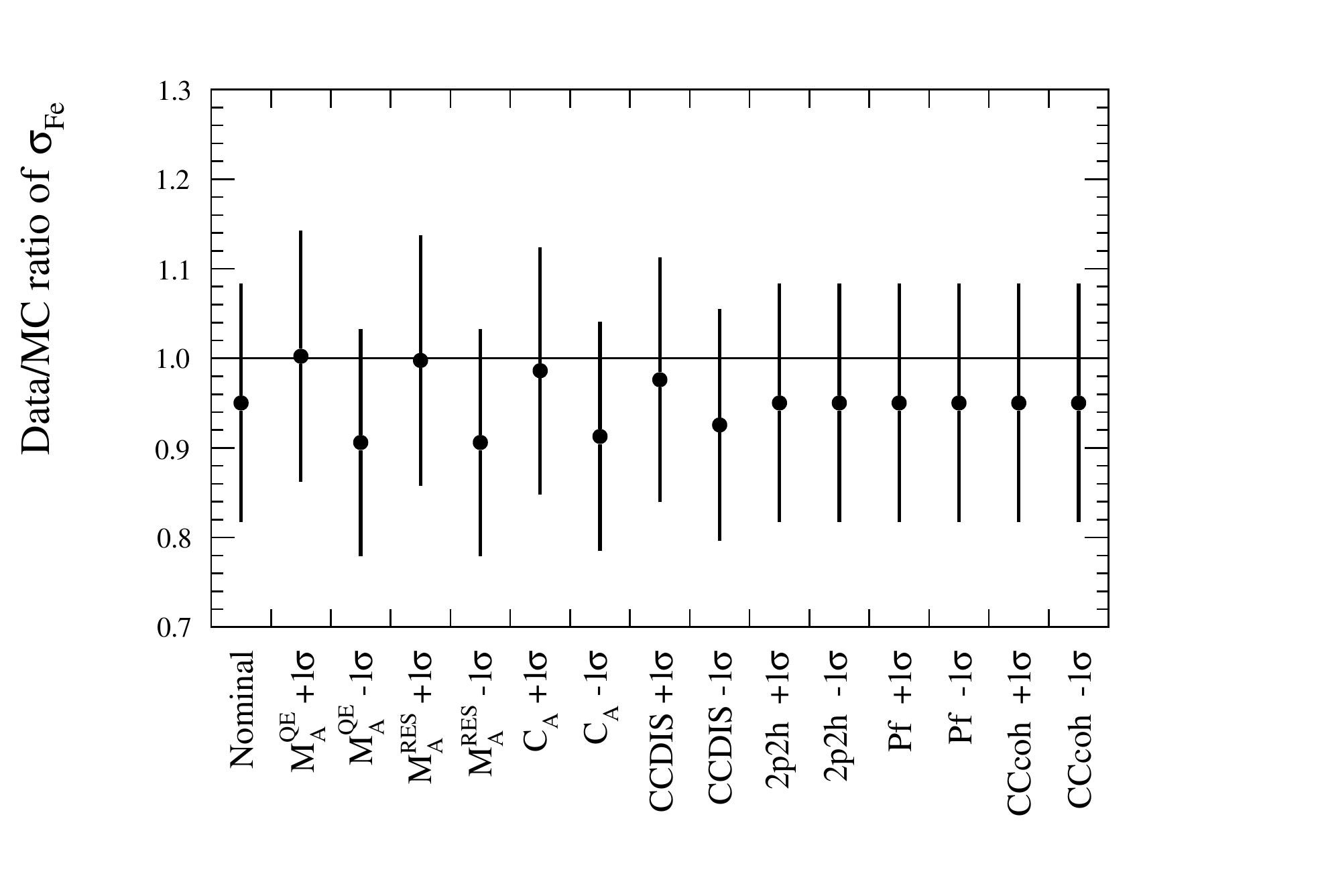}
\end{minipage}
\hspace*{-7mm}
\begin{minipage}{0.50\hsize}
        \centering
        \includegraphics[width=5.0cm, bb=20 0 405 350]{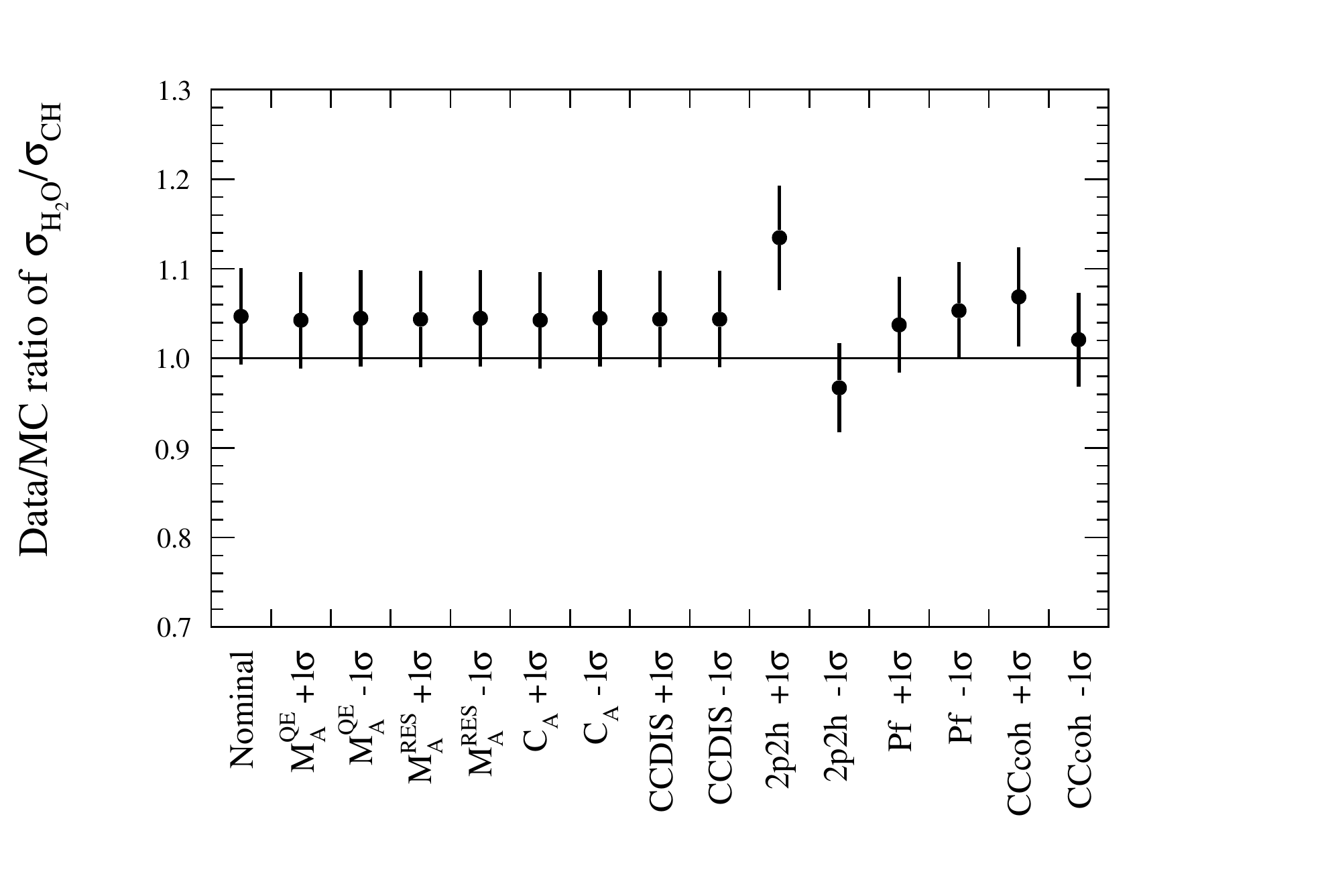}
\end{minipage}\\
\hspace*{-7mm}
\begin{minipage}{0.50\hsize}
        \centering
        \includegraphics[width=5.0cm, bb=20 0 405 350]{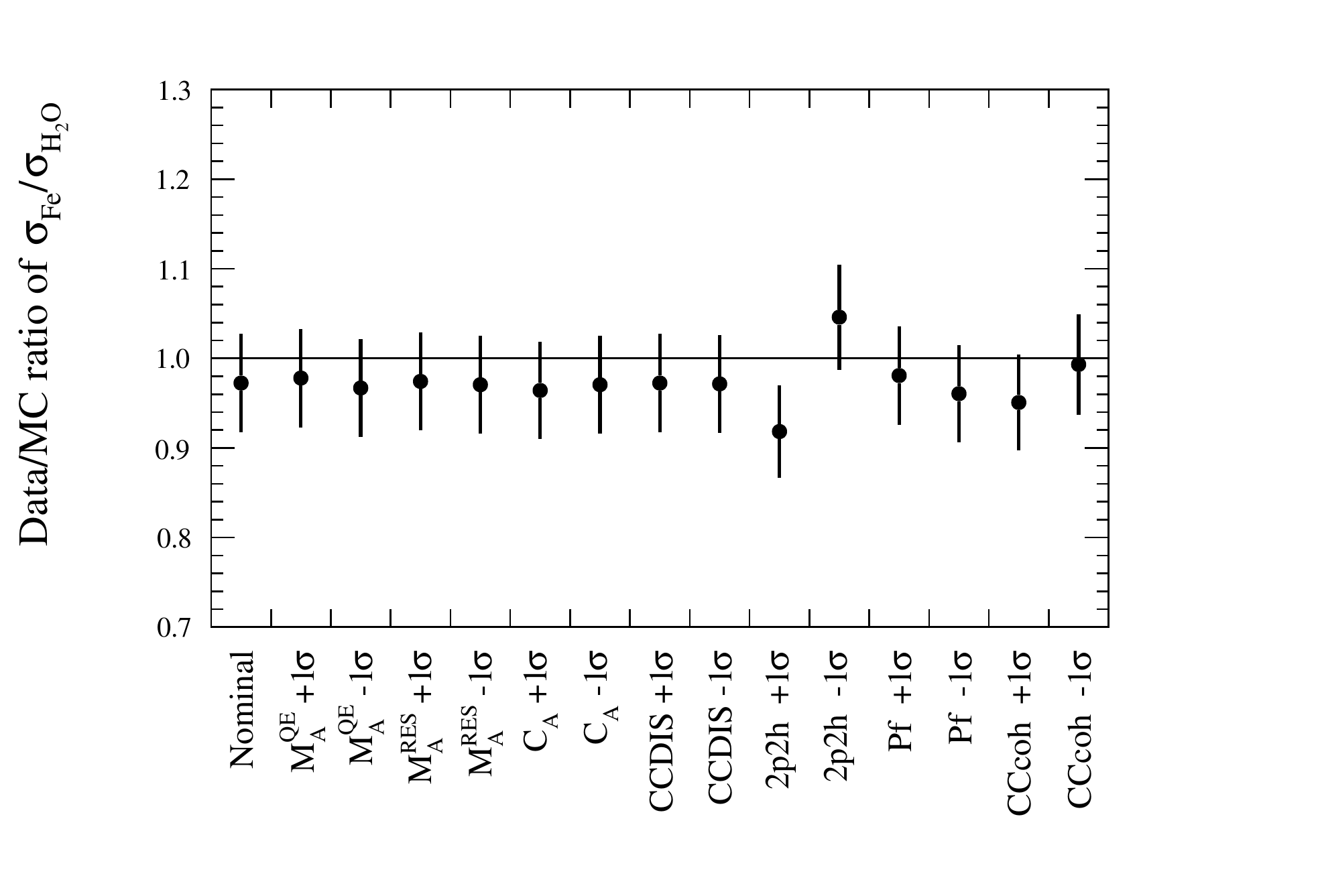}
\end{minipage}
\hspace*{-7mm}
\begin{minipage}{0.50\hsize}
        \centering
        \includegraphics[width=5.0cm, bb=20 0 405 350]{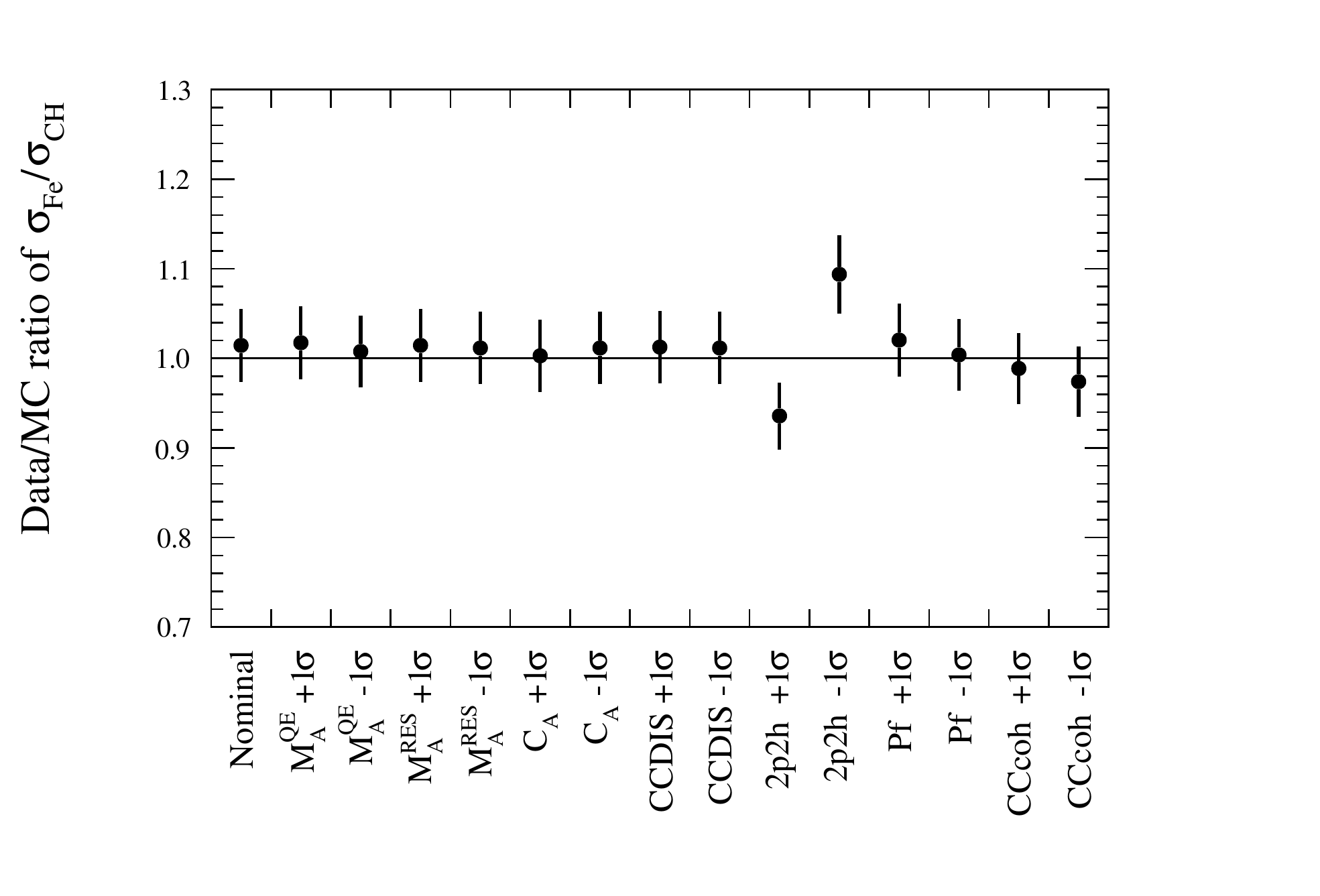}
\end{minipage}
\caption{ Ratio of the cross sections between data and NEUT predictions with the various cross section parameters listed in Table~\ref{ tab:neut_param }. Error bars show the sum of the statistical and systematics uncertainties of the measurement. }
  \label{ fig:neut_comparison }
\end{figure}

The errors of both the measured absolute cross section and cross section ratios are dominated by the systematic uncertainties. This is the most precise measurement to date of neutrino cross sections on water in this energy region and the first measurement of neutrino cross section ratios of water-to-hydrocarbon and water-to-iron.
Figure~\ref{ fig:result } shows the measured cross sections and their predictions by NEUT with nominal and varied parameters of the axial vector mass $M_{A}^{QE}$, normalization of 2p2h interaction, and Fermi momentum with 1$\sigma$, which have a relatively large effect on the cross section ratios in the parameters listed in Table~\ref{ tab:neut_param }. 
This is due to the fact that the variations of 2p2h normalization and Fermi momentum are only applied for $\rm H_{2}O$ but not CH and Fe as a conservative way to deal with our poor understanding of the target dependency of the neutrino interaction.
The predictions agree with data within the uncertainties. This measurement validates the neutrino interaction models on the water-target and difference between water, plastic, and iron and confirms the reliability of the T2K oscillation analysis.
Additional comparison of data and predictions with other parameters of the neutrino interaction listed in Table~\ref{ tab:neut_param } are summarized in Fig.~\ref{ fig:neut_comparison } and Table~\ref{ tab:neut_comparison }. All of the predictions agree with data within the estimated uncertainties. These results of the measurements and the neutrino flux at on-axis location are provided in text and ROOT format under the following link~\cite{cite:link}.


\begin{table}[!htb]
	\centering
	\caption{ Summary of the cross sections predicted by NEUT with the various parameter values listed in Table~\ref{ tab:neut_param } ($\times 10^{-38}~\rm{cm^2/nucleon}$ for the absolute cross sections). }
		\label{ tab:neut_comparison }
		\scalebox{0.85}{
		\begin{tabular}{ccccccc}
			\hline
			\hline
			Parameter                       &  $\sigma_{\rm H_{2}O}$  &  $\sigma_{\rm CH}$  &  $\sigma_{\rm Fe}$  &  $\sigma_{\rm H_{2}O}/\sigma_{\rm CH}$ & $\sigma_{\rm Fe}/\sigma_{\rm H_{2}O}$   & $\sigma_{\rm Fe}/\sigma_{\rm CH}$      \\
			\hline
			Nominal for $\rm H_{2}O,CH,H_{2}O/CH$     &  0.819              & 0.832           & -                   &     0.985   & -      & -       \\
			(RPA+RFG+MEC)                             &   \\                                                                                       
			\hline                                                                                                                                 
			Nominal for $\rm Fe,H_{2}O/Fe,Fe/CH$      &  0.860              & 0.875           & 0.904               &     0.982   & 1.052  & 1.034   \\
			(RFG+MEC)                                 &   \\                                                                                       
			\hline                                                                                                                                 
			$\rm M^{QE}_{A} - 1\sigma$                &  0.781              & 0.792           & 0.857               &     0.986   & 1.046  & 1.031   \\
			$\rm M^{QE}_{A} + 1\sigma$                &  0.855              & 0.869           & 0.948               &     0.984   & 1.058  & 1.041   \\
			$\rm M^{Res}_{A} - 1\sigma$               &  0.779              & 0.791           & 0.861               &     0.985   & 1.050  & 1.034   \\
			$\rm M^{Res}_{A} + 1\sigma$               &  0.859              & 0.873           & 0.948               &     0.984   & 1.054  & 1.037   \\
			$\rm C_{A5} - 1\sigma$                    &  0.789              & 0.800           & 0.871               &     0.986   & 1.061  & 1.046   \\
			$\rm C_{A5} + 1\sigma$                    &  0.853              & 0.866           & 0.941               &     0.984   & 1.054  & 1.037   \\
			$\rm Isospin \frac{1}{2} bg - 1\sigma$    &  0.806              & 0.818           & 0.889               &     0.985   & 1.051  & 1.035   \\
			$\rm Isospin \frac{1}{2} bg + 1\sigma$    &  0.835              & 0.848           & 0.923               &     0.985   & 1.053  & 1.037   \\
			$\rm CCother~shape - 1\sigma         $    &  0.797              & 0.809           & 0.880               &     0.985   & 1.052  & 1.036   \\
			$\rm CCother~shape + 1\sigma         $    &  0.841              & 0.854           & 0.928               &     0.985   & 1.053  & 1.037   \\
			$\rm p_{F}~C - 1\sigma                  $    &  0.819              & 0.835           & 0.904               &     0.981   & 1.052  & 1.028   \\
			$\rm p_{F}~C + 1\sigma                  $    &  0.819              & 0.825           & 0.904               &     0.993   & 1.052  & 1.045   \\
			$\rm p_{F}~O - 1\sigma                  $    &  0.824              & 0.832           & 0.904               &     0.991   & 1.043  & 1.034   \\
			$\rm p_{F}~O + 1\sigma                  $    &  0.812              & 0.832           & 0.904               &     0.976   & 1.065  & 1.034   \\
			$\rm Eb~C - 1\sigma                  $    &  0.819              & 0.831           & 0.904               &     0.986   & 1.052  & 1.035   \\
			$\rm Eb~C + 1\sigma                  $    &  0.819              & 0.833           & 0.904               &     0.984   & 1.052  & 1.032   \\
			$\rm Eb~O - 1\sigma                  $    &  0.818              & 0.832           & 0.904               &     0.984   & 1.054  & 1.034   \\
			$\rm Eb~O + 1\sigma                  $    &  0.820              & 0.832           & 0.904               &     0.986   & 1.051  & 1.034   \\
			$\rm MEC~norm~C - 1\sigma            $    &  0.819              & 0.764           & 0.904               &     1.072   & 1.052  & 1.121   \\
			$\rm MEC~norm~C + 1\sigma            $    &  0.819              & 0.900           & 0.904               &     0.910   & 1.052  & 0.959   \\
			$\rm MEC~norm~O - 1\sigma            $    &  0.754              & 0.832           & 0.904               &     0.906   & 1.139  & 1.034   \\
			$\rm MEC~norm~O + 1\sigma            $    &  0.884              & 0.832           & 0.904               &     1.063   & 0.978  & 1.034   \\
			$\rm CCcoh~norm~C - 1\sigma          $    &  0.819              & 0.809           & 0.904               &     1.013   & 1.052  & 1.061   \\
			$\rm CCcoh~norm~C + 1\sigma          $    &  0.819              & 0.855           & 0.904               &     0.958   & 1.052  & 1.007   \\
			$\rm CCcoh~norm~O - 1\sigma          $    &  0.800              & 0.832           & 0.904               &     0.962   & 1.076  & 1.034   \\
			$\rm CCcoh~norm~O + 1\sigma          $    &  0.838              & 0.832           & 0.904               &     1.007   & 1.030  & 1.034   \\
			$\rm CCcoh~norm~Fe - 1\sigma         $    &  0.819              & 0.832           & 0.896               &     0.958   & 1.043  & 1.023   \\
			$\rm CCcoh~norm~Fe + 1\sigma         $    &  0.819              & 0.832           & 0.913               &     0.958   & 1.062  & 1.044   \\
			\hline                                                                                                         
			\hline                                                                                                         
		\end{tabular}}
\end{table}

\section{Conclusion}

For the precise measurement of neutrino oscillation parameters, understanding of neutrino interactions with nuclei is essential. 
We reported measurements of the flux-integrated $\nu_{\mu}$ charged-current cross sections on water, hydrocarbon, iron, and their ratios in the T2K on-axis neutrino beam with a mean neutrino energy of 1.5~GeV in a restricted phase space for the kinematics of the induced muon with $\theta_{\mu}<$45$^{\circ}$ and $p_{\mu}>$0.4~GeV/$c$ in the laboratory frame. This is the most precise measurement to date of neutrino cross sections on water in this energy region and the first measurement of neutrino cross section ratios of water-to-hydrocarbon and water-to-iron. The results agree with current neutrino interaction models used in the T2K oscillation analysis within their uncertainties.

%
%

\end{document}